\documentclass[final,3p]{elsarticle}

\biboptions{numbers,sort&compress}

\usepackage{amssymb}
\usepackage{amsmath}
\usepackage{amsthm}
\usepackage{soul}
\usepackage{bm}
\usepackage{graphicx}
\usepackage{color}
\usepackage{scalerel}
\usepackage{subfig}
\usepackage{appendix}
\usepackage[scr=rsfs]{mathalpha} 
\usepackage{hyperref}

\allowdisplaybreaks[4]

\journal{}

\def\x{{\bm x}}

\def\bxi{{\bm{\xi}}}
\def\d{{\mathrm{d}}}
\def\u{{\bm u}}
\def\v{{\bm v}}

\def\W{{\bm W}}

\soulregister\ref7
\soulregister\citealt7

\begin{document}

\begin{frontmatter}

\title{Multiscale Kinetic Methods for Nonequilibrium Flow and Transport}

\author[inst1a,inst1b]{Zhaoli Guo}
\ead{zlguo@hust.edu.cn}

\affiliation[inst1a]{department={Institute of Multidisciplinary Research for Mathematics and Applied Science,},
                   organization={Huazhong University of Science and Technology},
                   city={Wuhan},
                   postcode={430074},
                   country={China}}                   
\affiliation[inst1b]{department={State Key Laboratory of Coal Combustion,},
                   organization={Huazhong University of Science and Technology},
                   city={Wuhan},
                   postcode={430074},
                   country={China}}                   

\author[inst2]{Kun Xu}
\ead{makxu@ust.hk}

\affiliation[inst2]{organization={Department of Mathematics, Hong Kong University of Science and Technology},
                   addressline={Clear Water Bay},
                   city={Hong Kong},
                   country={China}}

\begin{abstract}
Nonequilibrium flow and transport problems are inherently multiscale. Kinetic theory provides a fundamental physical basis for describing such phenomena, since it connects microscopic transport and interaction processes with emergent macroscopic behavior across regimes. In many situations, however, continuum descriptions lose validity in parts of the domain, whereas fully resolved kinetic descriptions become prohibitively expensive when numerical resolution remains tied to the smallest collision scales. Over the past two decades, a broad class of multiscale kinetic methods has therefore been developed to bridge rarefied, transitional, and continuum regimes in gas dynamics and other carrier-based transport systems. Existing reviews have clarified important parts of this field, including general numerical methods for kinetic equations, asymptotic-preserving methodology, and specific method families. 
This review adopts a different perspective by examining the subject through four interacting layers: numerical methods, computational strategies, asymptotic properties, and framework-level formulations. It surveys the main developments along these lines and emphasizes the common principles that connect them, including transport-interaction coupling, asymptotic consistency, scale-adaptive representation, and scale-dependent physical description. From this perspective, multiscale kinetic computation has evolved into a broader transport methodology for nonequilibrium systems across scales.
\end{abstract}

%


\end{frontmatter}

\section{Introduction}
\label{sec:introduction}

Nonequilibrium flow and transport phenomena involve a wide range of
spatial and temporal scales, and their effective description depends
on the scale of observation. In gas flows, molecular mean free paths
and relaxation times may be much smaller than the corresponding
macroscopic scales in some regions but comparable to them in others.
The near-equilibrium assumptions underlying classical continuum
descriptions may therefore be valid locally but fail elsewhere
\cite{ChapmanCowling1970,Cercignani1988}.
Near-continuum and strongly nonequilibrium regions can coexist within
a single flow. These issues arise in hypersonic aerodynamics, gas
transport in micro- and nanoscale devices, and vacuum systems
\cite{Bird1994,Guo_Xu_AIA2021_DUGKSReview}.
Related multiscale challenges also arise in radiative transfer,
phonon heat conduction, neutron transport, plasma dynamics, and
gas--particle systems
\cite{ZhuXu2021UGKSReview,Guo_Xu_AIA2021_DUGKSReview}.

Rarefied-gas dynamics has motivated much of the development of multiscale kinetic computation. The degree of gas rarefaction is commonly characterized by the Knudsen number $\mathrm{Kn}={\lambda}/{L}$,
where $\lambda$ is the molecular mean free path and $L$ is a characteristic macroscopic length. When $\mathrm{Kn}\ll 1$, frequent molecular collisions rapidly restore local thermodynamic equilibrium, and macroscopic descriptions based on the Euler or Navier--Stokes--Fourier (NSF) equations are usually adequate. As $\mathrm{Kn}$ increases, however, the stress and heat flux can no longer be represented reliably by Newton's law of viscosity and Fourier's law of heat conduction, and continuum models progressively lose predictive capability \cite{ChapmanCowling1970,Cercignani1988,ref:Struchtrup2004}. Rarefied-gas flow thus provides the classical context in which the essential multiscale difficulty becomes most transparent, i.e., molecular free transport, collisional relaxation, and macroscopic flow evolution may all influence the dynamics, but they operate on different characteristic scales and may dominate in different regimes.

Kinetic theory provides the corresponding mesoscopic description. For a dilute monatomic gas, the fundamental model is the Boltzmann equation \cite{ChapmanCowling1970,Cercignani1988}
\begin{equation}
	\partial_t f+\bm{\xi}\cdot\nabla f = Q_B(f,f),
	\label{eq:boltzmann_intro}
\end{equation}
where $f(\bm{x},\bm{\xi}, t)$ is the distribution function at time $t$, position $\bm{x}$, and molecular velocity $\bm{\xi}$, and $Q_B(f,f)$ denotes the full Boltzmann collision operator. The conservative macroscopic variables are defined by
\begin{equation}
	\bm{W}=(\rho,\rho\bm{u},\rho E)^T
	=\langle \bm{\psi} f\rangle,
	\label{eq:macro_variables_intro}
\end{equation}
where
\(\bm{\psi}=\left(1,\bm{\xi},\frac{1}{2}|\bm{\xi}|^2\right)^T\)
is the vector of collision invariants, and \(\rho\), \(\bm{u}\), and \(E\)
denote the density, flow velocity, and total specific energy, respectively.
Here \(\langle\cdot\rangle\) denotes integration over velocity space. The
specific internal energy is \(e=c_vT=E-|\bm{u}|^2/2\), where \(T\) is the
temperature and \(c_v\) is the specific heat at constant volume. The pressure
is given by \(p=\rho RT\), with \(R\) being the gas constant. The stress tensor
\(\bm{\sigma}\) and heat flux \(\bm{q}\) are defined as
\begin{equation}
	\bm{\sigma}=\langle \bm{C}\bm{C} f\rangle-p\bm{I},
	\qquad
	\bm{q}=\left\langle \frac{1}{2}|\bm{C}|^2\bm{C} f\right\rangle,
	\label{eq:stress_heat_intro}
\end{equation}
where \(\bm{C}=\bm{\xi}-\bm{u}\) is the peculiar velocity.

Taking moments of Eq.~\eqref{eq:boltzmann_intro} yields the exact conservative moment system
\begin{equation}
	\partial_t \bm{W}+\nabla\cdot \bm{\mathcal F}(f)=0,
	\label{eq:moment_system_intro}
\end{equation}
where the macroscopic flux is given by
\begin{equation}
	\bm{\mathcal F}(f)=\langle \bxi\bm{\psi} f\rangle =
	\left(
	\rho\bm{u},
	\rho\bm{u}\bm{u}+p\bm{I}+\bm{\sigma},
	(\rho E+p)\bm{u}+\bm{\sigma}\cdot\bm{u}+\bm{q}
	\right)^T .
	\label{eq:moment_system_flux}
\end{equation}
This system is exact but unclosed: the flux depends on the nonequilibrium moments $\bm{\sigma}$ and $\bm{q}$, which are determined by the kinetic distribution rather than by $\bm{W}$ alone. At local Maxwellian equilibrium, $\bm{\sigma}=0$ and $\bm{q}=0$, and Eq.~\eqref{eq:moment_system_intro} reduces to the Euler flux. In the near-equilibrium regime, the first-order kinetic correction gives the NSF constitutive laws. Away from local equilibrium, however, the same stress and heat-flux moments may be nonlocal, nonlinear, and history dependent, which is precisely why kinetic descriptions are needed.

For asymptotic and numerical discussions, it is often useful to introduce a dimensionless collisional stiffness or rarefaction parameter $\varepsilon$, leading to the scaled Boltzmann equation
\begin{equation}
	\partial_t f^{\varepsilon}+\bm{\xi}\cdot\nabla f^{\varepsilon}
	=\frac{1}{\varepsilon}Q_B(f^{\varepsilon},f^{\varepsilon}).
	\label{eq:scaled_boltzmann_intro}
\end{equation}
Here $\varepsilon$ may be regarded, depending on the nondimensionalization and asymptotic regime, as a scaled Knudsen number or collision-time parameter. Equation~\eqref{eq:scaled_boltzmann_intro} makes explicit the stiffness that appears when collisions occur on a much shorter scale than the macroscopic evolution. 

In principle, the Boltzmann equation provides a unified physical description from the free-molecular to the continuum regime. In practice, however, direct numerical solution is prohibitively expensive in many applications because the unknown depends on time, physical space, and velocity space, while the collision operator is nonlinear and costly to evaluate. The central numerical challenge is therefore not simply how to solve a kinetic equation faithfully on fully resolved kinetic scales, but how to compute reliably when the mesh size and time step are much larger than the mean free path and collision time. Classical deterministic discrete-velocity methods (DVMs) \cite{DimarcoPareschi2014,YangLiShu2022DVMReview} and the direct simulation Monte Carlo (DSMC) method \cite{Bird1994} remain foundational, but both encounter severe bottlenecks in near-continuum regimes when transport and collision are numerically decoupled. In such cases, the physical solution may vary smoothly on macroscopic scales, while the numerical method still has to resolve kinetic scales. This mismatch between physical smoothness and kinetic-scale resolution is the basic motivation for multiscale kinetic methods.

Over the past two decades, multiscale kinetic modeling and computation
have developed along several complementary directions, and the resulting
literature has been organized from different perspectives. General
surveys of kinetic numerics cover semi-Lagrangian, discrete-velocity,
spectral, asymptotic-preserving (AP), and hybrid methods
\cite{DimarcoPareschi2014}. A second body of work focuses on AP
formulations and on discretizations that remain consistent with the
asymptotic transition from kinetic to macroscopic models
\cite{Jin_1999_AP,Hu_Jin_Li_AP_2017,Jin2022AN}. More specialized reviews
and representative studies address individual method families, including
the unified gas-kinetic scheme (UGKS)
\cite{XuHuang2010UGKS,ZhuXu2021UGKSReview}, the discrete unified
gas-kinetic scheme (DUGKS)
\cite{GuoXuWang2013DUGKS,GuoWangXu2015DUGKS,Guo_Xu_AIA2021_DUGKSReview},
particle-based hybrid and multiscale methods
\cite{ZhangJohnPfeifferFeiWen2019ParticleReview}, the general synthetic
iterative scheme (GSIS) \cite{ZengZhangLiSuWu2026GSISRev},
all-Knudsen-number discrete-velocity methods
\cite{YangLiShu2022DVMReview}, and DUGKS-based formulations for
multiscale heat conduction \cite{ZhangGuo2025APS}. Together, these
studies establish much of the methodological background for modern
multiscale kinetic computation.

The present review adopts a different perspective. Rather than centering on a single solver family or a single asymptotic concept, it organizes multiscale kinetic computation into four related aspects. The first is the \emph{method} layer, concerned with how kinetic states are represented and evolved, for example through deterministic velocity-space discretization or stochastic particles. The second is the \emph{strategy} layer, concerned with how multiscale information is organized across levels, including hybrid, macro--micro, synthetic, and wave--particle formulations. The third is the \emph{property} layer, concerned with how multiscale fidelity is assessed, most notably through AP and unified-preserving (UP) viewpoints \cite{GuoLiXu2023UP}. The fourth is the \emph{framework} layer, concerned with the physical description itself, most clearly represented here by the unified gas-kinetic framework (UGKF) \cite{GuoZhuXu_AA2026}. These distinctions help clarify both the diversity of the field and the fact that different approaches often address different multiscale questions rather than directly competing on identical terms.

This layered perspective also broadens the scope of the subject. Although multiscale kinetic computation was developed first and most systematically for rarefied-gas dynamics, closely related ideas now reappear in radiative transfer, phonon transport, neutron transport, plasma kinetics, electron--phonon transport, gas--particle systems, and emerging kinetic viewpoints on turbulence. These extensions are therefore treated not as isolated peripheral developments, but as evidence that multiscale kinetic computation has evolved into a broader transport methodology for nonequilibrium systems across scales.

The remainder of the review is organized accordingly. Secs.~\ref{sec:deterministic_methods} and~\ref{sec:particle_methods} begin at the method level by examining deterministic and particle representations of kinetic states. Secs.~\ref{sec:hybrid_macro_micro} and~\ref{sec:wave_particle_collision} move to higher-level multiscale strategies, including hybrid, macro--micro, synthetic, and wave--particle organizations. Sec.~\ref{sec:ap_up} turns from construction to assessment through the AP and UP viewpoints. Sec.~\ref{sec:ugkf} raises the discussion to the framework level through UGKF, where the physical observation scale enters the description itself. Sec.~\ref{sec:extensions_other_transport} broadens the scope from rarefied-gas dynamics to other nonequilibrium transport systems. Sec.~\ref{sec:kinetic_turbulence} then discusses kinetic approaches to turbulence as a related but distinct extension, and Sec.~\ref{sec:summary_outlook} concludes with a summary and outlook. The literature is now too broad for any single review to be exhaustive. The present article is therefore selective but systematic. Its purpose is to clarify the physical and numerical principles that connect different multiscale developments, and to show how multiscale kinetic computation has evolved from a specialized response to rarefied-gas simulation into a broader methodology for nonequilibrium transport across scales.


\section{Deterministic kinetic methods}
\label{sec:deterministic_methods}
Deterministic kinetic methods provide the most direct numerical route from the mesoscopic description to concrete all-regime computation. Their role is not limited to velocity-space discretization. The central issue is how a discrete kinetic solver represents free transport, collision, and macroscopic asymptotics when the numerical mesh and time step are much larger than the mean free path and collision time.

This section focuses on deterministic methods designed explicitly for multiscale computation, rather than on deterministic kinetic numerics in general. We first use the classical discrete velocity method (DVM) as the baseline, because it makes the limitations of collisionless flux reconstruction and stiff relaxation most transparent. We then review UGKS and DUGKS as two representative direct-modeling approaches in which the interfacial evolution couples free transport and collision over the numerical time step. The section concludes with related deterministic branches, including improved DVMs, adaptive partitioning, and alternative multiscale flux or time-integration designs.

\subsection{Classical discrete velocity methods}
\label{subsec:classical_dvm}
Classical DVMs form the natural starting point for deterministic multiscale kinetic methods. They convert the velocity dependence of a kinetic equation into a finite set of discrete-velocity equations and then apply standard spatial and temporal discretizations in physical space. This construction is simple and flexible, but it also exposes the basic multiscale difficulty: if transport and collision are treated as essentially separate numerical processes, the method remains tied to kinetic scales even when the underlying solution varies mainly on macroscopic scales.

The formulation may start directly from the Boltzmann equation~\eqref{eq:boltzmann_intro}, or from a model kinetic equation in which the full collision integral is replaced by a one-argument operator. In this subsection we write such model equations in the generic form
\[
\partial_t f+\bm{\xi}\cdot\nabla f= Q(f).
\]
The simplest and most widely used relaxation prototype is the BGK model \cite{BGK1954},
\begin{equation}
	Q(f)=\frac{g-f}{\tau},
	\label{eq:bgk_model}
\end{equation}
where $g$ is the local equilibrium distribution and $\tau$ is the relaxation time. For a monatomic gas, $g$ is the Maxwellian equilibrium,
\begin{equation}
	g=g_M(\bm{\xi};\rho,\bm{u},T)
	\equiv
	\frac{\rho}{(2\pi R T)^{D/2}}
	\exp\!\left(
	-\frac{|\bm{\xi}-\bm{u}|^2}{2RT}
	\right),
	\label{eq:maxwellian_eq}
\end{equation}
with the macroscopic variables defined by Eqs.~\eqref{eq:macro_variables_intro} and \eqref{eq:stress_heat_intro}. More refined relaxation models, such as the Shakhov and ES-BGK equations, modify the target equilibrium so as to recover correct transport coefficients, especially the Prandtl number \cite{Shakhov1968,Holway1966ESBGK}.

In a DVM, the velocity space is replaced by a finite set of quadrature points $\{\bm{\xi}_\alpha\}_{\alpha=1}^{N_v}$. Denoting $f_\alpha(\bm{x},t)=f(\bm{x},\bm{\xi}_\alpha,t)$, one obtains
\begin{equation}
	\partial_t f_\alpha + \bm{\xi}_\alpha\cdot\nabla f_\alpha = Q_\alpha,
	\qquad \alpha=1,\ldots,N_v .
	\label{eq:dvm_discrete}
\end{equation}
The conservative variables are then evaluated from the same collision invariants as in Eq.~\eqref{eq:macro_variables_intro}, but with the velocity integral replaced by quadrature,
\begin{equation}
	\bm{W}
	\approx
	\sum_{\alpha=1}^{N_v}
	w_\alpha \bm{\psi}_\alpha f_\alpha,
	\qquad
	\bm{\psi}_\alpha=
	\left(
	1,\bm{\xi}_\alpha,\frac{1}{2}|\bm{\xi}_\alpha|^2
	\right)^T,
	\label{eq:dvm_moment}
\end{equation}
where $w_\alpha$ are the quadrature weights. Equation~\eqref{eq:dvm_moment} is therefore the discrete counterpart of the moment relation in Eq.~\eqref{eq:macro_variables_intro}.

For a finite-volume discretization, the physical domain is partitioned into control volumes (or cells) $V_i$ centered at $\bm{x}_i$, with volume $|V_i|$ for $i=1,2,\ldots,N_x$. The interface shared by $V_i$ and a neighboring cell $V_j$ is denoted by $ij$, with interface center $\bm{x}_{ij}$, outward unit normal $\bm{n}_{ij}$ with respect to $V_i$, and interface area $S_{ij}$. The cell-averaged discrete distribution, $f_{i,\alpha}(t)=|V_i|^{-1}\int{f(\x,\bxi_\alpha,t)} d\x$, then satisfies the generic update
\begin{equation}
	f_{i,\alpha}^{n+1}
	=
	f_{i,\alpha}^{n}
	-
	\frac{\Delta t}{|V_i|}
	\sum_{j\in N(i)}
	(\bm{\xi}_\alpha\cdot \bm{n}_{ij}) f_{ij,\alpha} S_{ij}
	+
	\Delta t\, Q_{i,\alpha}^*,
	\label{eq:dvm_fv_update}
\end{equation}
where the superscript $n$ denotes time $t_n$, $\Delta t=t_{n+1}-t_n$ is the time step, $f_{ij,\alpha}$ is the reconstructed interfacial distribution, and $Q_{i,\alpha}^*$ is the numerical cell-averaged collision term.

The multiscale limitation of a classical DVM is already visible in Eq.~\eqref{eq:dvm_fv_update}. If $f_{ij,\alpha}$ is obtained from a purely collisionless upwind reconstruction, the numerical flux represents free transport alone. This is suitable in highly rarefied regimes, but it introduces excessive dissipation and incorrect continuum-scale transport when collisions strongly shape the interfacial evolution. In addition, an explicit treatment of $Q_{i,\alpha}^*$ imposes a time-step restriction tied to the relaxation time. Thus the conventional DVM provides an accurate deterministic kinetic baseline in rarefied regimes, but it becomes inefficient, and often insufficiently accurate, as the flow approaches the continuum limit \cite{DimarcoPareschi2014,YangLiShu2022DVMReview}.

This observation explains the two main deterministic routes developed later. One retains the DVM structure and improves the treatment of stiff relaxation through implicit, semi-implicit, exponential, or AP discretizations. The other modifies the interface evolution itself so that transport and collision are coupled within the numerical flux. The latter route leads directly to UGKS and DUGKS, where the finite-volume update is built from a local multiscale kinetic evolution rather than from collisionless upwinding alone.

\subsection{Unified gas-kinetic scheme (UGKS)}
\label{subsec:ugks}

The UGKS, originally proposed by Xu and Huang, is one of the most representative deterministic multiscale kinetic methods for all-Knudsen-number flows \cite{XuHuang2010UGKS}. More specifically, UGKS is constructed following the direct-modeling philosophy: instead of assuming that one fixed governing equation should simply be discretized on all numerical meshes, the local flow evolution is modeled directly on the numerical space--time scale itself \cite{ref:XuBook,ZhuXu2021UGKSReview}.

Two ingredients are essential to this viewpoint. First, the method must preserve the fundamental conservation laws of mass, momentum, and energy at the discrete level. Second, the microscopic physics must be incorporated through a relaxation process toward equilibrium, so that dissipation and entropy production arise from kinetic evolution rather than from a purely numerical correction. In UGKS, this direct-modeling idea is realized through a multiscale interface flux, in which free transport and collision are coupled over a single numerical time step. UGKS should therefore be understood not merely as a discretization of a BGK-type equation, but as a scale-adaptive construction of flow physics on the discretization scale.

For the BGK model \eqref{eq:bgk_model}, the evolution of the cell-averaged distribution function and conservative variables in a control volume $V_i$ can be written as
\begin{equation}
	f_i^{n+1}
	=
	f_i^n
	-
	\frac{1}{|V_i|}
	\int_{t_n}^{t_{n+1}}
	\int_{\partial V_i}
	(\bm{\xi}\cdot \bm{n}) f\, ds\, dt
	+
	\frac{1}{|V_i|}
	\int_{t_n}^{t_{n+1}}
	\int_{V_i}
	\frac{g-f}{\tau}\, dV\, dt,
	\label{eq:ugks_f_update}
\end{equation}
and
\begin{equation}
	\bm{W}_i^{\,n+1}
	=
	\bm{W}_i^{\,n}
	-
	\frac{1}{|V_i|}
	\int_{t_n}^{t_{n+1}}
	\int_{\partial V_i}
	\int
	(\bm{\xi}\cdot \bm{n}) \bm{\psi} f\, d\bm{\xi}\, ds\, dt,
	\label{eq:ugks_W_update}
\end{equation}
where Eq. \eqref{eq:ugks_W_update} is obtained by taking the moments of Eq. \eqref{eq:ugks_f_update}, and the conservative property of the collision term is employed.
These equations show that the central issue is the construction of the time-dependent interface distribution function and, more specifically, its time average over one numerical time step.

The UGKS determines the interface distribution from the integral solution of the kinetic model equation along the characteristic line. For the interface centered at $\bm{x}_{ij}$ and time $t\in[t_n,t_{n+1}]$, the local analytic solution reads
\begin{equation}
	f(\bm{x}_{ij},\bm{\xi},t)
	=
	\frac{1}{\tau}
	\int_{t_n}^{t}
	g(\bm{x}',\bm{\xi},t')
	e^{-(t-t')/\tau}\, dt'
	+
	e^{-(t-t_n)/\tau}
	f_0(\bm{x}_{ij}-\bm{\xi}(t-t_n),\bm{\xi}),
	\label{eq:ugks_integral_solution}
\end{equation}
where $\bm{x}'=\bm{x}_{ij}-\bm{\xi}(t-t')$, $\tau$ is frozen locally over the integration interval, and $f_0$ is the reconstructed distribution function at $t_n$ around the interface. Equation~\eqref{eq:ugks_integral_solution} explicitly contains two physically distinct contributions, i.e., the equilibrium-driven hydrodynamic part and the free-transport kinetic part. Their relative importance is determined dynamically by the ratio $\Delta t/\tau$.

To derive a practical second-order scheme, both the initial distribution and the post-collision equilibrium state are approximated locally around the interface. For the initial distribution, UGKS adopts a piecewise linear reconstruction with left and right states,
\begin{equation}
	f_0(\bm{x},\bm{\xi})
	=
	\left(
	f_0^L(\bm{x}_{ij})+\Delta \bm{x}\cdot \nabla f_0^L
	\right)
	\left(1-H[\Delta \bm{x}\cdot \bm{n}_{ij}]\right)
	+
	\left(
	f_0^R(\bm{x}_{ij})+\Delta \bm{x}\cdot \nabla f_0^R
	\right)
	H[\Delta \bm{x}\cdot \bm{n}_{ij}],
	\label{eq:ugks_f0_recon}
\end{equation}
where $\Delta \bm{x}=\bm{x}-\bm{x}_{ij}$ and $H$ is the Heaviside function. For the equilibrium state, a first-order Taylor expansion around the interface is used:
\begin{equation}
	g(\bm{x},\bm{\xi},t) = g_0(\bm{\xi}) \left[
	1 + (1-H[\bar{x}])a^L\bar{x} + H[\bar{x}]a^R\bar{x}	+ b\bar{y} + c\bar{z} + A(t-t_n)
	\right],
	\label{eq:ugks_g_expansion}
\end{equation}
where $g_0$ is the interface equilibrium at $t_n$, $(\bar{x},\bar{y},\bar{z})$ are local coordinates centered at the interface, and the coefficients $a^L$, $a^R$, $b$, $c$, and $A$ are determined from the spatial and temporal derivatives of the macroscopic variables through compatibility relations \cite{XuHuang2010UGKS,ZhuXu2021UGKSReview}.

Substituting Eqs.~\eqref{eq:ugks_f0_recon}--\eqref{eq:ugks_g_expansion} into Eq.~\eqref{eq:ugks_integral_solution} yields an explicit expression for the interface distribution function. The corresponding time-averaged interface distribution for a discrete velocity $\bxi_\alpha$,
\begin{equation}
	\bar{f}_{ij,\alpha}
	=
	\frac{1}{\Delta t}
	\int_{t_n}^{t_{n+1}}
	f(\bm{x}_{ij},\bm{\xi}_\alpha,t)\, dt,
	\label{eq:ugks_time_avg_f}
\end{equation}
defines the numerical microscopic flux
\begin{equation}
	F_{ij,\alpha}
	=
	(\bm{\xi}_\alpha\cdot \bm{n}_{ij})
	\bar{f}_{ij,\alpha},
	\label{eq:ugks_micro_flux}
\end{equation}
and the macroscopic flux
\begin{equation}
	\bm{\mathcal F}_{ij}
	=
	\sum_\alpha w_\alpha
	\bm{\psi}_\alpha
	F_{ij,\alpha}.
	\label{eq:ugks_macro_flux}
\end{equation}

The updating of the distribution function inside each cell also adopts an implicit treatment of the collision term. With the trapezoidal discretization of the relaxation source,
\begin{equation}
	Q_{i,\alpha}^*
	=
	\frac{1}{2}
	\left[
	\frac{g_{i,\alpha}^{n+1}-f_{i,\alpha}^{n+1}}{\tau_i^{n+1}}
	+
	\frac{g_{i,\alpha}^n-f_{i,\alpha}^n}{\tau_i^n}
	\right],
	\label{eq:ugks_trapezoidal_collision}
\end{equation}
the updated distribution function can be written as
\begin{equation}
	f_{i,\alpha}^{n+1} = \frac{2\tau_i^{n+1}}{\Delta t+2\tau_i^{n+1}}
	\left[f_{i,\alpha}^n -\frac{\Delta t}{|V_i|} \sum_{j\in N(i)}F_{ij,\alpha} S_{ij} + \frac{\Delta t}{2}Q_{i,\alpha}^n +
	\frac{\Delta t}{2\tau_i^{n+1}}g_{i,\alpha}^{n+1} \right].
	\label{eq:ugks_implicit_update}
\end{equation}
Therefore, in UGKS the conservative variables and the gas distribution function are updated simultaneously, and the macroscopic state at the new time level is directly used in the implicit collision treatment. Specifically, the conservative update supplies the new macroscopic state, from which $g^{n+1}$ and $\tau^{n+1}$ are evaluated for the distribution update. The macroscopic and microscopic updates thus use the same interface flux moments.

The multiscale nature of UGKS follows immediately from Eq.~\eqref{eq:ugks_integral_solution}. When $\Delta t \ll \tau$, the exponential damping is weak and the interface solution is dominated by the transported initial distribution, so the scheme behaves like a kinetic solver for rarefied flows. When $\Delta t \gg \tau$, the nonequilibrium part is rapidly damped and the interface solution is dominated by the equilibrium contribution, so the scheme approaches the hydrodynamic limit. This direct coupling between transport and collision in the interface flux is the essential reason why UGKS can bridge rarefied and continuum regimes within a single framework.

Beyond the original explicit formulation, the UGKS family has developed along several closely related computational directions, with the common objective of extending the direct-modeling flux construction to practical large-scale multiscale simulations \cite{ZhuXu2021UGKSReview}. These developments do not change the essential mechanism of UGKS, namely, the interface flux is still constructed from a local kinetic evolution in which free transport and collision are coupled over a finite time interval. Rather, they aim to reduce the cost associated with stiffness, phase-space resolution, memory demand, and parallel implementation.

A first major direction is implicit acceleration. For steady-state computation, Zhu \emph{et al.} developed an implicit UGKS in which the macroscopic conservative variables are first updated implicitly to predict the equilibrium state, and the microscopic distribution function is then advanced by a fully implicit lower-upper symmetric Gauss-Seidel (LU-SGS) iteration \cite{ZhuZhongXu2016}. This macro--micro implicit strategy accelerates convergence in near-continuum regimes while retaining the kinetic description where it is needed. For unsteady flows, Zhu \emph{et al.} further proposed an implicit UGKS (IUGKS) that relaxes the restriction imposed by a global CFL time step \cite{ZhuZhongXu2019}. A key point in this extension is that the semi-discrete formulation must still use the time-averaged multiscale flux over a local physical time interval. If this time-averaged flux is replaced by a purely instantaneous or collisionless flux, the scheme loses the transport--collision coupling that distinguishes UGKS from a single-scale kinetic solver.

A second direction concerns adaptive reduction of the kinetic degrees of freedom. Since the nonequilibrium part of the distribution is significant only in localized regions for many multiscale flows, adaptive physical meshes, adaptive velocity-space discretizations, and hybrid continuous/discrete velocity descriptions have been introduced to concentrate kinetic resolution where it is most needed \cite{ZhuXu2021UGKSReview}. These techniques reduce the cost in near-equilibrium regions while preserving the original multiscale flux construction in strongly nonequilibrium zones. They are therefore particularly useful for flows in which rarefied or transitional structures occupy only a small portion of an otherwise continuum-dominated domain.

A third direction is memory reduction and parallel scalability. Because UGKS resolves both physical space and velocity space, its computational burden is determined not only by the number of physical cells, but also by the number of discrete velocities stored and updated in each cell. Memory-reduction strategies, physical- and velocity-space domain decomposition, and large-scale parallel implementations have therefore become essential components of practical UGKS algorithms \cite{ZhuXu2021UGKSReview}. Together with implicit and adaptive techniques, these developments have transformed UGKS from a conceptual all-regime kinetic scheme into a computational framework applicable to increasingly realistic multiscale flow simulations.

\subsection{Discrete unified gas-kinetic scheme (DUGKS)}
\label{subsec:dugks}

The DUGKS provides a compact finite-volume realization of the same multiscale idea underlying UGKS \cite{GuoXuWang2013DUGKS,GuoWangXu2015DUGKS}. It retains the central direct-modeling principle that the numerical flux should contain the coupled effect of free transport and relaxation over the numerical time scale. The difference is mainly in the construction of the interfacial state. Instead of using the full time-dependent integral solution as in UGKS, DUGKS reconstructs the distribution at the half time level from a discrete characteristic solution of the kinetic equation \cite{Guo_Xu_AIA2021_DUGKSReview}. This gives a simpler and more compact algorithm while preserving the multiscale coupling needed for all-regime computation.

With the discrete velocity set introduced in Sec.~\ref{subsec:classical_dvm}, the finite-volume update of the cell-averaged distribution can be written as
\begin{equation}
	f_{i,\alpha}^{n+1}-f_{i,\alpha}^{n}
	+
	\frac{\Delta t}{|V_i|}F_{i,\alpha}^{\,n+1/2}
	=
	\frac{\Delta t}{2}
	\left(Q_{i,\alpha}^{n+1}+Q_{i,\alpha}^{n}\right),
	\label{eq:dugks_f_update}
\end{equation}
where the collision term is integrated by the trapezoidal rule, and $F_{i,\alpha}^{\,n+1/2}$ denotes the net microscopic flux through the boundary of $V_i$ at the half time level. The key step is therefore the evaluation of the interface distribution at $t_{n+1/2}=t_n+h$, with $h=\Delta t/2$.

Integrating the kinetic equation along the characteristic line from $(\bm{x}_{ij}-\bm{\xi}_\alpha h,t_n)$ to $(\bm{x}_{ij},t_n+h)$ and approximating the collision term by the trapezoidal rule gives
\begin{equation}
	f(\bm{x}_{ij},\bm{\xi}_\alpha,t_n+h)
	-
	f(\bm{x}_{ij}-\bm{\xi}_\alpha h,\bm{\xi}_\alpha,t_n)
	=
	\frac{h}{2}
	\left[
	Q(\bm{x}_{ij},\bm{\xi}_\alpha,t_n+h)
	+
	Q(\bm{x}_{ij}-\bm{\xi}_\alpha h,\bm{\xi}_\alpha,t_n)
	\right].
	\label{eq:dugks_characteristic}
\end{equation}
The apparent implicitness is removed by introducing the half-step auxiliary distributions
\begin{equation}
	\bar{f}=f-\frac{h}{2}Q,
	\qquad
	\bar{f}^{+}=f+\frac{h}{2}Q.
	\label{eq:dugks_aux_half}
\end{equation}
Then Eq.~\eqref{eq:dugks_characteristic} becomes the explicit characteristic relation
\begin{equation}
	\bar{f}(\bm{x}_{ij},\bm{\xi}_\alpha,t_n+h)
	=
	\bar{f}^{+}(\bm{x}_{ij}-\bm{\xi}_\alpha h,\bm{\xi}_\alpha,t_n).
	\label{eq:dugks_explicit_half}
\end{equation}
Thus the interfacial state is not obtained from a collisionless upwind reconstruction. It is reconstructed after a half-step evolution in which transport and relaxation have already been combined through the auxiliary distribution.

To evaluate the right-hand side of Eq.~\eqref{eq:dugks_explicit_half}, $\bar{f}^{+}$ is reconstructed from the upwind cell. If $c=c(i,j,\alpha)$ denotes the donor cell determined by the sign of $\bm{\xi}_\alpha\cdot\bm{n}_{ij}$, a typical linear reconstruction is
\begin{equation}
	\bar{f}^{+}(\bm{x}_{ij}-\bm{\xi}_\alpha h,\bm{\xi}_\alpha,t_n)
	=
	\bar{f}_{c,\alpha}^{+,n}
	+
	(\bm{x}_{ij}-\bm{\xi}_\alpha h-\bm{x}_c)\cdot \nabla \bar{f}_{c,\alpha}^{+,n}.
	\label{eq:dugks_reconstruction}
\end{equation}
Because the collision term has vanishing moments with respect to the collision invariants, the conservative variables at the interface are obtained directly from $\bar f$ by the same quadrature rule as in Eq.~\eqref{eq:dvm_moment},
\begin{equation}
	\bm{W}(\bm{x}_{ij},t_n+h)
	=
	\sum_{\alpha} w_\alpha\bm{\psi}_\alpha\,
	\bar{f}(\bm{x}_{ij},\bm{\xi}_\alpha,t_n+h).
	\label{eq:dugks_interface_W}
\end{equation}
For the BGK model, the original distribution at the interface is then recovered as
\begin{equation}
	f(\bm{x}_{ij},\bm{\xi}_\alpha,t_n+h)
	=
	\frac{2\tau}{2\tau+h}
	\bar{f}(\bm{x}_{ij},\bm{\xi}_\alpha,t_n+h)
	+
	\frac{h}{2\tau+h}
	g_\alpha\!\left(\bm{W}(\bm{x}_{ij},t_n+h)\right),
	\label{eq:dugks_interface_f}
\end{equation}
where $\tau$ is evaluated from the interfacial state, and $g_\alpha$ denotes the equilibrium distribution at the discrete velocity $\bm{\xi}_\alpha$.

The net microscopic flux in Eq.~\eqref{eq:dugks_f_update} is then
\begin{equation}
	F_{i,\alpha}^{\,n+1/2}
	=
	\sum_{j\in N(i)}
	(\bm{\xi}_\alpha\cdot\bm{n}_{ij})
	f(\bm{x}_{ij},\bm{\xi}_\alpha,t_n+h)S_{ij}.
	\label{eq:dugks_flux}
\end{equation}
For the full time step, another pair of auxiliary distributions is introduced,
\begin{equation}
	\tilde{f}=f-\frac{\Delta t}{2}Q,
	\qquad
	\tilde{f}^{+}=f+\frac{\Delta t}{2}Q,
	\label{eq:dugks_aux_full}
\end{equation}
which converts the cell-centered update into the explicit form
\begin{equation}
	\tilde{f}_{i,\alpha}^{\,n+1}
	=
	\tilde{f}_{i,\alpha}^{+,n}
	-
	\frac{\Delta t}{|V_i|}F_{i,\alpha}^{\,n+1/2}.
	\label{eq:dugks_explicit_update}
\end{equation}
The conservative variables can be obtained directly from $\tilde{f}$ due to the conservative property of the collision operator,
\begin{equation}
	\bm{W}_i^{n+1}
	=
	\sum_{\alpha} w_\alpha\bm{\psi}_\alpha\,
	\tilde{f}_{i,\alpha}^{\,n+1}.
	\label{eq:dugks_W_update}
\end{equation}
This auxiliary-variable formulation is the source of the compactness of DUGKS: the implicit collision contribution is absorbed into modified distributions, while the transport update remains explicit in form.

The multiscale character of DUGKS follows from Eqs.~\eqref{eq:dugks_explicit_half}--\eqref{eq:dugks_interface_f}. The interfacial distribution is a relaxation-weighted combination of the reconstructed kinetic state and the local equilibrium. When $h\ll \tau$, the coefficient of $\bar f$ is dominant and the method behaves like a kinetic upwind solver. When $h\gg \tau$, the equilibrium contribution dominates and the flux approaches a hydrodynamic one. DUGKS therefore shares with UGKS the essential transport--collision coupling at the interface, although it realizes this coupling through a simpler discrete characteristic half-step rather than through the full integral solution. This point is also central to later AP/UP assessments of DUGKS in Sec.~\ref{sec:ap_up}.

Subsequent developments of DUGKS may be understood according to the numerical limitations they address, including discrete conservation, steady-state convergence, high-order and compact reconstruction, and extension to more complex kinetic models. These developments retain the same half-step characteristic mechanism, but modify how moments, reconstruction, or iteration are organized.

A first line concerns conservation enhancement. Liu \emph{et al.}\ proposed a conserved DUGKS for microchannel gas flows, in which the macroscopic conservative variables are updated through the corresponding macroscopic fluxes rather than only through moment quadrature of auxiliary distributions \cite{LiuCaoChenKongZheng2018ConservedDUGKS}. This improves conservation and long-time robustness. Chen \emph{et al.}\ further extended the conserved formulation to nonisothermal and compressible flows and introduced an unstructured discrete velocity space to reduce the number of velocity points in sparsely populated velocity regions \cite{ChenLiuWangZhong2019ConservedUDVS}. More recently, Zhang \emph{et al.}\ proposed a microscopically conservation-enforced DUGKS (MicroC-DUGKS), in which the conservation property of the collision term is enforced directly at the discrete level; this improves heat-flux evaluation and can reduce the required number of discrete velocities in near-continuum simulations \cite{ZhangLiFangGuo2026MicroC}.

A second line aims at implicit formulation and steady-state acceleration. Pan \emph{et al.}\ developed an implicit DUGKS (IDUGKS) for steady all-regime flows, where the macroscopic variables are first predicted implicitly and the microscopic distribution functions are then advanced by LU-SGS iterations \cite{PanZhongZhuo2019IDUGKS}. The same idea has been extended to more complex kinetic systems. For example, Zhang \emph{et al.}\ developed an implicit DUGKS for steady binary gas mixtures based on a BGK mixture model, using asynchronous microscopic--macroscopic LU-SGS iterations together with characteristic flux reconstruction for both species \cite{ZhangYueZhangSongGuo2025ImplicitMixtureDUGKS}. These methods show that the DUGKS flux construction can be embedded in implicit solvers without abandoning its multiscale interface evolution.

A third line improves accuracy, compactness, and reconstruction efficiency. Wu \emph{et al.}\ proposed a third-order DUGKS based on a two-stage time-stepping strategy and high-order flux reconstruction \cite{WuShiShuChen2018ThirdOrderDUGKS}. Bu \emph{et al.}\ developed a high-order DUGKS with multi-moment constrained conservative semi-Lagrangian reconstruction, in which point values and volume-integrated averages are evolved simultaneously \cite{BuLiFangGuo2025HDUGKS}. Wang \emph{et al.}\ proposed an optimized DUGKS in which the flux is evaluated from nodal distributions rather than only from interface-center values, reducing numerical dissipation and improving stability and efficiency \cite{WangLiangXu2023OptimizedDUGKS}. Zhong \emph{et al.}\ further developed compact DUGKS and compact steady DUGKS formulations, where reconstruction relies only on single-cell information; this compactness is attractive for strongly inhomogeneous transport and large-scale parallel computation \cite{ZhongGuoZhou2026CompactDUGKS}.

DUGKS has also been extended to kinetic models close to the original gas-dynamics formulation, including gas mixtures and dense-fluid or nanoscale transport. Shan \emph{et al.}\ applied DUGKS to strongly inhomogeneous dense-fluid systems by combining BGK relaxation with nonlocal collision effects and mean-field intermolecular interactions \cite{ShanWangZhangGuo2020InhomogeneousDUGKS}. Liu and Guo later developed a more efficient version for the same kinetic model, reducing the cost of the multiple nonlocal integrals from $\mathcal{O}(N N_\sigma)$ to $\mathcal{O}(N)$ and enabling genuinely two-dimensional nanoscale simulations with substantial speedup \cite{LiuGuo2026EfficientNanoDUGKS}, where $N$ and $N_\sigma$ are the grid numbers of the computational domain and integral range, respectively. Extensions of DUGKS-type ideas to phonon, neutron, plasma, and other transport systems are deferred to Sec.~\ref{sec:extensions_other_transport}, where the emphasis shifts from gas-dynamic algorithmic construction to cross-disciplinary portability.

Overall, DUGKS should be viewed as more than a simplified implementation of UGKS. Its discrete characteristic half-step, auxiliary-distribution formulation, and compatibility-based interface moments provide a compact way to realize transport--collision coupling in a finite-volume solver. The later developments reviewed above preserve this core structure while improving conservation, convergence, accuracy, compactness, and applicability to more complex kinetic models.

\subsection{Other deterministic multiscale schemes}
\label{subsec:other_deterministic_refined}

UGKS and DUGKS couple transport and collision directly in the interfacial evolution. Nevertheless, they should be viewed as one major class within a broader family of deterministic multiscale methods. A number of related schemes retain part of the DVM structure, introduce continuum information through macroscopic prediction, or redesign the time integration and flux reconstruction in a different form. These developments demonstrate that the central deterministic issue can be addressed through several distinct strategies, such as modifying the collision treatment, reducing the kinetic region or degrees of freedom, or constructing alternative multiscale fluxes.

For clarity, the discussion is organized into three groups. The first consists of improved DVM-type methods, which retain the conventional discrete-velocity update as much as possible but incorporate continuum-limit information into the flux or collision treatment. The second consists of adaptive deterministic reductions, where the kinetic solver is activated only in cells or variables for which nonequilibrium effects are significant. The third consists of alternative multiscale flux and time-integration designs, which pursue transport--collision coupling through semi-implicit, Lax--Wendroff-type, or exponential formulations rather than through the UGKS integral solution or the DUGKS half-step characteristic reconstruction.

\smallskip
\noindent\textbf{Improved DVM-type methods.}
\smallskip

A representative route parallel to UGKS and DUGKS is the improved discrete velocity method (IDVM). Its aim is to retain the algorithmic simplicity of the classical DVM while reducing its continuum-regime deficiencies. The key idea is to introduce a prediction step for the macroscopic equations, so that the equilibrium state at the new time level can be estimated and the stiff relaxation term can be treated implicitly \cite{YangShuYangChenDong2018}. At the flux level, the method blends a DVM-type kinetic flux with a Navier--Stokes-type macroscopic flux. A representative expression is
\begin{equation}
	\bm{\mathcal F}_{ij}
	=
	\beta\bm{\mathcal F}_{ij,\mathrm{DVM}}
	+
	(1-\beta)\bm{\mathcal F}_{ij,\mathrm{NS}},
	\qquad
	\beta=e^{-\Delta t/\tau},
	\label{eq:idvm_flux_refined}
\end{equation}
where $\bm{\mathcal F}_{ij,\mathrm{DVM}}$ denotes the kinetic DVM contribution, and $\bm{\mathcal F}_{ij,\mathrm{NS}}$ denotes the Navier--Stokes contribution. The exponential weight gives the desired limiting behavior, namely, the DVM contribution dominates when $\Delta t\ll \tau$, whereas the Navier--Stokes contribution dominates when $\Delta t\gg \tau$.

The strategy was later extended to implicit and memory-reduction DVMs for steady all-regime flows. In the memory-reduction formulation, the distribution functions over the whole discrete velocity space need not be stored simultaneously, and the memory requirement can be reduced to the same order as that of a conventional Euler or Navier--Stokes solver \cite{YangShuYangWu2018Memory}. Fully implicit and inner-iteration variants further improve the prediction of the equilibrium state and substantially accelerate convergence in near-continuum and continuum regimes \cite{HanYangLiWuDuShen2023}. These developments show that one can obtain practical all-regime performance without abandoning the DVM framework completely, provided that the continuum-limit response is incorporated into the collision and flux treatment.

\smallskip
\noindent\textbf{Adaptive deterministic reductions.}
\smallskip

A second route is to reduce the kinetic cost adaptively. In many all-regime flows, strong nonequilibrium occupies only a limited part of the domain, while large regions are close enough to local equilibrium to be described by continuum equations. Adaptive partitioning schemes exploit this structure by solving the kinetic equation only where the kinetic correction is needed and by using a continuum solver elsewhere.

A representative example is the adaptive-partitioning DUGKS (ADUGKS). Its cell classification is based on the coefficient of the free-transport contribution in the DUGKS characteristic reconstruction. With the local half time step denoted by $h_i$, the indicator is
\begin{equation}
	\beta_i=\frac{2\tau_i-h_i}{2\tau_i+h_i},
	\label{eq:adugks_beta_refined}
\end{equation}
which is not a collision-survival probability but an algebraic reconstruction coefficient that can be negative. When $\beta_i>0$, the cell retains a significant kinetic contribution and is updated by a modified DUGKS; otherwise it is evolved as a Navier--Stokes cell \cite{YangHanDingLiShuLiu2023}. Because this criterion is derived from the same characteristic formulation that underlies DUGKS, the switching is tied to the local transport--collision balance rather than to an externally prescribed Knudsen-number threshold. In this respect, ADUGKS is a deterministic adaptive-reduction method closely connected to DUGKS, while its broader philosophy also anticipates the adaptive kinetic--fluid strategies discussed later in Sec.~\ref{sec:hybrid_macro_micro}.

\smallskip
\noindent\textbf{Alternative multiscale flux and time-integration designs.}
\smallskip

A third route constructs multiscale fluxes through alternative time-integration or reconstruction strategies. The semi-implicit Richtmyer scheme of Chen \emph{et al.}  \cite{ChenGuoXuLi2020SIR} and the gas-kinetic Lax--Wendroff scheme (GKLWS) of Li \emph{et al.}\ are representative examples \cite{LiFangZhaoTaoMei2022}. Their common feature is that the interface state is not reconstructed by purely collisionless upwinding. Instead, relaxation toward equilibrium is built into the half-step or time-averaged interface distribution, while the overall algorithm remains simpler than the original UGKS construction.

In the semi-implicit Richtmyer framework, the half-step interfacial distribution at the interface centered at $\bm{x}_{ij}$ can be written as
\begin{equation}
	f_{ij,\alpha}^{\,n+1/2}	= \theta_{ij}^{\,n+1/2} f_{ij,\alpha}^{\,n}	
	+ \left(1-\theta_{ij}^{\,n+1/2}\right)g_{ij,\alpha}^{\,n+1/2}-h\,\theta_{ij}^{\,n+1/2}\,
	\bm{\xi}_\alpha\cdot\nabla f_{ij,\alpha}^{\,n},
	\qquad
	\theta_{ij}	=\frac{\tau_{ij}} {\tau_{ij}+h}.
	\label{eq:sir_halfstep_refined}
\end{equation}
Here $h=\Delta t/2$ denotes the half time step used in the interfacial reconstruction, as in the DUGKS. The parameter $\theta_{ij}$ controls the balance between the transported kinetic state and the equilibrium state: the expression approaches a kinetic transport form when $h\ll \tau_{ij}$, and an equilibrium-dominated flux when $h\gg \tau_{ij}$.

The GKLWS expresses the same principle in a Lax--Wendroff form, which starts
from a time expansion of the interface state. For the discrete velocity $\bm{\xi}_\alpha$, the interface distribution adopts its formal first-order expansion over the half step,
\begin{equation}
	f_{ij,\alpha}^{n+1/2}	= f_{ij,\alpha}^n	+
	h\left[	-\bm{\xi}_\alpha\cdot\nabla f_{ij,\alpha}^n + Q_{ij,\alpha}^n\right],
	\label{eq:gklws_halfstep_refined}
\end{equation}
which yields a second-order AP kinetic flux reconstruction \cite{LiFangZhaoTaoMei2022}. A later conservation-moment-based implicit GKLWS couples the discrete-velocity equations with the associated conservation-moment equations, leading to much faster convergence for steady all-Knudsen-number computations \cite{LiFangZhaoWangWen2024}.

Another recent development is the exponential-differencing AP-DVM proposed by Garmirian and Pfeiffer \cite{GarmirianPfeiffer2025}. Instead of using a Crank--Nicolson-type auxiliary-variable transformation as in DUGKS, the relaxation equation is integrated by exponential time differencing. For a relaxation frequency $\nu$ and a target distribution $f_{r}$, the modified distributions may be written as
\begin{equation}
	\hat f	=	e^{-\nu\Delta t}\tilde f	+	\left(1-e^{-\nu\Delta t}\right)f_{r},	\qquad
	f	=	\gamma\tilde f	+	(1-\gamma)f_{r},
	\qquad
	\gamma=\frac{1-e^{-\nu\Delta t}}{\nu\Delta t}.
	\label{eq:ed_dvm_positive_refined}
\end{equation}
The coefficients remain positive for all $\nu\Delta t$, which is attractive for deterministic AP-DVM construction and may also facilitate future coupling with particle-based BGK methods.

These alternative deterministic schemes show that the direct coupling of transport and collision can be pursued in several ways beyond the original UGKS and DUGKS constructions. They also prepare the transition from unified deterministic solvers to broader multiscale organizations, where kinetic, macroscopic, particle, or synthetic descriptions may be combined more explicitly.

\subsection{Section remarks}
\label{subsec:deterministic_remarks}

The deterministic developments reviewed in this section show that the main advance beyond the classical DVM lies not in velocity discretization alone, but in the systematic coupling of transport, collision, and macroscopic asymptotics at the discrete space--time scale. Classical DVM provides the baseline. UGKS and DUGKS represent two influential unified finite-volume realizations, in which the interface evolution incorporates both free transport and relaxation. Other deterministic schemes pursue the same objective through improved DVM formulations, adaptive reduction of kinetic cost, or alternative multiscale flux and time-integration designs.

Some of these ideas have also been extended to transport systems beyond neutral-gas dynamics. Those cross-disciplinary developments are deferred to Sec.~\ref{sec:extensions_other_transport}, where the emphasis shifts from numerical construction within gas dynamics to the broader portability of multiscale kinetic methodology. Sec.~\ref{sec:particle_methods} turns to particle and stochastic methods, in which the same multiscale objective is pursued through a fundamentally different representation of the kinetic state.

\section{Particle and stochastic methods}
\label{sec:particle_methods}

Particle and stochastic methods form the second major route at the method layer. Instead of resolving the distribution function on a fixed velocity grid, they represent kinetic information statistically by simulation particles, random processes, or weighted samples. This statistical representation leads to a different balance between physical fidelity, phase-space flexibility, statistical uncertainty, and near-continuum efficiency.

Historically, this route entered rarefied-gas dynamics through the direct simulation Monte Carlo (DSMC) method \cite{Bird1994}. Particle representations are especially natural in strongly nonequilibrium and high-dimensional phase-space regimes, where free flight, collision sampling, gas--surface interaction, internal energy exchange, mixtures, and chemical processes can be handled in a physically transparent manner. Their multiscale limitation appears in a form complementary to that of deterministic schemes. The latter must couple transport and collision in numerical fluxes or evolution operators, whereas particle methods must preserve the molecular transport picture while relaxing restrictions imposed by collision-time-scale steps, mean-free-path-scale cells, and statistical noise.

The development of modern particle methods can therefore be understood as a sequence of responses to these bottlenecks. Model-based stochastic particle methods use BGK-, Shakhov-, ES-BGK-, or Fokker--Planck-type equations and then redesign particle evolution to recover the correct coarse-step asymptotic behavior. Direct Boltzmann-based AP Monte Carlo methods keep the original collision physics but modify the stochastic treatment of stiff relaxation. Variance-reduction, information-preserving, and denoising strategies address the sampling uncertainty that becomes dominant in low-signal and near-equilibrium flows. The section begins with DSMC as the historical baseline, then reviews these model-based, direct Boltzmann-based, and noise-reduction developments, before closing with particle-centered hybrid extensions.

\subsection{Classical DSMC and multiscale challenges}
\label{subsec:dsmc_challenge}

The DSMC method is the most successful stochastic particle method for rarefied gas dynamics. It represents a large number of real molecules by simulation particles and solves the Boltzmann equation statistically by separating molecular free transport from stochastic binary-collision sampling over a small time step \cite{Bird1994}. Because the method follows particle trajectories in phase space, it is highly flexible for strongly nonequilibrium flows, high-Mach-number configurations, internal energy modes, gas mixtures, and complex gas--surface interactions. It has therefore become a benchmark solver for rarefied aerodynamics and microscale gas flows.

For the Boltzmann equation \eqref{eq:boltzmann_intro}, DSMC replaces direct evaluation of the collision integral by stochastic collision sampling. Over one numerical time step, the method adopts an operator-splitting structure. The free-transport step is
\begin{equation}
	\partial_t f^*+\bm{\xi}\cdot\nabla f^*=0,
	\label{eq:sec3_dsmc_transport}
\end{equation}
followed by the collision step
\begin{equation}
	\partial_t f=Q_B(f,f).
	\label{eq:sec3_dsmc_collision}
\end{equation}
This split treatment is physically justified when the time step is sufficiently smaller than the mean collision time and the cell size is sufficiently smaller than the local mean free path \cite{Bird1994}. Under these resolved kinetic conditions, free flight and collision may be treated independently without introducing excessive splitting error.

The same mechanism that makes DSMC effective in rarefied regimes also explains its near-continuum difficulty. Since the free-flight and collision substeps are decoupled, the numerical resolution must continue to follow collisional scales even when the physical solution varies mainly on macroscopic scales. The time step is then restricted by the mean collision time and the cell size by the local mean free path. As the Knudsen number decreases, the collision frequency rises and the cost becomes prohibitive. This difficulty is most severe in multiscale configurations where continuum, transitional, and rarefied regions coexist in the same computational domain.

A second obstacle is statistical noise. Macroscopic quantities are obtained from finite-particle sampling, so the relative uncertainty becomes large in low-signal, low-Mach-number, and near-equilibrium flows. In such regimes, the physically relevant nonequilibrium signal may be much smaller than the thermal fluctuation carried by the particles, and a very large number of samples is required to obtain accurate moments. Thus stochastic particle methods face two linked barriers. One is multiscale stiffness caused by collisional restrictions on spatial and temporal resolution, and the other is sampling uncertainty in macroscopic moments.

Most modern particle-based multiscale methods can be understood as attempts to weaken one or both barriers. Model-based stochastic particle methods based on BGK or Fokker--Planck equations mainly reduce the collisional stiffness by simplifying the collision dynamics and embedding continuum-limit transport behavior into particle evolution. Direct Boltzmann-based AP Monte Carlo methods pursue a similar objective while retaining the original Boltzmann collision operator. Variance-reduction and denoising methods primarily target statistical noise, but they are increasingly combined with multiscale transport corrections. This is the point from which the later subsections develop.

\subsection{Unified stochastic particle methods based on relaxation models}
\label{subsec:bgk_particle}
Model-based stochastic particle methods provide an early and systematic route for reducing the continuum-regime stiffness of DSMC while retaining a particle representation of the kinetic state. The starting point is to replace the full Boltzmann collision operator by a relaxation-type kinetic model, most commonly BGK, ESBGK, or Shakhov. The multiscale issue, however, is not solved by this replacement alone. If particle transport and relaxation are still advanced as two separated processes, the method can inherit the same coarse-step error that limits DSMC in near-continuum regimes. The essential development in this branch is therefore the redesign of particle evolution so that relaxation effects are incorporated into transport over one numerical step.

Early stochastic-particle BGK and ESBGK methods retained the DSMC-like splitting between free flight and relaxation. For a representative ESBGK particle solver, the split evolution may be written schematically as \cite{Burt2006}
\begin{equation}
	\partial_t f^*+\bm{\xi}\cdot\nabla f^*=0,
	\qquad
	\partial_t f=\nu_{ES}(f_G-f),
	\label{eq:sec3_spesbgk_split}
\end{equation}
where \(f_G\) denotes the Gaussian target distribution and \(\nu_{ES}\) is the ESBGK relaxation frequency, which is of order \(1/\varepsilon\) in the stiff scaling. The homogeneous relaxation step gives
\begin{equation}
	f(\bm{x},\bm{\xi},t_n+\Delta t)
	=
	e^{-\nu_{ES}\Delta t} f^*(\bm{x},\bm{\xi},t_n+\Delta t)
	+
	\left(1-e^{-\nu_{ES}\Delta t}\right)
	f_G^*(\bm{x},\bm{\xi},t_n+\Delta t),
	\label{eq:sec3_spesbgk_integral}
\end{equation}
and, in particle form, to the relaxation count
\begin{equation}
	N_s
	=
	\operatorname{int}
	\left[
	N_c\left(1-e^{-\nu_{ES}\Delta t}\right)
	\right].
	\label{eq:sec3_spesbgk_ns}
\end{equation}
Here \(N_c\) is the number of particles in a cell and \(N_s\) is the number selected for relaxation. This construction greatly simplifies the stochastic collision step and is useful for mixtures, high-speed nonequilibrium flows, and internal-energy extensions. Its limitation is that the free-flight step remains collisionless. Once \(\nu_{ES}\Delta t\) is no longer small, the split algorithm may generate incorrect continuum-limit transport coefficients, even though the underlying kinetic model has the desired near-equilibrium behavior \cite{FeiZhangLiLiu2020USPBGK}.

The unified stochastic particle (USP) methodology was introduced to correct this coarse-step defect. Its key idea is to move the leading near-equilibrium collisional contribution from the relaxation step into the particle migration step. In the original USP-ESBGK formulation, the split system is reorganized as
\begin{equation}
	\left(\partial_t+\bm{\xi}\cdot\nabla\right)f=J^*,
	\qquad
	\partial_t f=\nu_{ES}(f_G-f)-J^*,
	\label{eq:sec3_usp_correction}
\end{equation}
where \(J^*\) is a near-equilibrium correction. A representative approximation is
\begin{equation}
	J^*=\frac{\nu_{ES}}{\mbox{Pr}}P_c\left(g_M-f_{\mathrm{Grad}}\right),
	\qquad
	P_c=\exp\!\left(-\frac{\mathrm{Kn}_{\mathrm{GLL,max}}}{\mathrm{Kn}_c}\right),
	\label{eq:sec3_usp_Jstar}
\end{equation}
where \(g_M\) is the local Maxwellian, \(\mathrm{Kn}_{\mathrm{GLL,max}}\) is the maximum gradient-length local Knudsen number, and \(\mathrm{Kn}_c\) is a reference cutoff separating near-equilibrium and strongly nonequilibrium behavior \cite{FeiZhangLiLiu2020USPBGK}. A representative Grad approximation used in this construction is
\begin{equation}
	f_{\mathrm{Grad}}
	=
	g_M\left[
	1+\frac{\bm{\sigma}:\left(\bm{C} \bm{C}-\frac{1}{3}C^2\mathbf{I}\right)}{2p\theta}
	+\Pr\,\frac{2\,\bm{q}\cdot\bm{C}}{5p\theta}
	\left(\frac{C^2}{2\theta}-\frac{5}{2}\right)
	\right],
	\label{eq:sec3_usp_grad}
\end{equation}
with \(\theta=RT\). The factor \(P_c\) makes the correction active in near-continuum regions and inactive in strongly nonequilibrium regions. Thus the method reduces to the conventional stochastic-particle ESBGK algorithm in rarefied regimes, while adding a continuum correction when the local flow is close to equilibrium.

At the algorithmic level, the USP method still contains a migration step followed by a relaxation step, but neither is the same as in a purely split particle solver. In the migration stage, trapezoidal integration of \(J^*\) leads to the auxiliary distributions
\begin{equation}
	\tilde f^{*}=f^{*}-\frac{\Delta t}{2}J^{*}(f^{*}),
	\qquad
	\hat f^{n}=f^{n}+\frac{\Delta t}{2}J^{*}(f^{n}),
	\label{eq:sec3_usp_aux}
\end{equation}
so that particle migration can be written symbolically as
\begin{equation}
	\tilde f^{*}(\bm{x}^{n+1},\bm{\xi})=\hat f^{n}(\bm{x}^{n},\bm{\xi}),
	\qquad
	\bm{x}^{n+1}=\bm{x}^{n}+\bm{\xi}\,\Delta t.
	\label{eq:sec3_usp_transport_step}
\end{equation}
This transformation is closely related to the auxiliary-distribution idea used in DUGKS. In both cases, a trapezoidal correction is absorbed into modified distributions so that the resulting transport form is explicit while still carrying part of the collisional effect \cite{FeiZhangLiLiu2020USPBGK}. In particle implementation, the correction is represented by additional weighted particles associated with \(\pm\Delta t J^*/2\), so the migration step is no longer purely collisionless \cite{FeiZhangLiLiu2020USPBGK,FeiEfficientUSP2021}. These weighted particles provide the practical mechanism by which the near-equilibrium collisional contribution is embedded into free transport over one time step.

After the modified migration, the residual relaxation equation in Eq.~\eqref{eq:sec3_usp_correction} is solved stochastically. Its integral form over one step is
\begin{equation}
	f(\bm{x},\bm{\xi},t_n+\Delta t)
	=
	\beta f^*(\bm{x},\bm{\xi},t_n+\Delta t)
	+
	\nu_{ES}\int_0^{\Delta t}
	e^{-\nu_{ES}(\Delta t-s)}
	f_G(\bm{x},\bm{\xi},t_n+s)\,ds
	-
	\frac{1-\beta}{\nu_{ES}}J^*(\bm{x},\bm{\xi},t_n+\Delta t),
	\label{eq:sec3_usp_collision_integral}
\end{equation}
where \(\beta=e^{-\nu_{ES}\Delta t}\). Since the migration step has already incorporated the near-equilibrium part through \(J^*\), this residual relaxation acts primarily on the far-from-equilibrium remainder. In the usual particle realization, the relaxation probability is
\begin{equation}
	\mathcal{P}_r=1-\exp(-\nu_{ES}\Delta t),
	\qquad
	N_s=\operatorname{int}(N_c\mathcal{P}_r).
	\label{eq:sec3_usp_collision_probability}
\end{equation}
The selected particles are first resampled from a Maxwellian and then mapped to the ESBGK Gaussian target through
\begin{equation}
	\bm{C}=\mathbf{S}\bm{C}^{*},
	\label{eq:sec3_usp_gaussian_map}
\end{equation}
where \(\bm{C}^{*}\) is the resampled thermal velocity and \(\mathbf{S}\) is related to the ESBGK anisotropy tensor. The post-collision population therefore consists of unchanged particles, resampled particles, and the correction contribution induced by \(J^*\). This structure explains why the USP can retain the flexibility of particle methods while recovering the correct coarse-step Navier--Stokes transport more accurately than fully split stochastic-particle BGK or ESBGK methods. In the near-continuum regime, the resulting method recovers the desired Navier--Stokes asymptotics with second-order temporal accuracy \cite{FeiZhangLiLiu2020USPBGK}.

Subsequent developments mainly proceeded along three closely related directions. First, efficient implementations were designed so that the unified idea could be incorporated into existing DSMC-type particle infrastructures with minimal algorithmic overhead, thereby reducing the implementation gap between USP and conventional particle solvers while retaining the essential multiscale coupling between transport and relaxation \cite{FeiEfficientUSP2021}. Second, higher-order temporal and spatial versions were developed to reduce the residual truncation error once the leading continuum-limit dissipation had been removed; in these formulations, the asymptotic correction is retained while the basic particle algorithm remains close to the DSMC data structure \cite{FeiHighOrderUSP2022}. Closely related work includes second-order particle BGK integrators based on exponential or Crank--Nicolson-type treatments of the relaxation term, which emphasize temporal accuracy, positivity preservation, and compatibility with DSMC-like particle frameworks \cite{PfeifferEDBGK2022,GarmirianPfeifferCNSPBGK2025}. Third, the same philosophy was extended to more realistic kinetic descriptions, including kinetic models for polyatomic gases and gas mixtures, as well as hybrid particle--DSMC methods, for which particle representations remain particularly effective \cite{FeiUSPPolyMixture2026,FeiAAPMCMultiSpecies2025,FeiUSPBGKDSMC2021}. These developments indicate that the USP correction is no longer merely a special remedy for monatomic benchmark cases, but rather a transferable multiscale design principle, one that preserves the simplicity and flexibility of relaxation-model particles while embedding the leading coarse-step continuum transport directly into particle evolution. The BGK-based unified particle line is therefore better understood not as a single algorithm, but as a family of model-based multiscale particle solvers that, on the stochastic side, play a role analogous to that of unified deterministic kinetic schemes on the grid-based side.

\subsection{Particle Fokker--Planck methods}
\label{subsec:fp_particle}

In parallel with BGK-based stochastic particle methods, the Fokker--Planck (FP) route provides another important model-based path toward multiscale particle computation. Its central idea is to replace the discrete binary-collision process of the Boltzmann equation by a continuous drift--diffusion process in velocity space. As a result, collisions are no longer treated through explicit pair selection, and the particle dynamics can instead be represented by stochastic differential equations. This makes FP particle methods particularly attractive in low- and moderate-Knudsen-number regimes, where the cost of binary-collision handling becomes a major bottleneck \cite{JennyFP2010,GorjiJenny2014}.

A general FP kinetic model can be written as
	\begin{equation}
		\partial_t f+\bm{\xi}\cdot\nabla f
		=
		-\nabla_{\bm{\xi}}\cdot(\bm{A}f)
		+
		\frac{1}{2}\nabla_{\bm{\xi}}\nabla_{\bm{\xi}}:(\mathbf{D}f),
		\label{eq:sec3_fp_general}
	\end{equation}
	where \(\bm{A}\) is the drift coefficient and \(\mathbf{D}\) is the diffusion tensor in velocity space. Compared with the Boltzmann collision operator, Eq.~\eqref{eq:sec3_fp_general} models collisions as a continuous Markov process. The corresponding particle evolution is therefore equivalent to the Langevin system
	\begin{equation}
		d\bm{X}=\bm{\Xi}\,dt,
		\qquad
		d\bm{\Xi}=\bm{A}\,dt+\mathbf{B}\,d\bm{W},
		\label{eq:sec3_fp_sde}
	\end{equation}
	where \(\bm{X}\) and \(\bm{\Xi}\) denote the particle position and velocity stochastic processes, respectively; \(\mathbf{B}\) is a matrix satisfying \(\mathbf{B}\mathbf{B}^{T}=\mathbf{D}\), \(d\bm{W}\) denotes the increment of a Wiener process. In this way, instead of sampling pairwise collisions as in DSMC, one advances particle velocities through stochastic drift--diffusion updates.

	For the linear FP model, the drift and diffusion coefficients are
	\begin{equation}
		\bm{A}
		=
		-\frac{1}{\tau}\left(\bm{\xi}-\bm{u}\right),
		\qquad
		\mathbf{D}
		=
		\frac{2k_B T}{m\tau}\mathbf{I},
		\label{eq:sec3_fp_linear_model}
	\end{equation}
	so that the distribution relaxes toward the local Maxwellian. This formulation is computationally attractive because the stochastic update can be integrated very efficiently. However, like the classical BGK model, the original linear FP model gives an incorrect Prandtl number. This motivated cubic, entropic, and ellipsoidal-statistical (ES) FP models. For example, the ESFP formulation takes
	\begin{equation}
		\mathbf{D}
		=
		\frac{2k_B T}{m\tau}\mathbf{E},
		\qquad
		\mathbf{E}
		=
		(1-\nu)\mathbf{I}
		+
		\nu\frac{\mathbf{P}}{p}
		=
		\mathbf{I}
		+
		\nu\frac{\boldsymbol{\pi}}{p},
		\label{eq:sec3_esfp_diffusion}
	\end{equation}
	so that the transport coefficients become
	\begin{equation}
		\mu=\frac{p\tau}{2(1-\nu)},
		\qquad
		\Pr=\frac{3}{2(1-\nu)}.
		\label{eq:sec3_esfp_transport}
	\end{equation}
	Thus, by choosing \(\nu\) subject to positive definiteness of the diffusion tensor, one can recover the desired Prandtl number \cite{MathiaudMieussens2016ESFP}. Particle implementations and subsequent accuracy improvements are discussed in Refs.~\cite{KimUSPESFP2024,CuiMSP2025}.

	The multiscale issue, however, is not resolved by the kinetic model alone. A straightforward FP particle algorithm still separates physical-space streaming from velocity-space relaxation. A typical split update has the form
	\begin{equation}
		\bm{X}^{*}=\bm{X}^{n}+\bm{\Xi}^{n}\Delta t,
		\qquad
		\bm{\Xi}^{n+1}
		=
		\mathcal{R}_{\Delta t}
		\bigl(
		\bm{\Xi}^{*};\,
		\bm{U}^{n},\,T^{n},\,\mathbf{P}^{n},\ldots
		\bigr),
		\label{eq:sec3_fp_split}
	\end{equation}
where \(\mathcal{R}_{\Delta t}\) denotes the stochastic relaxation map and the local macroscopic fields are usually evaluated from cell averages. This construction is robust and easy to implement, but it introduces splitting and reconstruction errors when \(\Delta t\) and the cell size are no longer tied to the collision time and mean free path. Thus FP particle methods face the same structural question as BGK-type stochastic particles. The continuous drift--diffusion collision model must be combined with a particle evolution that preserves the correct near-continuum transport behavior on coarse numerical scales.

The multiscale development of FP particle methods follows essentially the same principle as that of USP-BGK. The leading coarse-step dissipation must be removed by embedding the correct near-continuum transport behavior directly into particle evolution. An early important step in this direction was to design the temporal discretization so that the numerical transport coefficients remain correct even when the time step is no longer restricted by the collisional time scale \cite{FeiMTDFPM2017}. A more systematic recent development is the construction of second-order FP particle methods. For monatomic gases, the unified stochastic particle FP (USP-FP) method achieves second-order temporal accuracy by reproducing the correct second-order relaxation behavior of viscous stress and heat flux \cite{KimUSPESFP2024}. At the spatial level, local macroscopic quantities are no longer taken directly as cell averages at particle positions. Instead, they are reconstructed through interpolation or polynomial approximation, for instance,
\begin{equation}
	\phi(\bm{x}_p)
	=
	\phi_j
	+
	(\bm{x}_p-\bm{x}_j)\cdot\nabla \phi_j
	+
	O(\Delta x^2),
	\label{eq:sec3_fp_linear_recon}
\end{equation}
where \(\phi\) denotes a local macroscopic quantity such as velocity, temperature, stress, or heat flux; $\x_p$ is the particle position, $\phi_j$ is the cell-averaged value of $\phi$ in cell $V_j$, and $\Delta x$ is the cell size. This step is important because an accurate stochastic relaxation update requires local fields at particle positions, not only cell-centered averages.

A closely related development is the multiscale stochastic particle (MSP) method based on the ESFP model  \cite{CuiMSP2025}. Schematically, one may view the update as
\begin{equation}
	f^{n+1}
	=
	\mathcal{R}^{\mathrm{exact}}_{\Delta t}\,
	\mathcal{T}_{\Delta t}f^n
	+
	\mathcal{E}_{\mathrm{corr}}(f^n,\Delta t,\Delta x),
	\label{eq:sec3_fp_msp}\end{equation}
where \(\mathcal{T}_{\Delta t}\) is the transport update, \(\mathcal{R}^{\mathrm{exact}}_{\Delta t}\) is the exact homogeneous FP relaxation, and \(\mathcal{E}_{\mathrm{corr}}\) compensates the leading splitting-induced error. This construction keeps the stochastic FP relaxation efficient while extending the admissible time step and cell size. From the multiscale viewpoint, it plays a role for FP particles analogous to that played by USP-BGK methods for relaxation-model particles.

Subsequent developments mainly followed two directions. One is the improvement of temporal and spatial accuracy under coarse discretization, through higher-order FP relaxation treatments and more accurate reconstruction of local macroscopic fields \cite{KimUSPESFP2024,CuiMSP2025}. The other is extension to more realistic kinetic models, including monatomic mixtures, polyatomic gases, or formulations with internal-energy transitions \cite{KimMixtureFP2025,HeppThesis2022,KimUSPFPM2025}. Overall, these developments show that the FP particle route is also better understood not as a single algorithm, but as a family of model-based multiscale particle solvers. Compared with BGK-based unified particle methods, they rely on continuous stochastic drift--diffusion rather than relaxation sampling; however, the underlying design principle is the same, namely to retain the efficiency of a simplified kinetic model while embedding the correct near-continuum transport behavior directly into particle evolution.

\subsection{AP Monte Carlo methods for the Boltzmann equation}
\label{subsec:apmc_boltzmann}
The BGK- and FP-based particle methods reviewed above improve multiscale efficiency by replacing the Boltzmann collision operator with simpler kinetic models and then designing particle algorithms that preserve the desired asymptotic behavior under coarse discretization. A conceptually different route is to retain direct Boltzmann fidelity. In this branch, the starting point is the scaled Boltzmann equation given by Eq.~\eqref{eq:scaled_boltzmann_intro}, where the parameter \(\varepsilon\) exposes the stiffness of the original collision operator. This scaled form can identify the collision stiffness that AP Monte Carlo methods must overcome. The numerical task is then to redesign the Monte Carlo discretization so that the method remains accurate and efficient when \(\varepsilon\) becomes small. The model-based route emphasizes collision simplification followed by multiscale particle design, whereas the direct Boltzmann route emphasizes asymptotic-preserving stochastic treatment of the original collision physics. These two directions are therefore complementary rather than competing.

The central idea of the direct Boltzmann route is to isolate the stiff near-equilibrium part of the collision dynamics and treat it more efficiently, while leaving only the genuinely nonequilibrium remainder to stochastic sampling. In other words, the method does not reduce stiffness by changing the kinetic model, but by changing the numerical treatment of the original collision physics \cite{FeiTRMCNS2023,FeiAAPMCMultiSpecies2025}. From the multiscale viewpoint, the objective is the same as in unified BGK- and FP-based particle methods. The leading coarse-step error should be removed and the correct continuum transport behavior should be recovered, but without replacing the Boltzmann collision operator by a surrogate relaxation model.

A representative line of development is the time-relaxed Monte Carlo (TRMC) family \cite{pareschi2001time}. Classical TRMC relaxes the raw collision-time restriction of DSMC in homogeneous problems. Recent multiscale extensions go further by incorporating the Navier--Stokes asymptotics directly into the collision treatment \cite{FeiTRMCNS2023}. In a schematic form,  the collision update associated with the Boltzmann operator may be
organized as
\begin{equation}
	Q_B(f,f)=Q^{\mathrm{NS}}(f)+Q^{\mathrm{neq}}(f),
	\label{eq:sec3_trmc_split_symbolic}
\end{equation}
where \(Q^{\mathrm{NS}}\) denotes the part treated deterministically so as to recover the desired near-continuum transport behavior, while \(Q^{\mathrm{neq}}\) represents the remaining nonequilibrium contribution to be handled stochastically. Equation~\eqref{eq:sec3_trmc_split_symbolic} should be understood in a structural rather than literal sense. It indicates that the stiff Chapman--Enskog content of the collision operator is extracted from direct random collision sampling and incorporated into a controlled asymptotic treatment.

This modification changes the role of Monte Carlo sampling in an essential way. In conventional DSMC, stochastic collisions must resolve both the dominant near-equilibrium relaxation and the comparatively small nonequilibrium remainder. In AP Monte Carlo methods, by contrast, the leading near-equilibrium part is treated through deterministic or semi-deterministic correction, so that stochastic sampling is reserved mainly for the residual nonequilibrium component. As a result, the method preserves direct Boltzmann fidelity in rarefied regimes while recovering the correct viscous and heat-conduction behavior much more accurately under coarse discretization \cite{FeiTRMCNS2023}.

This philosophy has also been extended to gas mixtures, where retaining the full Boltzmann description is particularly valuable because simplified mixture models often struggle to reproduce all transport coefficients consistently \cite{FeiAAPMCMultiSpecies2025}. In this context, the advantage of the direct Boltzmann route is not merely formal. It provides a way to combine multiscale asymptotic preservation with a more faithful treatment of interspecies collision physics, which is often difficult to achieve within a single simplified relaxation model.

A related development appears in diffusive-scaling particle methods, where kinetic and diffusive motions are blended within one time step so that the algorithm reduces automatically to a random-walk description in the high-collision limit while recovering the standard kinetic particle evolution in the opposite limit \cite{MortierKD2022}. More recent work further analyzes the discretization errors of DSMC itself and constructs multiscale corrections for both time-splitting and finite-cell collision errors \cite{YangMSMC2026}. This points toward stochastic particle methods with explicit temporal and spatial error control, rather than time-step relaxation alone.

Viewed as a whole, the direct Boltzmann-based AP Monte Carlo family forms a distinct branch of multiscale particle simulation. Compared with BGK- and FP-based unified particle methods, these approaches retain a closer connection to the original collision physics. Compared with conventional DSMC, they introduce asymptotic or error-controlled mechanisms that permit substantially coarser temporal and spatial discretizations. They therefore occupy an important middle ground between classical DSMC and model-based unified particle methods.

\subsection{Variance reduction and denoising}
\label{subsec:vr_denoise}

Besides collisional stiffness, statistical noise is the second major obstacle for stochastic particle simulation. In DSMC and related particle methods, macroscopic quantities are obtained from finite-particle sampling, so the relative uncertainty grows rapidly in low-signal problems. This difficulty is especially severe in low-Mach-number, low-speed, and weakly nonequilibrium flows, where the physically relevant deviation from equilibrium may be much smaller than the thermal fluctuation carried by the particles. As a result, even when the underlying particle method is asymptotically correct, the computation may still become impractical unless its statistical noise is effectively controlled \cite{SadrMEVRDSMC2023,YangDMP2025}.

From the multiscale point of view, variance reduction and denoising should therefore be regarded not as auxiliary post-processing tools, but as integral parts of modern particle-method design. The basic idea is to exploit the fact that many target flows remain close to a local equilibrium, so that the dominant equilibrium contribution may be treated analytically or through a correlated reference process, while stochastic sampling is reserved mainly for the smaller nonequilibrium signal. In this way, noise control becomes closely connected with the same near-equilibrium structure that underlies asymptotic-preserving particle design.

The most direct realization of this idea is the control-variate strategy used in the low-variance deviational simulation Monte Carlo (LVDSMC) \cite{Hadj2007} and the variance-reduced DSMC (VRDSMC) \cite{Hadj2010}. For example, in the VRDSMC, the moment of a velocity polynomial \(R(\bm{\xi})\) is regrouped as
\begin{equation}
	\int R(\bm{\xi}) f(\bm{\xi}|\bm{x},t)\,d\bm{\xi}
	=
	\int R(\bm{\xi})\bigl(1-w(\bm{\xi}|\bm{x},t)\bigr)f(\bm{\xi}|\bm{x},t)\,d\bm{\xi}
	+
	\int R(\bm{\xi}) f^{eq}(\bm{\xi}|\bm{x},t)\,d\bm{\xi},
	\label{eq:sec3_vr_control_variate}
\end{equation}
with
\begin{equation}
	w(\bm{\xi}|\bm{x},t)
	=
	\frac{f^{eq}(\bm{\xi}|\bm{x},t)}{f(\bm{\xi}|\bm{x},t)}.
	\label{eq:sec3_vr_weight}
\end{equation}
Here $f^{eq}$ denotes a chosen equilibrium or reference distribution. When \(f\) remains close to \(f^{eq}\), the first integral represents only a small nonequilibrium correction and can therefore be estimated with much lower variance than the full moment itself \cite{SadrMEVRDSMC2023}. The key point is that the large equilibrium contribution is no longer resolved through noisy particle sampling, but is supplied analytically.

This control-variate idea underlies the VRDSMC family and its later maximum-entropy variance-reduced DSMC (ME-VRDSMC) extension.
In ME-VRDSMC, the nonequilibrium DSMC simulation is supplemented by an auxiliary equilibrium simulation that shares the same random numbers. The two parts are kept correlated through importance weights, while the equilibrium contribution is evaluated analytically. To improve stability in the collision-dominated regime, kernel density estimation is combined with a cross-maximum-entropy reconstruction subject to moment constraints, so that mass, momentum, energy, and selected higher-order moments are preserved more accurately during weight resampling. In this sense, variance reduction here is no longer merely a statistical refinement, but already part of the kinetic representation itself.

Variance reduction has also been investigated specifically for particle Monte Carlo methods based on the Fokker--Planck equation, where the same objective must be achieved at the level of stochastic trajectories rather than collision sampling. 
A direct deviational formulation is less natural for FP dynamics, and Gorji \emph{et al.} proposed a different route based on parallel correlated stochastic processes \cite{GorjiVarianceFP2015}. In this approach, one evolves, together with the main stochastic process, an auxiliary process whose behavior is known or more easily controlled, and the macroscopic quantity of interest is evaluated from the correlated difference between the two. The resulting method plays a role analogous to that of a control variate, but it is formulated directly within the Langevin description of particle motion. This is conceptually important because it shows that noise reduction in multiscale particle methods is not limited to DSMC or Boltzmann-based formulations, but can also be built into FP-based stochastic dynamics.

Another approach augments each simulation particle with low-noise collective information. This idea originated from the information-preserving (IP) method \cite{IP2001} and has been developed further into the denoising multiscale particle (DMP) method \cite{YangDMP2025}. The basic perspective is that a particle should carry not only its molecular state \((\bm{x},\bm{\xi})\), but also additional information variables representing the local collective flow state. In the DMP framework, the kinetic description is generalized to the velocity--information joint distribution function	$f(\bm{x},\bm{\xi},t;\bm{\eta},\theta)$,
where \(\bm{\eta}\) and \(\theta\) are the information velocity and information temperature carried by each particle. Macroscopic quantities are then obtained not only from molecular variables, but also from these collective information variables, which are much less noisy in near-equilibrium flows.

To make this idea computationally practical, the information-preserving kinetic equation is modeled by an information-augmented Shakhov BGK equation,
\begin{equation}
	\partial_t f	+	\bm{\xi}\cdot\nabla f	+	\bm{G}^{com}_{\bm{\eta}}\cdot\nabla_{\bm{\eta}} f	+
	G^{com}_{\theta}\,\partial_{\theta} f  = \frac{f_t-f}{\tau},
	\label{eq:sec3_info_sbgk}
\end{equation}
with target distribution
\begin{equation}
	f_t	=	f^{S}	\delta(\bm{\eta}-\bm{u}^I) \delta(\theta-T^I),
	\label{eq:sec3_info_target}
\end{equation}
where $f^S$ is the Shakhov distribution. Here \(\bm{u}^I\) and \(T^I\) denote the information velocity and information temperature, respectively, which are low-noise collective variables carried by each particle and represent the locally averaged flow velocity and temperature associated with the information state, rather than the instantaneous molecular velocity \(\bm{\xi}\) and molecular thermal state of that particle.
In this model, the usual transport term advances the molecular state, while the additional terms involving \(\bm{G}^{com}_{\bm{\eta}}\) and \(G^{com}_{\theta}\) describe the evolution of the collective information carried by the particles. During transport, these information labels move with the particles; during relaxation, they are driven toward the local collective state. In this way, denoising is embedded directly into the kinetic evolution rather than imposed afterward.

The advantage of this formulation can also be interpreted quantitatively. In low- to moderate-signal flows, the DMP analysis shows that the relative uncertainty no longer scales primarily with the smallness of the physical signal itself, but instead becomes tied mainly to the local rarefaction level \cite{YangDMP2025}. This feature is important from the multiscale perspective, because it indicates that denoising and rarefaction-sensitive particle evolution can be organized within one unified framework. The DMP method therefore addresses both statistical noise and multiscale discretization.

Collectively, these developments show that noise control has become a core design principle in modern particle methods. Control-variate DSMC, correlated-process variance reduction for FP particles, and information-preserving or denoising particle formulations all exploit the same structural separation between a dominant equilibrium contribution and a smaller nonequilibrium signal. Particle sampling is then concentrated on the residual kinetic content. In this sense, variance reduction and denoising are not secondary statistical fixes, but part of the multiscale design of the kinetic representation itself.

\subsection{Particle-centered hybrid extensions}
\label{subsec:particle_hybrid_transition}

Particle-centered hybrid strategies form a natural closing topic for the present section. They retain particles as the primary carrier of kinetic information, but combine them with continuum solvers, simplified particle models, or deterministic velocity-space descriptions in order to reduce cost, noise, or regime-dependent inefficiency. In this sense, they remain closer to the particle-method lineage than to the macro--micro and synthetic formulations reviewed in Sec.~\ref{sec:hybrid_macro_micro}, even though they already point toward that broader class of multiscale couplings.

Several representative forms have been developed \cite{ZhangJohnPfeifferFeiWen2019ParticleReview,TeschnerReview2016}. DSMC--CFD hybrids use a continuum solver in near-equilibrium regions and reserve DSMC for strongly nonequilibrium zones. Particle--particle hybrids combine DSMC with BGK- or FP-based particle solvers so that different collision models are used in different regimes. Low-diffusion particle continuum methods provide particle-compatible continuum descriptions, while more recent stochastic BGK/DVM couplings use deterministic velocity-space information to reduce noise and limit the cost of a full deterministic discretization.

Examples include USPBGK--DSMC formulations, which use unified stochastic particle BGK evolution in near-continuum regions and DSMC-like collision treatment in strongly nonequilibrium regions \cite{FeiUSPBGKDSMC2021}. Particle FP/DSMC couplings exploit the complementary advantages of continuous drift--diffusion and binary-collision sampling \cite{FP-DSMC}. The low-diffusion method has also been revisited as a deterministic particle-based continuum solver compatible with DSMC-type frameworks \cite{MirzaLD2017}. In addition, stochastic particle BGK/DVM couplings in velocity space aim to combine the low-noise character of deterministic velocity discretization with the flexibility of particle sampling \cite{GarmirianPfeifferCoupling2026}.

These methods are not reviewed in detail here, because their detailed classification overlaps with the broader hybrid literature. Their role in the present section is to mark the point at which particle methods begin to merge with other multiscale organizations. Sec.~\ref{sec:hybrid_macro_micro} therefore shifts the emphasis from particle-centered representation to macro--micro decomposition, high-order/low-order coupling, synthetic equations, and related frameworks in which macroscopic or moment equations become active components of the kinetic computation.

\subsection{Section remarks}
\label{subsec:particle_remarks}
The developments reviewed in this section show that particle-based multiscale simulation has evolved far beyond classical DSMC. Model-based particle methods replace the Boltzmann collision operator with simplified BGK, ES-BGK, Shakhov, or Fokker--Planck descriptions, and then redesign particle evolution so that the correct near-continuum transport behavior is retained under coarse space--time discretization. Direct Boltzmann-based AP Monte Carlo methods pursue the same multiscale objective without replacing the original collision physics. Variance-reduction, information-preserving, and denoising methods address the statistical barrier by separating the dominant near-equilibrium contribution from the smaller nonequilibrium signal.

These directions address different limitations of classical particle simulation. Unified BGK- and FP-type particles mainly reduce the collisional stiffness inherited from operator splitting. AP Monte Carlo methods reduce stiffness while preserving a closer connection to the Boltzmann collision operator. Variance-reduction and denoising strategies reduce the sampling cost that becomes severe in low-signal or near-equilibrium regimes. Particle-centered hybrid methods then combine these ideas with continuum, simplified-particle, or deterministic velocity-space descriptions.

The common theme is that modern particle methods no longer rely on a purely collisionless transport step followed by an independently sampled collision step. Instead, transport, relaxation, asymptotic behavior, and statistical representation are reorganized within the particle framework. In this sense, particle and stochastic methods provide the stochastic counterpart of deterministic multiscale kinetic schemes. Their shared goal is to preserve kinetic fidelity in rarefied regimes while extending practical accuracy and efficiency toward the near-continuum limit.

\section{Hybrid and synthetic strategies}
\label{sec:hybrid_macro_micro}
The preceding two sections reviewed multiscale kinetic methods mainly from the viewpoint of representation, namely, deterministic velocity-space discretization and stochastic particle sampling. The present section shifts the emphasis to the organization of kinetic and macroscopic descriptions within one multiscale computation. 
The central issue is how information should be exchanged between kinetic and macroscopic levels so that the method remains accurate, efficient, and robust across regimes.

This shift is important because multiscale stiffness is also an information-transfer difficulty. A purely kinetic solver may contain the correct nonequilibrium physics, but hydrodynamic information can propagate inefficiently through high-dimensional relaxation or iterative updates in continuum and near-continuum regimes. Macroscopic equations transmit large-scale information efficiently, but they lose validity in kinetic boundary layers, strongly nonequilibrium regions, and nonlocal transport regimes. Hybrid, macro--micro, and synthetic strategies arise from this tension. They use macroscopic structure where it is reliable and useful, while retaining kinetic information where it is physically necessary.

The discussion proceeds from equation-level decomposition to solver-level acceleration. We begin with micro--macro formulations, where the distribution is separated into equilibrium and nonequilibrium components and the resulting system provides a systematic route to AP discretization. We then consider adaptive kinetic--fluid coupling and particle micro--macro schemes, where the microscopic correction is localized or represented statistically. We next discuss high-order/low-order (HOLO) coupling, moment acceleration, and GSIS-type synthetic equations, in which an auxiliary macroscopic system becomes an active component of the nonlinear solver. The final subsections discuss deterministic--stochastic macro-guided couplings and reduced-order extensions.

\subsection{Micro--macro AP decomposition}
\label{subsec:mm_general}
The natural starting point for decomposition-based multiscale methods is the observation that, in collisional regimes, the solution of a kinetic equation is close to a local equilibrium determined by a finite set of conservative moments. This motivates a decomposition of the distribution function into a macroscopic equilibrium part and a microscopic nonequilibrium correction. The macroscopic component carries the conservative wave structure and the limiting fluid dynamics, whereas the microscopic component contains the nonequilibrium information responsible for viscous, thermal, kinetic-boundary-layer, and other non-fluid effects. In the continuum fluid regime, this microscopic part is damped by collisions and becomes asymptotically determined by the gradients of the macroscopic variables.

The value of this decomposition is not only formal. It separates the conservative moment dynamics from the collision-damped kinetic correction and therefore provides a natural framework for AP discretization. In the fluid limit, the microscopic component becomes small and can often be treated implicitly, locally, or with reduced resolution. Away from equilibrium, the same component retains the kinetic information needed to describe non-fluid effects. Micro--macro methods therefore reorganize the kinetic equation without replacing it by a macroscopic closure.

The macro--micro theory of Liu and Yu \cite{ref:LiuYu2004}
provides a theoretical basis for this approach by separating the
macroscopic and microscopic components of the Boltzmann equation
and identifying their distinct dissipative roles. Let $f_0$ be the
local Maxwellian with the same conservative moments as $f$.
In the projection notation used here, the distribution is split as
\begin{equation}
	f=f_0+f_1,\qquad
	f_0=P_0 f,\qquad
	f_1=P_1 f,
	\label{eq:sec4_liuyu_proj}
\end{equation}
where $P_0$ is the orthogonal projection onto the macroscopic collision-invariant subspace associated with the local Maxwellian, and $P_1=I-P_0$ is the corresponding microscopic projection.
These projections depend on the local Maxwellian, and the
microscopic component satisfies $\langle\bm{\psi}f_1\rangle=0$.
For the unscaled Boltzmann equation, the resulting macro--micro Boltzmann system may be written schematically as 
\begin{equation}
	\begin{aligned}
		\partial_t f_0
		+P_0\bigl[\bm{\xi}\cdot\nabla(f_0+f_1)\bigr]
		&=0,\\
		\partial_t f_1
		+P_1\bigl[\bm{\xi}\cdot\nabla(f_0+f_1)\bigr]
		&=Lf_1+N(f_1),
	\end{aligned}
	\label{eq:sec4_liuyu_system}
\end{equation}
where $Lh=Q_B(f_0,h)+Q_B(h,f_0)$ is the collision operator
linearized about $f_0$, and $N(h)=Q_B(h,h)$ is the quadratic
microscopic term. Equation \eqref{eq:sec4_liuyu_system} shows that conservative information is carried by the projected macroscopic component, while the microscopic component is controlled by collisional dissipation. Their coupling accounts for the dissipative corrections
to the macroscopic dynamics. This formulation provides the conceptual basis for many later numerical micro--macro schemes.

For numerical AP methods, the same decomposition is commonly
written in terms of the local equilibrium associated with the
conservative variables. In the fluid-dynamic scaling, we write
\begin{equation}
	f=f_M+f_m,\qquad
	f_M=g_M(\bm W),\qquad
	\langle\bm{\psi}f_m\rangle=0,
	\label{eq:sec4_mm_basic}
\end{equation}
where $f_M=f_0$ and $f_m=f_1$. The projection operators remain
$P_0$ and $P_1$, evaluated at $f_M$. In smooth fluid regimes
away from initial and boundary layers, $f_m$ is typically of
order $\varepsilon$ and is asymptotically determined by gradients
of the macroscopic variables.

For the scaled Boltzmann equation, taking conservative moments
and applying $P_1$ gives
\cite{Bennoune_Lemou_Mieussens_2008}
\begin{equation}
	\partial_t\bm W
	+\nabla\cdot\bm{\mathcal F}_E(\bm W)
	+\nabla\cdot
	\langle\bm{\xi}\,\bm{\psi}f_m\rangle
	=0,
	\label{eq:sec4_mm_macro}
\end{equation}
\begin{equation}
	\partial_t f_m
	+P_1(\bm{\xi}\cdot\nabla f_m)
	=
	\frac{1}{\varepsilon}
	\bigl[Lf_m+N(f_m)\bigr]
	-P_1(\bm{\xi}\cdot\nabla f_M),
	\label{eq:sec4_mm_micro}
\end{equation}
where $\bm{\mathcal F}_E(\bm W)=\langle\bm{\xi}\,\bm{\psi}f_M\rangle$. For the normalized BGK collision operator, the corresponding
microscopic collision term reduces to $Lf_m=-f_m$ and $N=0$. Equations~\eqref{eq:sec4_mm_macro} and
\eqref{eq:sec4_mm_micro} express the same projected kinetic
system in moment--microscopic form and provide a basis for AP
discretization. The conservative variables are evolved through
a macroscopic balance law corrected by the microscopic flux,
while the microscopic equation retains the stiff collisional
dynamics that determine the nonequilibrium correction in the
fluid regime.

A representative first-order semi-implicit time discretization
of this system is given by Bennoune \emph{et al.}
\cite{Bennoune_Lemou_Mieussens_2008}, in which the microscopic and macroscopic updates read
\begin{equation}
	\frac{f_m^{n+1}-f_m^n}{\Delta t}
	+P_1^n(\bm{\xi}\cdot\nabla f_m^n)
	=
	\frac{1}{\varepsilon}
	\bigl[L^n f_m^{n+1}+N(f_m^n)\bigr]
	-P_1^n(\bm{\xi}\cdot\nabla f_M^n),
	\label{eq:sec4_mm_discrete_fm}
\end{equation}
\begin{equation}
	\frac{\bm W^{n+1}-\bm W^n}{\Delta t}
	+\nabla\cdot\bm{\mathcal F}_E(\bm W^n)
	+\nabla\cdot\langle \bxi\,\bm{\psi}f_m^{n+1}\rangle=0 .
	\label{eq:sec4_mm_discrete_W}
\end{equation}
where the linearized collision term is treated implicitly, while
the quadratic and transport terms are evaluated explicitly.
For the normalized BGK model, this reduces to the update
with $L^n f_m^{n+1}=-f_m^{n+1}$ and $N=0$. This pair shows concretely how the AP property is achieved. The stiff relaxation acts only on the microscopic correction, while the macroscopic subsystem remains in conservative form. For the scaling displayed here, the limit  $\varepsilon\to 0$ at fixed mesh and time step gives a consistent scheme for the limiting Euler or Navier--Stokes equations, depending on the scaling and on the order of the asymptotic expansion \cite{Bennoune_Lemou_Mieussens_2008,LemouMieussens2008,ref:Lemou2010}.

This reformulation is attractive for several reasons. First, the macroscopic subsystem is built into the solver from the outset, so asymptotic correctness is enforced structurally rather than recovered only after resolving the kinetic scale. Second, conservation of mass, momentum, and energy is naturally tied to the macroscopic equation whenever these quantities are physically conserved. Third, the microscopic correction becomes small in the fluid regime, which provides a basis for reduced, localized, or particle-based representations of the nonequilibrium part.

The same philosophy extends beyond the simplest single-species BGK model. Multispecies and more general collisional models can be treated within a micro--macro framework, although additional source terms associated with interspecies exchange, multiple relaxation mechanisms, or more complicated linearized operators must be handled \cite{ref:JinShi2010,ref:GambaJinLiu2019}. These developments show that micro--macro decomposition is not merely a convenient reformulation for an isolated model problem, but a general strategy for designing AP solvers. The next issue is how this global decomposition can be made more computationally efficient when kinetic nonequilibrium is spatially localized.

\subsection{Adaptive kinetic--fluid coupling}
\label{subsec:mm_adaptive}
The micro--macro formulation in Sec.~\ref{subsec:mm_general} evolves the
nonequilibrium correction throughout the computational domain. This global
representation is mathematically clean and asymptotically robust, but it may be
unnecessarily expensive in multiscale flows where nonequilibrium effects are
localized. Large portions of the domain may remain close to local equilibrium,
while kinetic boundary layers, shocks, rarefied zones, or other non-fluid
structures occupy only limited regions. This observation motivates adaptive
kinetic--fluid coupling, in which a macroscopic description is used as the
background evolution and kinetic correction is activated only where the local
state requires it.

A representative formulation of this idea was developed by Degond \emph{et al.} \cite{ref:DegondDimarcoMieussens2010}. Their method starts from a
micro--macro decomposition, but interprets the microscopic correction as a
localized kinetic upscaling term. In the notation of
\eqref{eq:sec4_mm_basic}, this idea may be expressed schematically as
\begin{equation}
	f \approx f_M+\chi(\bm{x},t) f_m,\qquad 0\le \chi(\bm{x},t)\le 1,
	\label{eq:sec4_localized_mm}
\end{equation}
where $\chi$ is a transition function identifying fluid, buffer, and kinetic regions. When $\chi=0$, only the macroscopic equilibrium part is retained; when $\chi=1$, the full micro--macro representation is active; intermediate values define a buffer region. Unlike static domain decomposition, the topology of these regions is allowed to change during the computation, allowing kinetic regions to move, merge, split, appear, or disappear as the flow evolves.

A useful feature of this framework is that the transition function is updated by a discrete localization rule. If $\beta_i^n$ denotes a local breakdown
indicator and $\beta_{\mathrm{thr}}^{\ast}\le \beta_{\mathrm{thr}}$ are two thresholds, a typical update can be written as
	\begin{equation}
		\chi_i^{n+1}=
		\begin{cases}
			1, & \beta_i^n\ge \beta_{\mathrm{thr}},\\[2mm]
			0, & \beta_i^n<\beta_{\mathrm{thr}}^{\ast},\\[2mm]
			\dfrac{\beta_i^n-\beta_{\mathrm{thr}}^{\ast}}
			{\beta_{\mathrm{thr}}-\beta_{\mathrm{thr}}^{\ast}},
			& \beta_{\mathrm{thr}}^{\ast}\le \beta_i^n\le \beta_{\mathrm{thr}} .
		\end{cases}
		\label{eq:sec4_chi_update}
	\end{equation}
Thus each cell is assigned automatically to a kinetic, fluid, or buffer region.	It was also
shown that the breakdown parameter may be replaced by a local Knudsen-number criterion \cite{ref:DegondDimarcoMieussens2010}, although a more accurate option is to estimate the
mismatch between the macroscopic and kinetic updates directly from the
decomposed equations.

This dynamic-localization viewpoint is conceptually important because it shows that micro--macro coupling need not be interpreted as a rigid global
decomposition. Instead, the nonequilibrium component acts as a kinetic upscaling correction retained only where it is needed. In the formulation of
Degond  \emph{et al.}  \cite{ref:DegondDimarcoMieussens2010}, the transition function is introduced together with kinetic, fluid, and buffer
zones, and the resulting method preserves positivity and uniform flows while reducing the expensive kinetic region to a small part of the full domain in
many unsteady tests.

The relation to classical hybrid methods also deserves clarification. Although
the method involves regions in which either the fluid or kinetic description is
dominant, these regions are not prescribed once and for all at the beginning of
the computation. They are generated and updated from the evolving state of the
decomposed system itself. For this reason, such methods are more naturally
viewed as decomposition-based adaptive kinetic--fluid methods than as classical
static domain couplings.

More recent work has extended this philosophy to diffusive scaling and dynamic
domain adaptation. Laidin \cite{ref:Laidin2023} considered
adaptive kinetic--fluid hybrid methods in which the micro--macro decomposition
is used to treat the interface itself, so that information exchange between
kinetic and fluid regions is expressed in the same macro--micro formulation as
the bulk solver. This is advantageous when the effective Knudsen number varies
strongly in space and when the kinetic region is not known in advance. Instead
of coupling two unrelated models across an externally prescribed interface, one
interprets both descriptions as different realizations of the same decomposed
system.

These adaptive kinetic--fluid methods form an intermediate class between classical AP reformulations and more elaborate hybrid or synthetic-coupling strategies. They retain the clear macro--micro structure of the decomposition while introducing a dynamic localization mechanism to reduce computational cost.

\subsection{Particle micro--macro schemes}
\label{subsec:mm_particle}
Another important extension of the micro--macro viewpoint is to represent the
microscopic correction by particles while keeping the macroscopic component on
an Eulerian mesh. This direction is especially attractive when a full
phase-space discretization of the microscopic equation is too expensive, but
one still wishes to preserve the asymptotic structure and the noise-reduction
advantages of a micro--macro formulation. Unlike the particle methods reviewed
in Sec.~\ref{sec:particle_methods}, the particles do not approximate the full
kinetic distribution. They carry only the nonequilibrium correction, whereas
the dominant equilibrium part is advanced by a finite-volume or finite-difference
solver for the macroscopic variables. Since this correction becomes small as the
solution approaches equilibrium, the particle burden and the sampling noise can
both be substantially reduced in the fluid limit.
This construction also differs from the adaptive kinetic--fluid coupling in
Sec.~\ref{subsec:mm_adaptive}. There, the microscopic correction is localized in space, whereas here the same correction is represented by a different numerical carrier. The resulting method preserves the AP structure of the decomposed equations,
avoids a full deterministic phase-space discretization of the microscopic part,
and uses particles mainly for the genuinely kinetic residual.

A representative example is the particle micro--macro scheme developed by
Crestetto \emph{et al.}~\cite{ref:CCL2018} for collisional
kinetic equations in the diffusion scaling. To avoid confusion with the
gas-dynamic notation used in the preceding sections, we use \(\v\) for the
velocity variable in this model. The model problem is
\begin{equation}
	\partial_t f+\frac{1}{\varepsilon}\v\cdot\nabla f
	=\frac{1}{\varepsilon^2}(\rho\mathcal{M}-f),
	\qquad \rho=\langle f\rangle,
	\label{eq:sec4_diffusive_kinetic}
\end{equation}
where $\mathcal{M}(\v)$ is a fixed normalized equilibrium profile satisfying
$\langle \mathcal{M}\rangle=1$ and $\langle \v\mathcal{M}\rangle=0$.

For this scalar diffusive model, the general macro--micro decomposition \eqref{eq:sec4_mm_basic}
reduces to
\[
f_M=\rho\mathcal{M},\qquad f_m=f-\rho\mathcal{M},
\qquad \langle f_m\rangle=0 .
\]
Following the projection idea in \eqref{eq:sec4_liuyu_proj} , this decomposition
can be written through the simplified equilibrium projection associated with
the fixed profile \(\mathcal{M}\),
\begin{equation}
	\Pi\phi=\langle\phi\rangle\mathcal{M},
	\qquad
	(I-\Pi)\phi=\phi-\langle\phi\rangle\mathcal{M}.
	\label{eq:sec4_particle_mm_projection}
\end{equation}
Thus \(f_M=\Pi f\), \(f_m=(I-\Pi)f\), and the microscopic component has zero
density moment. Applying \(\Pi\) and \(I-\Pi\) to
\eqref{eq:sec4_diffusive_kinetic} gives the equivalent micro--macro system
\begin{equation}
	\partial_t\rho+\frac{1}{\varepsilon}\nabla\cdot
	\langle \v f_m\rangle=0,
	\label{eq:sec4_particle_mm_rho}
\end{equation}
and
\begin{equation}
	\partial_t f_m+\dfrac{1}{\varepsilon}(I-\Pi)
	\bigl[\v\cdot\nabla(\rho\mathcal{M}+f_m)\bigr]
	=-\frac{1}{\varepsilon^2}f_m .
	\label{eq:sec4_particle_mm_micro}
\end{equation}
Using \(\langle v\mathcal{M}\rangle=0\), the microscopic equation \eqref{eq:sec4_particle_mm_micro}
can also be written as
\begin{equation}
	\partial_t f_m
	+
	\frac{1}{\varepsilon}\mathcal{M}\v\cdot\nabla\rho
	+
	\frac{1}{\varepsilon}(I-\Pi)(\v\cdot\nabla f_m)
	=
	-\frac{1}{\varepsilon^2}f_m .
	\label{eq:sec4_particle_mm_micro2}
\end{equation}
This form makes explicit that the stiff relaxation acts only on the microscopic
correction, while the macroscopic density is advanced through the microscopic
flux \(\langle \v f_m\rangle\). By separating the macroscopic density-gradient
term from the residual microscopic transport, it also provides a convenient
formulation for a particle representation of \(f_m\).

The main difficulty is that both transport and relaxation are stiff in the
diffusion scaling. A direct particle discretization of
\eqref{eq:sec4_particle_mm_micro2} would therefore still inherit kinetic-scale
restrictions. Crestetto \emph{et al.}~\cite{ref:CCL2018} removed this difficulty
by reformulating the microscopic equation into a nonstiff evolution with
time-step-dependent coefficients. In first-order form, the reformulated equation
may be written schematically as
\begin{equation}
	\partial_t f_m
	=-\frac{1-e^{-\Delta t/\varepsilon^2}}{\Delta t}f_m
	-\varepsilon\frac{1-e^{-\Delta t/\varepsilon^2}}{\Delta t}
	\mathcal{G}(\rho,f_m),
	\label{eq:sec4_particle_mm_reform}
\end{equation}
where $\mathcal{G}(\rho,f_m)$ contains the transport contribution. This
reformulation makes the particle transport part uniformly stable and is the key
step that allows a practical AP particle micro--macro algorithm.

From a methodological standpoint, particle micro--macro schemes are significant
because they separate two roles that are coupled in a full kinetic solver. The
dominant equilibrium dynamics can be advanced efficiently on an Eulerian mesh
through macroscopic variables, while the residual nonequilibrium correction is
represented by particles without requiring a full deterministic phase-space
discretization. Particles are therefore used only for the microscopic component
\(f_m\), rather than for the whole distribution.
This changes the numerical role of particles. In DSMC and many model-particle
methods, particles carry both the large equilibrium component and the much
smaller nonequilibrium signal, so near-equilibrium computations suffer from
large sampling noise relative to the quantity of interest. In particle
micro--macro schemes, the equilibrium contribution is supplied by the
macroscopic solver, while particles represent only the genuinely nonequilibrium
residual. This reduces statistical noise in near-equilibrium regimes, lowers
the effective dimension of the kinetic correction, and makes the particle burden
decrease as the fluid limit is approached.

This philosophy has also been extended to gas mixtures. For a two-species BGK
model, Crestetto \emph{et al.}~\cite{ref:CKP2020} used the same species-wise
decomposition,
\[
f_s=f_{M,s}+f_{m,s},\qquad s=1,2,
\]
with vanishing microscopic moments
\begin{equation}
	\int \bm{\psi} f_{m,s}\,d\bxi=0,
	\qquad s=1,2 .
	\label{eq:sec4_mixture_zero_mom}
\end{equation}
A representative species equation then takes the form
\begin{equation}
	\partial_t f_{M,1}+\partial_t f_{m,1}
	+\bxi\cdot\nabla f_{M,1}+\bxi\cdot\nabla f_{m,1}
	=-\frac{1}{\varepsilon_1}\nu_{12}n_1 f_{m,1}
	+\frac{1}{\widetilde{\varepsilon}_1}\nu_{12}n_2
	\bigl(M_{12}-f_{M,1}-f_{m,1}\bigr),
	\label{eq:sec4_mixture_species1}
\end{equation}
with an analogous equation for the second species. Here $\varepsilon_1$ and $\widetilde{\varepsilon}_1$ are the dimensionless scaling parameters. The microscopic part is
solved by particles and the fluid part by a finite-volume scheme. Since the
particles represent only the nonequilibrium correction, the method reduces
particle noise and lowers the computational cost as the solution approaches the
fluid regime.

Particle micro--macro schemes therefore form a bridge between the particle
methods of Sec.~\ref{sec:particle_methods} and the macro-guided strategies
developed in the present section. They retain the flexibility of particles, but
only after the equilibrium structure has been separated from the kinetic state.
More broadly, they show that the macroscopic subsystem can be an active
component of the solver rather than merely an asymptotic limit recovered
\emph{a posteriori}. This viewpoint leads naturally to methods in which the
macroscopic equations are used not only for decomposition, localization, or
variance reduction, but also for nonlinear acceleration and solver
preconditioning.

\subsection{HOLO coupling and moment acceleration}
\label{subsec:holo_gsis}
While micro--macro and particle micro--macro methods reduce the cost of the nonequilibrium correction by decomposition, localization, or stochastic representation, another line of development starts from a different numerical observation. In highly collisional regimes, the main computational difficulty often lies not only in asymptotic stiffness, but also in the slow convergence of nonlinear or source iterations for the high-dimensional kinetic system. This motivates the introduction of a lower-order macroscopic system that is solved together with the original higher-order kinetic equation. The lower-order system carries the dominant large-scale information and provides closures, preconditioning data, or nonlinear-elimination guidance to the kinetic solver, while the high-order solver supplies the kinetic information needed to keep the coupled method accurate away from equilibrium.

The generic algebraic structure of HOLO methods is well summarized in ~\cite{Chacon2017HOLO}. If
$X$ denotes the low-order algebraic unknowns and $Y$ the high-order ones, the fully coupled nonlinear problem and its HO/LO decomposition may be written
abstractly as
\begin{equation}
{\mathcal R}(X,Y)=0,\qquad
{\mathcal R}^{\mathrm{HO}}(X,Y)=0,\qquad
{\mathcal R}^{\mathrm{LO}}(X,Y)=0 .
	\label{eq:sec4_holo_split}
\end{equation}
Here \(\mathcal R^{\mathrm{HO}}\) represents the high-order kinetic residual, whereas \(\mathcal R^{\mathrm{LO}}\) represents the lower-order moment or macroscopic residual. The essential HOLO step is nonlinear elimination. 
The high-order subsystem is formally solved as \(Y= H(X)\), and this relation is inserted into the low-order residual,
\begin{equation}
	G^{\mathrm{LO}}(X)
	=
	\mathcal R^{\mathrm{LO}}\bigl(X, H(X)\bigr)=0 .
	\label{eq:sec4_holo_elim}
\end{equation}
Thus the nonlinear iteration is driven by the lower-order system, while the high-order solve supplies the closures or correction data required by the
low-order equations. This nonlinear-elimination viewpoint is the core complexity-reduction idea behind HOLO algorithms \cite{Chacon2017HOLO}.

An early and transparent gas-kinetic example is the moment-based acceleration method of Taitano \emph{et al.}~\cite{Taitano2014HOLO} for a neutral-gas
Boltzmann transport equation with a BGK collision operator. Starting from the one-dimensional BGK model, the low-order moment system is obtained by taking the velocity moments of the high-order kinetic equation,
\begin{equation}
	\partial_t \bm W^{\mathrm{LO}}
	+
	\nabla\cdot
	\bm\Phi^{\mathrm{LO}}
	\bigl(
	\bm W^{\mathrm{LO}};
	\bm{P}^{\mathrm{HO}}, \bm{Q}^{\mathrm{HO}}
	\bigr)
	=
	\bm\Gamma^{\mathrm{HO}},
	\qquad
	\bm W^{\mathrm{LO}}=(\rho,\rho \u,\rho E)^T .
	\label{eq:sec4_taitano_moments}
\end{equation}
Here \(\bm{P}^{\mathrm{HO}}\) and \(\bm{Q}^{\mathrm{HO}}\) denote the
pressure and heat-flux moment closures supplied by the high-order kinetic
solution, while \(\bm\Gamma^{\mathrm{HO}}\) denotes the discrete consistency
corrections. The low-order equations propagate the conservative macroscopic
information and, at the same time, receive closure and consistency data from the
kinetic solve.

In the corresponding HOLO formulation, the high-order kinetic equation is
evolved with a Maxwellian constructed from the low-order moments, while the
high-order kinetic solution supplies the closure moments and consistency
corrections appearing in Eq.~\eqref{eq:sec4_taitano_moments}. Schematically,
the high-order kinetic equation may be written as
\begin{equation}
	\partial_t f^{\mathrm{HO}}
	+
	\bxi\cdot\nabla f^{\mathrm{HO}}
	=
	\frac{1}{\tau^{\mathrm{LO}}}
	\bigl(g^{\mathrm{LO}}-f^{\mathrm{HO}}\bigr),
	\label{eq:sec4_taitano_holo}
\end{equation}
where \(f^{\mathrm{HO}}\) denotes the high-order kinetic distribution and
\(g^{\mathrm{LO}}\) is the Maxwellian constructed from the low-order moments.
Together with the low-order moment system
\eqref{eq:sec4_taitano_moments}, Eq.~\eqref{eq:sec4_taitano_holo}
summarizes the essential HOLO coupling. The low-order variables determine the
Maxwellian and relaxation data used in the high-order solve, while the
high-order solution supplies the moment closures \(\bm{P}^{\mathrm{HO}}\) and
\(\bm{Q}^{\mathrm{HO}}\), together with the consistency correction
\(\bm\Gamma^{\mathrm{HO}}\), required by the low-order system.
The role of \(\bm\Gamma^{\mathrm{HO}}\) is to enforce agreement between the
high-order kinetic update and the low-order moment equations at nonlinear
convergence. This explicit treatment of discrete consistency is one of the
lasting contributions of moment-based acceleration methods
\cite{Taitano2014HOLO,Chacon2017HOLO}.

Recent developments have further clarified the relation between HOLO methods and micro--macro decomposition. A notable example is the work of Hauck \emph{et al.}~\cite{ref:HauckLaiuSchnake2025}, where a HOLO method is combined with a micro--macro decomposition for fully implicit time-stepping of the BGK model. In that formulation, the distribution is decomposed as in \eqref{eq:sec4_mm_basic}, and the smallness of $f_m$ in highly collisional regimes enables compressed storage of the kinetic perturbation away from initial and boundary layers. At the same time, the discontinuous Galerkin discretization in phase space naturally provides the moments required for HO/LO consistency. This work shows that the same equilibrium--nonequilibrium decomposition that supports AP analysis can also support nonlinear elimination,
moment-based preconditioning, and memory reduction.

Historically, micro--macro decomposition was developed mainly as a route to asymptotic-preserving reformulation, whereas HOLO and moment-acceleration methods were developed mainly to accelerate nonlinear or source iterations for kinetic solves. Recent work shows that this separation is no longer rigid. Once the kinetic solution is written as a macroscopic equilibrium part plus a reduced nonequilibrium correction, the resulting structure becomes useful not only for asymptotic analysis, but also for solver compression, nonlinear elimination, and moment-based preconditioning.

\subsection{General synthetic iterative schemes (GSIS)}
\label{subsec:gsis}
GSIS shares the use of coupled kinetic and macroscopic equations with HOLO and moment-acceleration methods. The relation between GSIS and HOLO-type methods has been discussed explicitly by Zeng \emph{et al.}~\cite{ZengSuWu2023UnsteadyGSIS,ZengZhangLiSuWu2026GSISRev}. At a broad level, both approaches solve a mesoscopic kinetic equation together with a lower-dimensional macroscopic system. The distinction lies in how the macroscopic equations are constructed and coupled to the kinetic solve. In many HOLO formulations, the low-order system mainly provides a discretely consistent accelerator, preconditioner, closure carrier, or nonlinear-elimination guide for the high-order kinetic solve. In GSIS, the synthetic macroscopic equations play a more active role by explicitly separating the NSF constitutive terms from higher-order corrections supplied by the kinetic iterate. They are constructed to propagate slow hydrodynamic information efficiently across the domain and to feed this information back into the kinetic iteration. Thus, the macroscopic equations are not merely auxiliary moment balances or closure approximations, but form an essential multiscale mechanism for accelerating kinetic convergence while preserving the correct near-continuum behavior.

This distinction is most important in the near-continuum regime. Conventional
iterative schemes for kinetic equations converge slowly there because
large-scale hydrodynamic information is transmitted only indirectly through
repeated local updates of the distribution function. Frequent molecular
collisions rapidly relax the local distribution, but the propagation of density,
velocity, and temperature perturbations across the domain remains inefficient
when it relies only on molecular streaming. The GSIS addresses this difficulty by coupling the
kinetic equation to synthetic macroscopic equations in which the
Navier--Stokes--Fourier (NSF) constitutive structure is made explicit \cite{SuZhuWangZhangWu2020JCP}.

To make the construction concrete, consider the linearized Boltzmann equation
used in the original GSIS formulation. The distribution is written as a small
perturbation around a global Maxwellian,
\begin{equation}
	f(\x,\bxi,t)=f_{\mathrm{eq}}(\bxi)+\alpha h(\x,\bxi,t),
	\qquad
	|\alpha h/f_{\mathrm{eq}}|\ll 1 ,
	\label{eq:sec4_gsis_linearization}
\end{equation}
where \(f_{\mathrm{eq}}\) is the normalized equilibrium distribution and
\(h\) is the perturbation distribution. The perturbation \(h\) satisfies the
linearized Boltzmann equation
\begin{equation}
	\partial_t h+\bxi\cdot\nabla h=L(h,f_{\mathrm{eq}}),
	\label{eq:sec4_gsis_lbe}
\end{equation}
where \(L\) denotes the linearized collision operator. Taking velocity moments
of \eqref{eq:sec4_gsis_lbe} gives the exact but unclosed macroscopic balance
equations used as the starting point of GSIS. In the standard linearized notation used in GSIS,
\(\varrho\), \(\theta\), and \(\bm{U}\) denote the density, temperature, and velocity
perturbations, respectively, while \(\bm{\sigma}\) and \(\bm{q}\) denote the stress
deviator and heat flux perturbations. With the Einstein summation convention, the moment system reads
\begin{equation}
	\begin{aligned}
		&\partial_t\varrho+\nabla\cdot\bm{U}=0,\\
		&2\partial_t\bm{U}+\nabla\varrho	+\nabla\theta
		+\nabla\cdot\bm{\sigma}=0,\\
		&\frac{3}{2}\partial_t\theta+\nabla\cdot\bm{q} +\nabla\cdot\bm{U}=0 .
	\end{aligned}
	\label{eq:sec4_gsis_moments}
\end{equation}
The system \eqref{eq:sec4_gsis_moments} is exact at the moment level, but it is
not closed because \(\sigma_{ij}\) and \(q_i\) are still determined by the
kinetic solution. If these quantities are evaluated only from the current
kinetic iterate, the macroscopic equations do not recover the NSF transport
mechanism until the kinetic solution is already close to convergence. GSIS
modifies this situation by rewriting the stress and heat flux in the synthetic
form
\begin{equation}
	\sigma_{ij}=-2\delta_{\mathrm{rp}}^{-1}
	\mathring{S}_{ij}
	+\mathrm{HoT}\,\sigma_{ij},
	\qquad
	q_i=-\frac{5}{4\Pr}\delta_{\mathrm{rp}}^{-1}
	\partial_i \theta
	+\mathrm{HoT}\,q_i,
	\label{eq:sec4_gsis_constitutive}
\end{equation}
where \(\delta_{\mathrm{rp}}\) is the rarefaction parameter, and
\[
\mathring{S}_{ij} = \frac{1}{2}\left(\partial_j U_i+\partial_i U_j\right)-\frac{1}{3}\partial_k U_k \delta_{ij}.
\]
The terms \(\mathrm{HoT}\,\sigma_{ij}\) and \(\mathrm{HoT}\,q_i\) are high-order rarefaction corrections obtained from the kinetic solution. In the iterative implementation, these high-order terms are evaluated from the current kinetic iterate and vanish in the NSF limit. The Newton law of viscosity and Fourier law of heat conduction are therefore embedded directly in the macroscopic iteration, while the kinetic iterate supplies only the non-NSF rarefaction corrections. This is the defining synthetic step of GSIS
\cite{SuZhuWangZhangWu2020JCP,ref:GSISFastAP2020}.

The coupling between the kinetic and synthetic levels can be described as a
moment-consistent feedback loop. At each outer iteration, a conventional
kinetic update first produces an intermediate distribution \(h^{(k+1/2)}\).
The moments and high-order terms are then computed from this kinetic iterate.
The synthetic macroscopic equations are solved to obtain updated macroscopic
fields. Finally, the distribution function is corrected so that its low-order
moments agree with the updated macroscopic solution. In schematic form,
\begin{equation}
	h^{(k+1)}=h^{(k+1/2)}+\Delta h^{(k+1)},
	\label{eq:sec4_gsis_feedback}
\end{equation}
where \(\Delta h^{(k+1)}\) is constructed from the differences between the
synthetic macroscopic moments and the moments of \(h^{(k+1/2)}\). In the
linearized implementation, this correction is formed from increments of
density, velocity, temperature, stress, and heat flux multiplied by
equilibrium-weighted low-order basis functions of \(\bxi\). The synthetic
system therefore does not replace the kinetic equation. It guides the kinetic
iteration by rapidly transmitting the hydrodynamic components that would
otherwise propagate slowly through repeated streaming--collision updates.

The resulting algorithm has two important consequences. First, the synthetic
equations provide a direct channel for long-range hydrodynamic information
transfer, leading to much faster convergence than conventional kinetic
iterations in the continuum and near-continuum regimes. Second, the NSF limit
can be preserved accurately even when the spatial cells and time steps are
chosen according to macroscopic rather than kinetic scales. The fast-convergence
and asymptotic-preserving properties were analyzed by Su \emph{et al.}\
\cite{ref:GSISFastAP2020}. Thus GSIS addresses convergence acceleration and
asymptotic consistency within the same kinetic--macroscopic iteration.

The first general GSIS for steady rarefied gas flows was introduced by
Su \emph{et al.}~\cite{SuZhuWangZhangWu2020JCP}. In that work, the
synthetic macroscopic equations were solved together with the Boltzmann
equation and were shown to recover the Navier--Stokes behavior while explicitly
containing Newton's law for stress and Fourier's law for heat conduction. This
construction removes the mean-free-path restriction on the spatial cell size
without requiring the time-dependent multiscale interface flux used in
UGKS-type methods. Since the synthetic iteration is formulated through moment
equations and kinetic high-order corrections, it is not tied to one particular
collision model and can be extended to more complicated gas kinetic
descriptions.

The GSIS idea was subsequently generalized in several directions. Zhu
\emph{et al.}~\cite{ref:GSIS_Zhu2021} extended it to nonlinear gas kinetic
equations and showed, through Fourier stability analysis and canonical
benchmarks, that the scheme remains rapidly convergent and accurate over a wide
range of rarefaction levels. Zeng \emph{et al.}\
\cite{ZengSuWu2023UnsteadyGSIS} then developed unsteady GSIS
formulations, where the challenge is not only the slow convergence of each
implicit time step but also the possible loss of the NSF limit when spatial
cells are too coarse. In that setting, the synthetic iteration preserves the
NSF behavior even when the time step is chosen from hydrodynamic rather than
collisional scales.

Further developments show that GSIS has become a general methodological
framework rather than a solver for one specific model problem. The method has been extended to rarefied gas mixtures
\cite{ref:GSIS_Mixture2024}, and was further broadened to polyatomic gases with rotational degrees of freedom
~\cite{ZengYuanZhangLiWu2023Polyatomic}. Liu \emph{et al.}\
\cite{ref:GSIS_Boundary2024} introduced a generalized boundary treatment for
multiscale simulations with complex gas--surface interaction. The review \cite{ZengZhangLiSuWu2026GSISRev} summarizes extensions to
nonlinear, polyatomic, multispecies, unsteady, moving-boundary, and optimization
problems, indicating the increasing scope of the GSIS framework.

A noteworthy feature of GSIS is its numerical flexibility. Because the kinetic
equation and the synthetic equations are solved as distinct but strongly
coupled subsystems, they may be discretized by different numerical methods and
even by different orders of accuracy. This is useful in practice: the
mesoscopic equation, which is expensive because of its phase-space dimension,
can be treated by a relatively simple kinetic discretization, while the
synthetic equations can benefit from mature CFD-type high-order discretizations.
In this respect, GSIS inherits the spirit of HOLO coupling while moving toward
a more practical multiscale solver architecture.

The synthetic-iteration idea has also spread beyond rarefied gas dynamics. In phonon transport, Liu \emph{et al.}\
\cite{Liu_Zhang_Yuan_GSIS_Phonon_2022} developed a GSIS for the phonon Boltzmann equation under Callaway's dual-relaxation model, where synthetic
macroscopic equations are coupled to the kinetic equation to accelerate the iteration in the small-Knudsen-number regime. This extension suggests that the
synthetic-equation viewpoint is not tied to one specific gas kinetic model. Rather, it provides a general strategy for accelerating mesoscopic solvers in
multiscale transport problems \cite{Liu_Zhang_Yuan_GSIS_Phonon_2022,ZengZhangLiSuWu2026GSISRev}.

\subsection{Deterministic--stochastic coupling}
\label{subsec:det_stoch_synthetic}
The preceding subsections show that macroscopic information can enter a kinetic
computation in several ways. In deterministic solvers, it appears through
micro--macro variables, low-order moment equations, or synthetic equations. The
same idea can also be used in stochastic simulation, where the goal is not only
to accelerate convergence but also to reduce variance, particle number, or
coarse-step bias. Deterministic--stochastic coupling in this sense is broader
than GSIS--DSMC alone. It includes particle micro--macro schemes, cellwise
hybrid representations, moment-guided Monte Carlo methods, limit-guided
multilevel Monte Carlo, and GSIS-type synthetic acceleration of particle
solvers. In this broader view, moment-guided Monte Carlo and related stochastic hybrids
are naturally placed alongside AP and kinetic--fluid coupling strategies \cite{DimarcoPareschi2014}.
	
The particle micro--macro schemes introduced in Sec.~\ref{subsec:mm_particle} are already deterministic--stochastic hybrids. The macroscopic component is
evolved by an Eulerian solver, while the microscopic correction \(f_m\) is represented by particles. Their main purpose is AP reformulation and
noise/cost reduction rather than synthetic acceleration, but they establish the basic principle that a deterministic macroscopic subsystem can reduce the
stochastic burden by removing the dominant equilibrium contribution from the particle representation~\cite{ref:CCL2018}.

An early deterministic--stochastic hybrid approach was proposed by Dimarco and Pareschi \cite{ref:DimarcoHybrid2008}. Their starting
point was neither a static kinetic--fluid domain decomposition nor a pure
particle method, but a cellwise hybrid representation in which the numerical
solution is written as a convex combination of a particle contribution and a
deterministic equilibrium contribution. For a discrete-velocity formulation, the
post-relaxation state is expressed as
\begin{equation}
	f_\alpha^{r}(x,t)=\bigl(1-\beta^{r}(x,t)\bigr)f_{\alpha}^{r,p}(x,t)
	+\beta^{r}(x,t)E_\alpha^{r}(x,t),
	\qquad
	\beta^{r}(x,t)=e^{-t/\varepsilon}\beta(x,0)+1-e^{-t/\varepsilon}.
	\label{eq:sec4_dimarco2008_hybrid}
\end{equation}
Here $\alpha$ denotes the discrete velocity index, the superscript $r$ denotes the
relaxation stage, $f_\alpha^{r,p}$ is the particle contribution, $E_\alpha^r$ is the
deterministic equilibrium contribution, and $\beta^r$ is the deterministic
fraction. The quantity \(\beta(x,0)\) is the
deterministic fraction at the beginning of the relaxation substep. In the near-equilibrium regime, $\beta^r$ tends to one, so the method
progressively degenerates toward
a deterministic fluid-dynamic description; away from equilibrium, a
non-negligible particle component is retained in order to preserve kinetic
fidelity. The same work also introduced componentwise variants that maximize
the deterministic fraction after transport, making clear that the hybridization
is performed within each cell through an equilibrium/nonequilibrium
decomposition, rather than by prescribing different solvers in different
spatial regions \cite{ref:DimarcoHybrid2008}.

The later fluid-solver-independent hybrid BGK method of Dimarco and Pareschi
\cite{ref:DimarcoFSI2010} may be viewed as a systematic generalization of this
cellwise hybrid construction. In that formulation, the kinetic and macroscopic
models are again solved over the whole domain and combined through a hybrid
representation, but the equilibrium component is no longer advanced by a particular
kinetic discretization. Instead,  it is evolved by a macroscopic solver for the
limiting fluid system, which can be chosen from a broad class of finite-volume
or finite-difference schemes. An important numerical ingredient is a
moment-matching transformation, which uses the macroscopic update to correct
the moments of the transported equilibrium samples.

For clarity, consider a one-dimensional particle velocity sample
\(\{\xi_j\}_{j=1}^J\). Let
\[
\mu_1^p=\frac{1}{J}\sum_{j=1}^J \xi_j,
\qquad
\mu_2^p=\frac{1}{J}\sum_{j=1}^J \xi_j^2
\]
be the empirical first and second velocity moments of the particle ensemble,
and let \(\mu_1^{mac}\) and \(\mu_2^{mac}\) denote the corresponding target
moments obtained from the macroscopic solver. The corrected particle velocities
are defined by

\begin{equation}
	\xi_j^{\ast}=\mu_1^{mac}+\lambda(\xi_j-\mu_1^p),
	\qquad
	\lambda=
	\left(
	\frac{\mu_2^{mac}-(\mu_1^{mac})^2}
	{\mu_2^p-(\mu_1^p)^2}
	\right)^{1/2},
	\qquad j=1,\dots,J .
	\label{eq:sec4_mgmc_match}
\end{equation}
Then the corrected ensemble satisfies
\[
\frac{1}{J}\sum_{j=1}^J \xi_j^{\ast}=\mu_1^{mac},
\qquad
\frac{1}{J}\sum_{j=1}^J (\xi_j^{\ast})^2=\mu_2^{mac}.
\]
Thus the transformation rescales the particle fluctuations and shifts the
particle mean so that the momentum and energy moments of the transported
equilibrium samples follow the macroscopic update. This gives a precise realization of the idea that macroscopic
information should guide the particle ensemble. In this sense, the
fluid-solver-independent construction turns the earlier cellwise hybrid
representation into a more general macro-guided stochastic framework
\cite{ref:DimarcoFSI2010}.

The moment-guided Monte Carlo (MGMC) method of Degond \emph{et al.}
\cite{ref:MGMC2011} adopts the same philosophy in a more explicitly stochastic
formulation. The solution of the moment equations is used to guide the evolution of
the particle distribution and to suppress fluctuations through moment matching.
As emphasized both in the original paper and in later reviews, MGMC improves
efficiency in near-equilibrium regimes, but it is essentially Euler-limit
preserving rather than NSF-preserving
\cite{ref:MGMC2011,DimarcoPareschi2014,ZengZhangLiSuWu2026GSISRev}.

Another branch of stochastic multiscale coupling uses the asymptotic limiting
model itself to bias or correct the Monte Carlo simulation. For the same class
of diffusive relaxation models represented by
\eqref{eq:sec4_diffusive_kinetic}, L{\o}vbak and Samaey
\cite{ref:APMLMC2023} introduced an asymptotic-preserving multilevel Monte
Carlo strategy. At the particle level, the method uses a kinetic--diffusion
surrogate, written here in one spatial dimension as
\begin{equation}
	\partial_t f+\frac{\varepsilon v}{\varepsilon^2+\Delta t}\partial_x f
	=\frac{\widetilde v^2\Delta t}{\varepsilon^2+\Delta t}\partial_{xx} f
	+\frac{1}{\varepsilon^2+\Delta t}\bigl(\rho\mathcal{M}(v)-f\bigr),
	\label{eq:sec4_apmlmc_surrogate}
\end{equation}
where $\widetilde v^2=\int{v^2{\cal M}(v)\,\d v}$ is the fixed variance of the normalized zero-mean equilibrium velocity distribution. This formulation yields stable coarse simulations even when $\Delta t\gg \varepsilon^2$. The corresponding multilevel
estimator for a quantity of interest $\Phi$ can be written as
\begin{equation}
	\widehat Y(t^{\ast})=\sum_{\ell=0}^{L}\widehat Y_{\ell}(t^{\ast}),
	\qquad
	\widehat Y_{\ell}(t^{\ast})
	=\frac{1}{P_{\ell}}\sum_{p=1}^{P_{\ell}}
	\Bigl[
	\Phi(X^{N_{\ell}}_{p,\Delta t_{\ell}},V^{N_{\ell}}_{p,\Delta t_{\ell}})
	-\Phi(X^{N_{\ell-1}}_{p,\Delta t_{\ell-1}},V^{N_{\ell-1}}_{p,\Delta t_{\ell-1}})
	\Bigr],
	\label{eq:sec4_apmlmc_telescopic}
\end{equation}
with the convention that the second term is omitted for $\ell=0$. Here
$P_\ell$ is the number of correlated samples on level $\ell$,
$\Delta t_\ell$ is the time step, $N_\ell$ denotes the corresponding number of
time steps up to the observation time $t^\ast$, and
$(X^{N_\ell}_{p,\Delta t_\ell},V^{N_\ell}_{p,\Delta t_\ell})$ denotes the
particle position--velocity state of sample $p$ on level $\ell$. The
diffusive-limit model provides a low-cost biased estimate, and finer kinetic
simulations are used only to correct that bias \cite{ref:APMLMC2023}. Although
this class of methods is not built from a micro--macro decomposition in the
strict sense of \eqref{eq:sec4_mm_basic}, it belongs to the same macro-guided
stochastic tradition: a lower-dimensional asymptotic description is used to
reduce the cost of Monte Carlo simulation in highly collisional regimes.

Against this broader background, GSIS--DSMC should be regarded as one
particularly powerful member of a larger family rather than as the sole
deterministic--stochastic coupling strategy \cite{ref:GSIS_DSMC2025}. What distinguishes the GSIS-based
approach is that the auxiliary macroscopic equations are not merely used for
moment matching, bias correction, or reduction of particle number. Instead, they are
constructed as synthetic equations that actively propagate slow hydrodynamic
information and thereby accelerate convergence to steady or quasi-steady
solutions. In this sense, GSIS--DSMC stands to MGMC and AP-biased Monte Carlo
methods in much the same way that GSIS stands to ordinary HOLO acceleration. In both cases, the macroscopic subsystem is elevated from an auxiliary correction procedure to a genuine multiscale propagation mechanism.

A representative example is the GSIS--DSMC strategy of Luo \emph{et al.}\
\cite{ref:GSIS_DSMC2025} for boosting the convergence of DSMC. The key idea is
to retain a stochastic kinetic solver for the Boltzmann equation, but to embed
it in a deterministic--stochastic coupling framework where macroscopic
synthetic equations are solved simultaneously and provide reciprocal feedback
to the particle simulation. Importantly, this is done without physical domain
partitioning. Instead, the macroscopic properties updated by GSIS are used as
continuous guidance for the particle evolution, both through direct linear
transformation of particle information and through modification of the sampling
distribution used in the stochastic collision step.

In the planar Fourier-flow example studied in Ref.~\cite{ref:GSIS_DSMC2025}, the synthetic equations transmit the slowly
varying heat-flux and stress information, while the stochastic solver supplies the high-order stress and heat-flux corrections through time-averaged particle
data. Thus the synthetic equations play the same structural role as in deterministic GSIS, but the high-order solver is now stochastic rather than grid based. This development extends the HOLO/GSIS paradigm beyond purely deterministic discretizations and provides a stochastic route to fast near-continuum convergence.

A closely related implementation-oriented development is the direct intermittent
GSIS--DSMC (DIG) framework \cite{LuoWu2024DIG}. DIG applies the
solution of the macroscopic synthetic equations only intermittently, typically
after a number of Monte Carlo steps, through a linear transformation of particle
velocities. The purpose is to retain the NSF-limit asymptotic-preserving and
fast-convergence properties of GSIS while making only limited modifications to
a standard DSMC code. This intermittent strategy has also been extended toward more complex gas models by incorporating
synthetic equations for internal-energy relaxation together with the usual
translational dynamics \cite{ZengZhangLiSuWu2026GSISRev}.

Considered jointly, the stochastic side of macro-guided multiscale methodology
contains three closely related ideas. Particle micro--macro decomposition uses
particles only for the microscopic residual after the equilibrium component has
been separated. Moment-guided and limit-guided Monte Carlo methods use
macroscopic or asymptotic information to reduce variance, stiffness, particle
number, or sampling cost. GSIS-type synthetic acceleration goes one step
further by turning the macroscopic subsystem into an explicit propagation
mechanism for slow hydrodynamic information. This broader perspective avoids
identifying deterministic--stochastic coupling with GSIS alone, while still
recognizing that GSIS presently offers one of the most systematic routes toward
fast NSF-consistent stochastic simulation in the near-continuum regime.

\subsection{Reduced-order and low-rank extensions}
\label{subsec:sec4_reducedOrder}
The preceding subsections have reviewed macro-guided strategies in which
macroscopic, moment, or synthetic equations are coupled to a kinetic solver to
improve asymptotic consistency, localization, stochastic variance reduction, or
nonlinear acceleration. Recent reduced-order and low-rank extensions fit
naturally at the end of this discussion. They do not constitute a separate
kinetic--fluid coupling paradigm. Rather, they extend the same
decomposition-based principle toward the compression of high-dimensional kinetic
degrees of freedom. The common idea is to separate structures that carry the
dominant macroscopic information from kinetic corrections that require more
detailed phase-space resolution.

One such direction is the use of reduced-order structures within a
micro--macro framework. Peng \emph{et al.}~\cite{ref:Peng2024} developed a
micro--macro decomposed reduced basis method for the time-dependent radiative
transfer equation. In that work, reduced spaces are constructed separately for
the macroscopic and microscopic components, so that the reduced-order surrogate
respects the equilibrium structure of the decomposed system. In the notation of
the present section, this method may be viewed as a reduced-order realization of
the same micro--macro separation used in AP schemes, rather than as a new
decomposition of the distribution function itself. The reduced model is therefore
not obtained by compressing the full distribution in an undifferentiated manner. 
Instead, it exploits the fact that the equilibrium-dominated component and the nonequilibrium
correction play different asymptotic roles and may be approximated in different
low-dimensional spaces. This example shows that micro--macro decomposition is now
influencing not only AP schemes and nonlinear acceleration, but also
reduced-order modeling.

A related direction is the use of low-rank velocity-space representations in
kinetic solvers for multiscale BGK-type models. Galindo-Olarte
\emph{et al.}~\cite{ref:QiuNakao2025} proposed a nodal discontinuous Galerkin
formulation in which the physical-space discretization is retained in full
rank, while low-rank compression is applied only in velocity space. For each
physical element or degree of freedom, the velocity-space coefficient matrix is
represented in the low-rank form
\begin{equation}
	\mathsf C_i^p(t)=\Theta_i^p(t)\,S_i^p(t)\,\Psi_i^p(t)^T .
	\label{eq:sec4_qiu_lowrank}
\end{equation}
Here \(i\) denotes the physical cell or element index, \(p\) denotes the local
nodal degree of freedom in the discontinuous Galerkin discretization, and
\(\mathsf C_i^p(t)\) is the corresponding matrix of velocity-space coefficients
at time \(t\). This notation is local to the low-rank representation. The
matrices \(\Theta_i^p(t)\) and \(\Psi_i^p(t)\) contain low-rank basis vectors in
the two velocity directions, while \(S_i^p(t)\) is a small core matrix containing
the rank-dependent coefficients. The rank may be chosen adaptively in the
numerical implementation. The velocity moments are then evaluated through the
low-rank representation and advanced together with auxiliary moment equations.
This provides a concrete example of how velocity-space compression, nodal DG
discretization, IMEX time integration, and macro-assisted multiscale evolution
can be combined in one kinetic solver.

These reduced-order and low-rank extensions are not central to the historical
development of hybrid kinetic--fluid methods, but they are important for
understanding the present direction of the field. The same structural idea that
underlies micro--macro AP schemes, namely the
separation of equilibrium-dominated macroscopic information from
high-dimensional kinetic corrections, can also be
used to design reduced bases, compressed phase-space representations, and
auxiliary moment-coupled solvers. In this sense, decomposition-based multiscale
methodology is becoming connected not only with asymptotic analysis and
iteration acceleration, but also with model reduction and high-dimensional
compression.

\subsection{Section remarks}
The methods reviewed in this section share a common principle. They
use macroscopic information to guide, constrain, accelerate, or partially replace
the kinetic description. In micro--macro AP methods, the macroscopic limit and
the kinetic correction are separated at the equation level, so that the numerical
scheme remains consistent across kinetic and fluid regimes. In hybrid
kinetic--fluid methods, the macroscopic model is used locally in near-continuum
regions, while the kinetic solver is retained where nonequilibrium effects
remain important. In synthetic iterative schemes, macroscopic or moment equations
are not used to replace the kinetic equation, but to accelerate its convergence
by transmitting the slow hydrodynamic error modes more efficiently. In
macro-guided stochastic methods, moment equations and deterministic corrections
reduce particle noise and improve near-continuum behavior. Reduced-order and
low-rank extensions further show that the same separation principle can be used
to compress high-dimensional kinetic representations.

Considered jointly, these developments indicate that hybrid and synthetic
strategies are not merely domain-decomposition techniques. They represent a
broader macro-guided layer of multiscale kinetic methodology. The kinetic
equation remains the most detailed description, but macroscopic equations,
moments, reduced bases, or synthetic corrections provide additional structures
that improve efficiency and asymptotic consistency. The key feature is that the
macroscopic component is introduced as an auxiliary organizing mechanism. It
guides the kinetic solver from outside or alongside the distribution function.

This viewpoint also exposes a limitation of the macro-guided strategy. The
multiscale organization is often imposed through an external coupling between
kinetic and macroscopic descriptions, through regional switching, or through
auxiliary equations. Although highly effective, such approaches do not always
alter the kinetic representation according to the local observation scale,
relaxation process, or collision history. This limitation motivates a further
layer of multiscale methodology, where the transition between particle-like and
continuum-like behavior is built directly into the representation and evolution
of the distribution function.

\section{Wave--particle direct modeling}
\label{sec:wave_particle_collision}
Sec.~\ref{sec:hybrid_macro_micro} considered macro-guided strategies, where a kinetic
description is coupled to macroscopic, moment, reduced, or synthetic equations.
The present section turns to a different organizing principle. Instead of using
an auxiliary macroscopic system to guide the kinetic solver, wave--particle and
collision-history-based methods reorganize the kinetic representation itself.
The distribution is decomposed into components such as uncollided and collided
parts, pre-collision and post-collision parts, or particle and wave parts, and
each component is represented according to its transport and relaxation behavior
over the numerical time step.

This viewpoint has both transport-theory and direct-modeling origins. First-collision and collided--uncollided decompositions were introduced to separate sharply anisotropic free-streaming components from smoother scattered components. In the unified gas-kinetic framework, the same principle becomes a conservative finite-volume construction. UGKS couples transport and collision in the interface evolution, while the unified gas-kinetic particle (UGKP) \cite{Li_UGKP_Photon_2018} and unified gas-kinetic wave-particle (UGKWP) \cite{LiuZhuXu2020UGKWP} represent the kinetic state partly by particles and partly by analytic or wave-like components.

The section proceeds from first-collision source ideas and time-dependent
collision hybrids to UGKP and UGKWP under the UGKS framework. It then reviews
UGKWP-family developments, simplified or related wave--particle branches, and
the kinetic population representation of UGKWP, which clarifies the structure
behind the practical algorithm.

\subsection{First-collision roots}
\label{subsec:sec5_first_collision}

Early forms of collision-based decomposition appeared in transport calculations
for localized-source problems. The key difficulty is that a sharply localized
source generates a highly anisotropic free-streaming component, which is hard
to represent accurately by deterministic angular discretizations and may lead
to severe ray effects. A natural remedy is to separate the solution into an
uncollided contribution, which has not yet undergone scattering, and a collided
or scattered contribution, which is smoother in angle and space. The singular
free-streaming part can then be treated by an analytic, ray-tracing, or Monte
Carlo description, while the scattered part is handled by a deterministic
transport solver.

A representative formulation is the first-collision source method. In the
notation of neutral-particle transport, the angular flux may be decomposed
schematically as
\begin{equation}
	\Psi=\Psi^{u}+\Psi^{c},
	\label{eq:sec5_first_collision_basic}
\end{equation}
where $\Psi^{u}$ denotes the uncollided, or first-flight, contribution and
$\Psi^{c}$ denotes the collided, or scattered, contribution. For steady neutral-particle transport, $\Psi^{u}$ can often be evaluated
analytically or by ray tracing/Monte Carlo, and its first-collision
contribution then acts as a distributed source for the deterministic solution
of $\Psi^{c}$. The practical motivation is that discrete ordinates methods perform poorly for
strongly anisotropic free-streaming fields generated by localized sources, but
become much more effective once scattering has redistributed the source in
angle and space.

This viewpoint was stated clearly in the first-collision source method of
Alcouffe~\cite{ref:Alcouffe1985}, where localized sources embedded in
collision-dominated media were identified as a natural setting for the method.
In such problems, the first-collision contribution supplies an auxiliary source
for the deterministic transport calculation, thereby mitigating ray effects
without abandoning the advantages of discrete ordinates for the scattered
field. In modern terminology, this may be regarded as an early instance of
collision-history-based decomposition, although its original purpose was
mainly the reduction of angular-discretization artifacts rather than the
construction of a general multiscale kinetic framework. The later
balance-preserving refinement by Alcouffe \emph{et al.}~\cite{ref:AlcouffeOdellBrinkley1990}
further clarified how first-collision source ideas can be incorporated
consistently into discrete $S_N$ transport solvers.

The significance of these early transport works lies less in their detailed
implementation than in the organizing principle they introduced. The transport
solution can be decomposed according to collision history, and different
numerical descriptions can then be assigned to the resulting components.
Initially motivated by ray-effect reduction, this idea later evolved into a
broader representation-level strategy for multiscale transport.

\subsection{Time-dependent collision hybrids}
\label{subsec:sec5_collision_hybrid}

The next stage in this line of development was the extension from steady
first-collision source corrections to genuinely time-dependent
collision-based hybrid methods. In this formulation, the collided--uncollided
separation is no longer used only to treat a localized source. Instead, the
kinetic evolution itself is repartitioned over each time step according to
whether particles have undergone collision or scattering during that interval.
Thus collision history becomes a dynamic numerical organizing variable rather
than a fixed source decomposition.

A representative formulation was proposed by Hauck and McClarren
\cite{ref:HauckMcClarren2013} for time-dependent linear kinetic transport.
Their hybrid method consists of three basic operations, namely partitioning the
kinetic equation into collisional and noncollisional components, applying
different numerical methods to the two components, and repartitioning the
kinetic distribution after each time step. In contrast to classical spatial domain
decomposition, this is a decomposition of the transport dynamics rather than of
the physical domain. The uncollided component carries the highly anisotropic
free-streaming part of the solution and is therefore evolved with a
high-fidelity discretization, whereas the collided component can be treated
with a lower-fidelity discretization because scattering tends to smooth the
distribution in angle and space.

This collision-based perspective was later extended to the BGK equation by
Shin \emph{et al.}~\cite{ShinHauckMcClarren2024}. For a hyperbolically scaled
BGK model with source term, the distribution is decomposed over each time
interval \(t\in(t_n,t_{n+1})\) as
\begin{equation}
	f=f_u+f_c ,
	\label{eq:sec5_bgk_split}
\end{equation}
where \(f_u\) and \(f_c\) denote the uncollided and collided components,
respectively. A representative split system is
\begin{equation}
	\partial_t f_u+\bm{\xi}\cdot\nabla f_u
	+\frac{1}{\varepsilon}f_u=S,
	\label{eq:sec5_bgk_uncollided}
\end{equation}
\begin{equation}
	\partial_t f_c+\bm{\xi}\cdot\nabla f_c
	+\frac{1}{\varepsilon}f_c
	=\frac{1}{\varepsilon}g(f_u+f_c),
	\label{eq:sec5_bgk_collided}
\end{equation}
with \(f_c(\bm{x},\bm{\xi},t_n)=0\) at the beginning of the time step and
\(f_u(\bm{x},\bm{\xi},t_n)=f(\bm{x},\bm{\xi},t_n)\) obtained from the
recombined solution at the previous time level. Here \(S\) denotes a prescribed
source. The uncollided equation is therefore a
damped transport equation with source, while the collided equation represents
the relaxation-generated component. Their sum recovers the original BGK
equation over the time interval.

An important numerical consequence of this decomposition is that the two
components can be evolved with different temporal strategies. In the BGK hybrid
method of Shin \emph{et al.}~\cite{ShinHauckMcClarren2024}, the uncollided
equation is treated implicitly, while the collided moments are evolved with a second-order predictor-corrector method whose transport part is explicit. As a result, the time step is constrained by the fluid-wave speed rather than by the
largest kinetic velocity in the computational velocity domain, which gives the
method a practical advantage over standard IMEX micro--macro formulations in
high-speed or broad-velocity-range applications.

At the same time, this collision-based hybridization remains primarily a
strategy for assigning different numerical fidelities to different components
of the kinetic solution. A related use of collision-history information appears in
the conservative finite-volume evolution of UGKP and UGKWP, as discussed next.

\subsection{Unified gas-kinetic particle and wave-particle method}
\label{subsec:sec5_ugkwp_framework}
In UGKP and UGKWP \cite{LiuZhuXu2020UGKWP}, collision-history information enters the
conservative finite-volume evolution inherited from UGKS. Free transport, collision, and
hydrodynamic response are coupled in the numerical flux over the time step. The
wave--particle representation organizes the local
transport process through particle tracking and analytic wave contributions.

This construction follows the direct-modeling philosophy of UGKS
\cite{XuHuang2010UGKS}. UGKP and UGKWP are therefore not external
couplings between a particle solver and a continuum solver. They are particle
and wave--particle realizations of the same multiscale evolution model that
underlies UGKS. The conservative variables are still updated through
finite-volume fluxes, while the kinetic state is represented according to the
local transport--collision balance. Collisionless transport is naturally
carried by particles, whereas strongly collisional contributions can be retained
analytically through an equilibrium or wave component.

This connection can be understood from the UGKS integral solution and the
corresponding conservative update reviewed in Sec.~\ref{subsec:ugks}. The
time-dependent interface distribution in Eq.~\eqref{eq:ugks_integral_solution}
contains a transported initial part and an equilibrium-generated relaxation
part, and the conservative update in Eq.~\eqref{eq:ugks_W_update} is built from
the associated time-integrated flux. The time-averaged interface distribution
and the corresponding microscopic and macroscopic fluxes are given in
Eqs.~\eqref{eq:ugks_time_avg_f}--\eqref{eq:ugks_macro_flux}. Thus UGKS already
contains a collision-history interpretation at the flux level. The transported
part corresponds to molecules that remain collisionless up to the interface
time, whereas the relaxation integral represents the collisional contribution.
UGKP and UGKWP retain this conservative multiscale evolution, but replace the
deterministic velocity-space representation by a particle or wave--particle
representation.

From the standpoint of methodological development, UGKP and UGKWP should be
distinguished. The UGKP method first appeared
as an explicitly named multiscale particle realization in photon and radiative
transfer problems \cite{Li_UGKP_Photon_2018,ref:UGKP_TRT2020}. In those works,
a finite-volume macroscopic update is coupled to particle tracking of
nonequilibrium transport, so that the method recovers diffusion-type behavior
in the collision-dominated limit and particle free transport in the optically
thin limit. This radiative branch is important historically because it
establishes UGKP as a genuine multiscale particle methodology, rather than as a
minor modification of Monte Carlo transport.

For gas dynamics, the same logic was subsequently embedded in the UGKS
framework and then refined into UGKWP \cite{LiuZhuXu2020UGKWP}. In UGKP, the distribution is still
advanced according to the multiscale interface evolution represented by
Eq.~\eqref{eq:ugks_integral_solution}, but the velocity distribution is carried
by simulation particles rather than by a deterministic discrete-velocity mesh.
Each simulation particle is assigned a free-transport time by sampling the
first-collision time. For a local constant relaxation time, the survival
probability up to time \(t\) is
\begin{equation}
	P_{\mathrm{free}}(t)=e^{-t/\tau},
	\label{eq:sec5_ugkp_survival_distribution}
\end{equation}
the free-transport time within one step is sampled as
\begin{equation}
	t_f=\min\{-\tau\ln\eta,\Delta t\},
	\qquad
	\eta\sim\mathscr{U}(0,1),
	\qquad
	\Delta t=t_{n+1}-t_n .
	\label{eq:sec5_first_collision_time_sampling}
\end{equation}
Particles with \(t_f=\Delta t\) remain collisionless over the whole time step,
whereas particles with \(0<t_f<\Delta t\) are treated as collisional and are
removed after their first collision
\cite{Li_UGKP_Photon_2018,ref:UGKP_TRT2020,LiuZhuXu2020UGKWP,ref:UGKWP_II_2019}.
The trajectory of particle \(k\) during its free-transport stage is
\begin{equation}
	\bm{x}_k^{\,*}=\bm{x}_k^n+\bm{\xi}_k t_{f,k},
	\label{eq:sec5_ugkp_particle_streaming}
\end{equation}
where \(\bm{x}_k^n\) and \(\bm{\xi}_k\) are the particle position and velocity
at the beginning of the step.

As a finite-volume particle method, UGKP has a coupled macro--micro update. On
the microscopic level, particles are streamed up to their sampled
free-transport time. On the macroscopic level, the conservative update combines
the free-transport contribution tallied from tracked particles and the
collisional contribution computed analytically from the equilibrium part of the
UGKS-type integral evolution. Let \(\mathscr P_{i,\mathrm{in}}^{\mathrm{fr}}\) and
	\(\mathscr P_{i,\mathrm{out}}^{\mathrm{fr}}\) denote incoming and outgoing
	boundary-crossing events during free transport. The net particle contribution
to the conservative update is
\begin{equation}
			\bm W_i^{\mathrm{fr}}
			=\sum_{k\in\mathscr P_{i,\mathrm{in}}^{\mathrm{fr}}}\boldsymbol\phi_k
			-\sum_{k\in\mathscr P_{i,\mathrm{out}}^{\mathrm{fr}}}\boldsymbol\phi_k,\qquad
			\boldsymbol\phi_k
			=\left(m_k,m_k\bm\xi_k,\frac12m_k|\bm\xi_k|^2\right)^T.
	\label{eq:sec5_ugkp_particle_tally}
\end{equation}

The collisional part of the transport process is not followed by individual
post-collision particles. Instead, its flux is computed analytically in the
same spirit as UGKS. The macroscopic conservative update may therefore be
written as
\begin{equation}
	\bm{W}_i^{n+1}
	=
	\bm{W}_i^{n}
	-\frac{1}{|V_i|}\sum_{j\in N(i)}\bm{\mathcal F}_{ij}^{eq}|S_{ij}|
	+\frac{\bm{W}_{i}^{fr}}{|V_i|},
	\label{eq:sec5_ugkp_macro_update}
\end{equation}
where \(\bm{\mathcal F}_{ij}^{eq}\) denotes the collisional or equilibrium flux
contribution generated from the integral solution. After the total conservative
variables are updated, the contribution of surviving collisionless particles is
subtracted to obtain the hydrodynamic part:
\begin{equation}
	\bm{W}_i^{p,n+1}
	=\frac{1}{|V_i|}\sum_{k\in V_i^{n+1}}\boldsymbol{\phi}_k,
	\qquad
	\bm{W}_i^{h,n+1}
	=\bm{W}_i^{n+1}-\bm{W}_i^{p,n+1}.
	\label{eq:sec5_ugkp_wh_update}
\end{equation}
Here \(\bm W_i^{p,n+1}\) is the particle-carried part and
\(\bm W_i^{h,n+1}\) is the hydrodynamic part associated with particles that
have collided and have been removed. At the beginning of the next time step,
this hydrodynamic part is sampled again from the local equilibrium. Thus UGKP
is already a multiscale particle realization of UGKS, in which the free transport is
resolved kinetically, while the collisional transport is retained analytically and
through conservative macroscopic evolution.

UGKWP improves UGKP precisely at the resampling stage \cite{LiuZhuXu2020UGKWP}. In UGKP, the entire
hydrodynamic part \(\bm W^h\) is resampled as particles at the beginning of the
next step, even though many of these particles will collide almost immediately.
The key observation of the UGKWP method is that only the collisionless fraction of the
hydrodynamic part needs to be represented by particles. If \(\bm W_i^h\) denotes
the hydrodynamic wave part in cell \(V_i\), then only
\begin{equation}
	\bm{W}_i^{hp}=e^{-\Delta t/\tau_i}\bm{W}_i^h
	\label{eq:sec5_ugkwp_collisionless_resampling}
\end{equation}
is resampled as particles for the next step, while the remaining fraction
\((1-e^{-\Delta t/\tau_i})\bm W_i^h\) stays in the analytic wave
representation and contributes to the flux deterministically.
 
In the original UGKWP formulation, the free-transport flux of
the unsampled collisional wave part can be written as
\begin{align}
	\bm{\mathcal F}_{ij}^{fr,wave}
	&=\bm{\mathcal F}_{ij,UGKS}^{fr}(\bm{W}^h)
	-\bm{\mathcal F}_{ij,DVM}^{fr}(\bm{W}^{hp}) \nonumber\\
	&=\int (\bm{\xi}\cdot\bm{n}_{ij}) \bm{\psi}
	\Bigl[
	(q_4-\Delta t e^{-\Delta t/\tau})g_0^h
	+\Bigl(q_5+\frac{\Delta t^2}{2}e^{-\Delta t/\tau}\Bigr)
	\bm{\xi}\cdot\nabla{g}^h
	\Bigr]\,d\bm{\xi}.
	\label{eq:sec5_ugkwp_wave_flux_formula}
\end{align}
Here \(\bm{\mathcal F}_{ij,UGKS}^{fr}(\bm W^h)\) is the free-transport flux that would be
obtained from the UGKS integral solution for the hydrodynamic state \(\bm W^h\),
\(\bm{\mathcal F}_{ij,DVM}^{fr}(\bm W^{hp})\) is the free-transport flux represented by
the sampled particle fraction, \(g_0^h\) is the Maxwellian determined by
\(\bm W^h\) at $t_n$, \(g^h\) denotes its spatial reconstruction, and
the time-integration coefficients for locally frozen $\tau$ are $q_4=\tau(1-e^{-\Delta t/\tau})$, $q_5=\tau\Delta t\,e^{-\Delta t/\tau}
-\tau^2 (1-e^{-\Delta t/\tau})$. Correspondingly, the conservative update in UGKWP becomes
\begin{equation}
	\bm{W}_i^{n+1}
	=
	\bm{W}_i^n
	-\frac{1}{|V_i|}\sum_{j\in N(i)}
	\bigl(\bm{\mathcal F}_{ij}^{eq}+\bm{\mathcal F}_{ij}^{fr,wave}\bigr)|S_{ij}|
	+\frac{\bm{W}_{i}^{fr,p}}{|V_i|},
	\label{eq:sec5_ugkwp_flux_update}
\end{equation}
where \(\bm W_i^{fr,p}\) is the free-transport contribution tallied from the
sampled particles. A one-step UGKWP update can therefore be summarized as
follows: (1) decompose the current state into a particle part \(\bm W^p\) and a
wave part \(\bm W^h\); (2) sample only the collisionless fraction of the wave part, stream these newly sampled particles for the
full time step,  and advance previously retained particles up to their sampled first-collision times; (3) tally the particle
free-transport flux; (4) evaluate the unsampled wave flux and the equilibrium flux
analytically, and update the conservative variables; and (5) reconstruct the
wave--particle decomposition for the next step. This replacement of repeated
resampling by analytic wave evolution is the essential improvement of UGKWP
over UGKP.

The multiscale consequence is immediate. In the highly rarefied regime,
\(e^{-\Delta t/\tau}\) is close to one, particle transport dominates, and
UGKWP behaves as a particle method. In the continuum regime,
\(e^{-\Delta t/\tau}\) becomes very small, so that only a small fraction of the
hydrodynamic part is resampled as particles and the method reduces to a
gas-kinetic hydrodynamic solver. In the transition regime, wave and particle
representations coexist inside each control volume. This adaptive
wave--particle balance is the main reason why UGKWP is efficient in
near-continuum and transitional flows while retaining kinetic fidelity in
rarefied regimes.

\subsection{UGKWP-family developments}
\label{subsec:sec5_ugkwp_family}

With the basic UGKP/UGKWP construction in place, the methodology rapidly
developed into a broader family of wave--particle multiscale methods. The
earliest developments focused on geometric generality and large-scale gas-flow
applications. The unstructured-mesh formulation extended the wave--particle
framework from structured benchmark configurations to complex geometries, where
cell volumes, face normals, and local time scales may vary strongly across the
computational domain~\cite{ref:UGKWP_II_2019}. Subsequent three-dimensional
UGKWP solvers further demonstrated that the same wave--particle mechanism can
be embedded in finite-volume infrastructures suitable for practical aerospace
flows with large variations of local Knudsen number
\cite{ref:UGKWP_3D_2020,ref:UGKWP_AIAA2024}. 
These developments are important
not only as implementation advances, but also as evidence that the
collision-based wave--particle decomposition is compatible with the
unstructured-grid and large-scale parallel computations required in
engineering-scale multiscale simulations.

Across these developments, the central mechanism remains the same. During one
numerical time step, a particle either survives as a free-transport particle or
is absorbed into the collisional contribution. The survival factor
\(e^{-\Delta t/\tau}\), introduced in
Eq.~\eqref{eq:sec5_ugkwp_collisionless_resampling}, determines the fraction of
the state represented by particles, while the complementary collisional part is
retained analytically in the wave representation. Thus the particle component
carries the kinetic free-transport information, whereas the wave component
evolves collectively through macroscopic variables and gas-kinetic fluxes. This
same mechanism underlies later extensions and adaptive refinements of the
UGKWP family.

A second direction is the extension of the UGKWP idea beyond monatomic gas
dynamics. Representative examples include photon transport and plasma-related transport
\cite{ref:UGKWP_Photon2020,Liu_Xu_UGKWP_MixturePlasma_2021}.
In these applications, the detailed carrier physics differs from that of
neutral monatomic gases, but the same representation principle is retained.
Ballistic or weakly interacting components are represented by particles, while
strongly interacting or collision-dominated components are retained through a
wave, analytic, or macroscopic contribution. The detailed physical models and
cross-disciplinary implications of these extensions are discussed later in
Sec.~\ref{sec:extensions_other_transport}.

A third direction concerns more complex molecular physics within gas dynamics
itself. For diatomic gases, UGKWP was first extended to translational--rotational nonequilibrium \cite{ref:UGKWP_Diatomic2021}, and later to
translational--rotational--vibrational nonequilibrium \cite{ref:UGKWP_DiatomicRV2024}. In these formulations, the wave--particle
framework is retained, but the relaxation model is enriched by internal-mode
relaxation processes. A schematic form of the kinetic model may be written as
\begin{equation}
	\partial_t f+\bm{\xi}\cdot\nabla f
	=
	\frac{g_t-f}{\tau}
	+\frac{g_{tr}-g_t}{Z_r\tau}
	+\frac{g_M-g_{tr}}{Z_v\tau},
	\label{eq:sec5_diatomic_relaxation}
\end{equation}
where \(g_t\), \(g_{tr}\), and \(g_M\) denote successive equilibrium states
associated with translational, translational--rotational, and full
translational--rotational--vibrational equilibrium, respectively. The
parameters \(Z_r\) and \(Z_v\) are the rotational and vibrational collision
numbers, and \(\tau\) is the translational relaxation time. 
The wave--particle decomposition is then applied to a relaxation process with
multiple time scales. Free-transport particles still represent the kinetic
transport part of the distribution, whereas the wave component carries the
strongly collisional contribution associated with translational and internal
mode relaxation. Thus the UGKWP principle is preserved, but the local
equilibrium structure and the resampling procedure must account for rotational
and vibrational nonequilibrium.

The same philosophy has also been extended beyond single-phase gas transport.
For gas--particle two-phase flow, the gas phase is evolved by a gas-kinetic
Navier--Stokes solver, while the dispersed solid-particle phase is represented
by UGKWP \cite{Yang_Shyy_Xu_GKS_UGKWP_POF2022,ref:UGKWP_Multiphase2022_CiCP}.
This produces a coupled GKS--UGKWP framework that connects the
Eulerian--Lagrangian description in the dilute or collisionless particle limit
with the Eulerian--Eulerian two-fluid description in the dense or
collision-dominated particle limit. The wave--particle decomposition is
therefore not restricted to molecular gas transport; it can also serve as an
adaptive representation for particulate phases whose local dynamics changes
between trajectory-crossing and hydrodynamic behavior.

Efficiency-oriented refinements have become another important part of the
UGKWP family. The original wave--particle decomposition is mainly controlled by
the ratio \(\Delta t/\tau\), or equivalently by a cell Knudsen number. This is
physically meaningful, but it may still assign too many particles to regions
that are locally rarefied yet close to equilibrium. This observation motivated
adaptive wave--particle decompositions in which the temporal criterion is
supplemented by additional gradient-based or resolution-based indicators
\cite{ref:AUGKWP_2023,ref:AUGKWP_CF2026}. A schematic expression is
\begin{equation}
	f=f^w+f^p,\qquad
	\omega_p=
	\mathcal{A}\!\left(
	\frac{\Delta t}{\tau},
	\mathrm{Kn}_{\nabla},
	\mathrm{Kn}_{\ell}
	\right),
	\label{eq:sec5_adaptive_weight}
\end{equation}
where \(f^w\) and \(f^p\) denote the wave and particle components,
respectively, \(\omega_p\) is the particle fraction or particle weight, and
\(\mathcal A\) is an adaptive criterion. Here \(\mathrm{Kn}_{\nabla}\) denotes
a gradient-based nonequilibrium indicator, while \(\mathrm{Kn}_{\ell}\) denotes
a resolution- or length-scale-based Knudsen indicator. Such criteria make the
particle fraction sensitive not only to the collision time relative to the
numerical time step, but also to whether nonequilibrium transport actually
needs to be resolved by particles. As a result, the particle burden can be
substantially reduced in near-equilibrium regions without changing the
underlying UGKWP transport mechanism.

In this broader view, these developments show that UGKWP has evolved from a single
gas-flow algorithm into a broader family of collision-based wave--particle
methods. The core gas-dynamic line has been extended to unstructured meshes,
three-dimensional engineering flows, diatomic gases with internal
nonequilibrium, and adaptive wave--particle decompositions, while related
extensions to radiative, plasma, and gas--particle transport demonstrate the
portability of the same representation principle. What remains common
throughout is that the kinetic solution is decomposed according to local
transport, relaxation, and nonequilibrium scales, and the balance between wave
and particle components is adjusted accordingly. The following subsection
discusses simplified and related branches that develop this idea in less
UGKS-dependent forms.

\subsection{Simplified wave--particle branches}
\label{subsec:sec5_related_branches}

In parallel with the UGKWP-family extensions, several simplified or related
wave--particle branches have emerged. These methods share the same basic
motivation, namely that the numerical representation should change continuously between
continuum-like and rarefied descriptions according to the local transport
scale. They differ, however, in how directly they inherit the UGKS
integral-solution flux and the finite-volume direct-modeling structure of
UGKWP. Some methods simplify the UGKWP construction by replacing the full UGKS-based
wave flux with a model-competition or CFD-coupled flux, while others build an
analogous wave--particle representation from the DUGKS discrete characteristic
evolution.

A representative simplified branch is the simplified unified wave--particle
(SUWP) method of Liu \emph{et al.}~\cite{ref:SUWP_PRE2020}. Rather than
retaining the full UGKS-type reconstruction of the wave and particle
contributions, SUWP extracts from the BGK integral solution a quantified
model-competition (QMC) mechanism. In this formulation, the numerical flux is
written schematically as a weighted combination of a particle-based rarefied
flux and a continuum flux:
\begin{equation}
	\bm{\mathcal F}^{\mathrm{SUWP}}
	=
	\omega_r\bm{\mathcal F}^{p}
	+
	(1-\omega_r)\bm{\mathcal F}^{c}.
	\label{eq:sec5_qmc_flux}
\end{equation}
Here \(\bm{\mathcal F}^{p}\) denotes the flux associated with the rarefied particle
solver, \(\bm{\mathcal F}^{c}\) denotes the continuum flux supplied by the macroscopic
solver, and \(\omega_r\in[0,1]\) is the weight determined by the
QMC mechanism, which measures the relative contribution of the rarefied particle
model and the continuum model over the numerical time step. In the continuum limit \(\omega_r\) becomes small and the
continuum flux dominates; in the rarefied limit \(\omega_r\) approaches one and
the particle flux dominates. Thus the method preserves the central
wave--particle idea while replacing the full UGKS wave flux by a simpler
model-competition formula.

This simplified line was followed by the simple hydrodynamic-particle method
(SHPM) \cite{ref:SHPM_POF2022} and the simplified
hydrodynamic-wave particle method (SHWPM) \cite{ref:SHWPM_AMM2023}. In SHPM, the continuum contribution is provided
directly by a conventional CFD solver, whereas the particle component captures
the free-transport nonequilibrium contribution. SHWPM then introduces a
simplified wave flux that can be reconstructed from the inviscid and viscous
fluxes of the CFD solver, further reducing the number of sampled particles in
near-continuum regimes. These methods are attractive because they can be
implemented using familiar Navier--Stokes solvers and DSMC-type particle
tracking. At the same time, they remain conceptually related to UGKWP in that continuum and particle
descriptions coexist within each control volume, but their continuum component
is supplied more directly by conventional CFD fluxes rather than by the full
UGKS wave construction.

The simplified branch has also been extended to more complex molecular models.
For diatomic gases with rotational nonequilibrium, Yang \emph{et al.}\
\cite{ref:SUWP_DiatomicRykov_Acta2026} derived a QMC mechanism from the
integral solution of the Rykov model and incorporated a two-temperature
continuum description into the SUWP framework. For rotational--vibrational
nonequilibrium, Yang \emph{et al.}~\cite{ref:SUWP_DiatomicRV_AST2026}
extended the QMC mechanism to a three-temperature formulation, with the macroscopic
part supplied by a kinetic inviscid flux or Navier--Stokes-type solver and the
microscopic part represented by collisionless DSMC particles. These extensions
show that the simplified wave--particle idea can accommodate internal-mode
relaxation while remaining closer to the coupling of a conventional continuum
solver with a collisionless particle solver than to the original UGKS-based
UGKWP construction. The convergence-accelerated local-time-stepping SUWP method
of Yang \emph{et al.}~\cite{ref:SUWP_LTS_TAML2026} further illustrates that
this simplified branch is developing its own acceleration techniques for
three-dimensional steady hypersonic nonequilibrium simulations.

Another related branch is the discrete unified gas-kinetic wave--particle
(DUGKWP) method \cite{ref:DUGKWP_PRE2023}. Its
construction is inspired by UGKWP, but the numerical evolution is built from
the discrete characteristic solution used in DUGKS rather than from the
UGKS integral solution. In this sense, DUGKWP bridges the UGKWP and DUGKS formulations. It shares with UGKWP the objective of combining deterministic and
stochastic descriptions and of recovering a hydrodynamic solver in the
continuum limit. At the same time, its flux construction and time evolution
belong more naturally to the DUGKS branch of multiscale schemes.

The purpose of this classification is not to merge these methods into a single
algorithmic family, but to identify the representation principle they share. The original UGKWP line remains the
most systematic wave--particle realization of the UGKS direct-modeling
philosophy. The simplified and related branches demonstrate that the
wave--particle idea can also be simplified, discretized differently, or coupled
more directly to conventional CFD solvers. What they share is the central
insight that multiscale transport can be represented through a collision-based
coexistence of continuum-like wave evolution and rarefied particle motion,
with the balance between the two determined by the local transport regime.

\subsection{Kinetic representation of UGKWP}
\label{subsec:sec5_kinetic_representation}
The algorithmic developments reviewed above can be given a more precise
kinetic interpretation. The kinetic reinterpretation of UGKWP by Guo
\emph{et al.}~\cite{GuoZhuXu_CiCP2026} clarifies the relation between UGKP
and UGKWP, and also makes more explicit the distinction between UGKWP
and other collision-based hybrid methods. In particular, in the BGK
hybrid method of Shin \emph{et al.}~\cite{ShinHauckMcClarren2024}, the
collided--uncollided equations and their stepwise reinitialization are
introduced directly as part of the hybrid formulation.
Although the resulting continuous system recovers the BGK equation,
the component equations and initial conditions are not derived there
from its integral solution. In the UGKWP reinterpretation, by contrast,
the uncollided and collided populations are first identified from the
integral solution of the BGK equation, from which their kinetic
equations and initial conditions follow directly. This analysis is
then extended to resolve the particle and wave populations involved
in UGKWP and their repartitioning at each time step. The population
equations and their initialization thus follow from an explicit
solution structure, rather than serving as the starting point of
the numerical hybridization.

At the first level, underlying UGKP, the BGK solution is decomposed
into uncollided and collided populations. Let
\[
\Lambda(t,s;\bm{x},\bm{\xi})
=\int_s^t \frac{1}{\tau(\bm{x}-\bm{\xi}(t-r),r)}\,dr
\]
be the accumulated relaxation frequency along the characteristic ending at
\((\bm{x},\bm{\xi},t)\). The mild solution can be written as
\begin{equation}
	f(\bm{x},\bm{\xi},t)
	=e^{-\Lambda(t,t_0;\bm{x},\bm{\xi})}
	f(\bm{x}-\bm{\xi}(t-t_0),\bm{\xi},t_0)
	+\int_{t_0}^{t}
	\frac{1}{\tau(\bm{x}',s)}
	e^{-\Lambda(t,s;\bm{x},\bm{\xi})}
	g(\bm{W}(\bm{x}',s),\bm{\xi})\,ds,
	\label{eq:sec5_bgk_mild_fu_fc}
\end{equation}
where \(\bm{x}'=\bm{x}-\bm{\xi}(t-s)\), \(\tau\) is the relaxation time not necessarily a local constant. 
Equation~\eqref{eq:sec5_bgk_mild_fu_fc} identifies an uncollided population
and a collided population:
\begin{equation}
	\begin{aligned}
		f_u(\bm{x},\bm{\xi},t)
		&=
		e^{-\Lambda(t,t_0;\bm{x},\bm{\xi})}
		f(\bm{x}-\bm{\xi}(t-t_0),\bm{\xi},t_0),\\
		f_c(\bm{x},\bm{\xi},t)
		&=
		\int_{t_0}^{t}
		\frac{1}{\tau(\bm{x}',s)}
		e^{-\Lambda(t,s;\bm{x},\bm{\xi})}
		g(\bm{W}(\bm{x}',s),\bm{\xi})\,ds,
		\qquad
		\bm{x}'=\bm{x}-\bm{\xi}(t-s).
	\end{aligned}
	\label{eq:sec5_fu_fc_definitions}
\end{equation}
Thus \(f=f_u+f_c\), where \(f_u\) is the population that has not yet collided
during the observation interval and \(f_c\) is the population generated by
relaxation. Importantly, these two populations and their initial data are not
introduced as an independent modeling assumption. They are identified directly from the two
terms in the integral solution \eqref{eq:sec5_bgk_mild_fu_fc}. The uncollided
component inherits the initial distribution transported along characteristics,
whereas the collided component has zero initial value and is generated entirely
by the relaxation source.

The corresponding sub-kinetic equations are
\begin{equation}
	\begin{cases}
		\partial_t f_u+\bm{\xi}\cdot\nabla f_u
		=-\dfrac{1}{\tau}f_u,\\[0.5ex]
		f_u(\bm{x},\bm{\xi},t_0)=f(\bm{x},\bm{\xi},t_0),
	\end{cases}
	\qquad
	\begin{cases}
		\partial_t f_c+\bm{\xi}\cdot\nabla f_c
		=-\dfrac{1}{\tau}\bigl(f_c-g(\bm W,\bm{\xi})\bigr),\\[0.5ex]
		f_c(\bm{x},\bm{\xi},t_0)=0 .
	\end{cases}
	\label{eq:sec5_ugkp_twopop_system}
\end{equation}
These equations follow directly by differentiating the two terms in
Eq.~\eqref{eq:sec5_fu_fc_definitions} along characteristics. Their initial
conditions are likewise inherited from the integral representation. The
uncollided population starts from the full distribution at \(t_0\), while the
collided population starts from zero and is created only by the relaxation
source. The sum of the two equations in Eq. \eqref{eq:sec5_ugkp_twopop_system} reconstructs the original BGK equation. The moments of $f_u$ and $f_c$ give
\begin{equation}
	\bm W_\ell=\int \bm{\psi} f_\ell\,d\bm{\xi},
	\qquad
	\bm{\mathcal F}_\ell=\int \bm{\xi}\,\bm{\psi} f_\ell\,d\bm{\xi},
	\qquad \ell=u,c,
	\label{eq:sec5_population_moments}
\end{equation}
and, for a velocity-independent relaxation time,
\begin{equation}
	\begin{cases}
		\partial_t \bm W_u+\nabla\!\cdot \bm{\mathcal F}_u
		=-\dfrac{1}{\tau}\bm W_u,\\[0.5ex]
		\bm W_u(\bm{x},t_0)=\bm W(\bm{x},t_0),
	\end{cases}
	\qquad
	\begin{cases}
		\partial_t \bm W_c+\nabla\!\cdot \bm{\mathcal F}_c
		=\dfrac{1}{\tau}\bm W_u,\\[0.5ex]
		\bm W_c(\bm{x},t_0)=0 .
	\end{cases}
	\label{eq:sec5_ugkp_moment_system}
\end{equation}
This moment system makes explicit the exchange between the uncollided and
collided populations and gives the kinetic representation underlying UGKP
\cite{GuoZhuXu_CiCP2026}.

The second level, corresponding to UGKWP, is more refined. As emphasized by
Guo \emph{et al.}~\cite{GuoZhuXu_CiCP2026}, UGKWP is not obtained by simply
splitting \(f=f_u+f_c\) into additional collision classes. Instead, the gas
system is first represented by two computational phases, namely, a discrete particle
phase and a continuum hydrodynamic-wave phase. Each phase is then further
decomposed according to its collision behavior over one time step. In the most
detailed formulation, the distribution is written as
\begin{equation}
	f=f_{pf}+f_{pr}+f_{pc}+f_{wf}+f_{wr}+f_{wc}.
	\label{eq:sec5_six_population}
\end{equation}
Here the first subscript denotes the computational phase, with \(p\) for
particle and \(w\) for wave, while the second subscript denotes the collision
status, with \(f\) for free transport over the whole time step, \(r\) for particles
or molecules that are removed after their first collision, and \(c\) for
collided populations. Thus \(f_{pf}\) and \(f_{wf}\) are collisionless particle
and wave populations, \(f_{pr}\) and \(f_{wr}\) are pre-collision populations
removed after first collision, and \(f_{pc}\) and \(f_{wc}\) are collided
populations.

For \(t_n<t\le t_{n+1}\), the six-population kinetic system can be written as
\begin{subequations}
	\label{eq:sec5_six_population_system}
	\begin{align}
		&\begin{cases}
			\partial_t f_{pf}+\bm{\xi}\cdot\nabla f_{pf}=0,\\[0.5ex]
			f_{pf}(\bm{x},\bm{\xi},t_{n}^+)
			=\beta f_u(\bm{x},\bm{\xi},t_{n}^-),
		\end{cases}
		\label{eq:sec5_six_population_system_a}\\[1ex]
		&\begin{cases}
			\partial_t f_{pr}+\bm{\xi}\cdot\nabla f_{pr}
			=-\dfrac{1}{\tau}(f_{pr}+f_{pf}),\\[0.5ex]
			f_{pr}(\bm{x},\bm{\xi},t_{n}^+)
			=(1-\beta)f_u(\bm{x},\bm{\xi},t_{n}^-),
		\end{cases}
		\label{eq:sec5_six_population_system_b}\\[1ex]
		&\begin{cases}
			\partial_t f_{pc}+\bm{\xi}\cdot\nabla f_{pc}
			=-\dfrac{1}{\tau}\bigl(f_{pc}-g_p\bigr),\\[0.5ex]
			f_{pc}(\bm{x},\bm{\xi},t_{n}^+)=0,
		\end{cases}
		\label{eq:sec5_six_population_system_c}\\[1ex]
		&\begin{cases}
			\partial_t f_{wf}+\bm{\xi}\cdot\nabla f_{wf}=0,\\[0.5ex]
			f_{wf}(\bm{x},\bm{\xi},t_{n}^+)
			=\beta f_c(\bm{x},\bm{\xi},t_{n}^-)
			\approx g(\beta\bm W_c(\bm{x},t_{n}^-),\bm{\xi}),
		\end{cases}
		\label{eq:sec5_six_population_system_d}\\[1ex]
		&\begin{cases}
			\partial_t f_{wr}+\bm{\xi}\cdot\nabla f_{wr}
			=-\dfrac{1}{\tau}(f_{wr}+f_{wf}),\\[0.5ex]
			f_{wr}(\bm{x},\bm{\xi},t_{n}^+)
			=(1-\beta)f_c(\bm{x},\bm{\xi},t_{n}^-),
		\end{cases}
		\label{eq:sec5_six_population_system_e}\\[1ex]
		&\begin{cases}
			\partial_t f_{wc}+\bm{\xi}\cdot\nabla f_{wc}
			=-\dfrac{1}{\tau}\bigl(f_{wc}-g_w\bigr),\\[0.5ex]
			f_{wc}(\bm{x},\bm{\xi},t_{n}^+)=0 .
		\end{cases}
		\label{eq:sec5_six_population_system_f}
	\end{align}
\end{subequations}
Here \(t_{n}^-\) and \(t_{n}^+\) denote the values immediately before and after
the repartitioning at time \(t_n\), respectively, and
\begin{equation}
	\beta=e^{-\Delta t/\tau}, \qquad \Delta t=t_{n+1}-t_n ,
	\label{eq:sec5_beta_definition}
\end{equation}
for a local constant relaxation time during the step. The phase equilibria are
\begin{equation}
	g_\alpha(\bm{x},\bm{\xi},t)
	=\frac{\rho_\alpha}{\rho}\,g(\bm W(\bm{x},t),\bm{\xi}),
	\qquad \alpha=p,w,
	\label{eq:sec5_phase_equilibria}
\end{equation}
where \(\rho_\alpha\) is the density of phase \(\alpha\) and
\(\rho=\rho_p+\rho_w\). The approximation
\(f_{wf}(t_{n}^+)\approx g(\beta\bm W_c,\bm{\xi})\) expresses the practical
resampling of the collisionless part of the wave population from the local
equilibrium determined by the scaled conservative state \(\beta\bm W_c\). This
six-population system is the most detailed kinetic representation of UGKWP \cite{GuoZhuXu_CiCP2026}.

For practical implementation, the six-population system is recombined into a
four-population one:
\begin{equation}
	f=f_{pu}+f_{wf}+f_{wr}+f_c,
	\qquad
	f_{pu}=f_{pf}+f_{pr},
	\qquad
	f_c=f_{pc}+f_{wc}.
	\label{eq:sec5_four_population}
\end{equation}
Here \(f_{pu}\) denotes all original uncollided particle populations, \(f_{wf}\)
the collisionless wave population sampled as particles, \(f_{wr}\) the
pre-collision wave population removed after first collision, and \(f_c\) the
combined collided population. The corresponding four-population equations are
\begin{subequations}
	\label{eq:sec5_four_population_system}
	\begin{align}
		&\begin{cases}
			\partial_t f_{pu}+\bm{\xi}\cdot\nabla f_{pu}
			=-\dfrac{1}{\tau}f_{pu},\\[0.5ex]
			f_{pu}(\bm{x},\bm{\xi},t_{n}^+)
			=f_u(\bm{x},\bm{\xi},t_{n}^-),
		\end{cases}
		\label{eq:sec5_four_population_system_a}\\[1ex]
		&\begin{cases}
			\partial_t f_{wf}+\bm{\xi}\cdot\nabla f_{wf}=0,\\[0.5ex]
			f_{wf}(\bm{x},\bm{\xi},t_{n}^+)
			=\beta f_c(\bm{x},\bm{\xi},t_{n}^-)
			\approx g(\beta\bm W_c(\bm{x},t_{n}^-),\bm{\xi}),
		\end{cases}
		\label{eq:sec5_four_population_system_b}\\[1ex]
		&\begin{cases}
			\partial_t f_{wr}+\bm{\xi}\cdot\nabla f_{wr}
			=-\dfrac{1}{\tau}(f_{wr}+f_{wf}),\\[0.5ex]
			f_{wr}(\bm{x},\bm{\xi},t_{n}^+)
			=(1-\beta)f_c(\bm{x},\bm{\xi},t_{n}^-),
		\end{cases}
		\label{eq:sec5_four_population_system_c}\\[1ex]
		&\begin{cases}
			\partial_t f_c+\bm{\xi}\cdot\nabla f_c
			=-\dfrac{1}{\tau}\bigl(f_c-g(\bm W,\bm{\xi})\bigr),\\[0.5ex]
			f_c(\bm{x},\bm{\xi},t_{n}^+)=0 .
		\end{cases}
		\label{eq:sec5_four_population_system_d}
	\end{align}
\end{subequations}

These population systems can therefore be viewed as a hierarchy of kinetic
representations. The two-population system \eqref{eq:sec5_ugkp_twopop_system} corresponds to the
kinetic representation of UGKP. By further resolving the transported and collision-associated
components, the six-population system \eqref{eq:sec5_six_population_system} gives a more detailed interpretation of
UGKWP. After collecting the populations that are treated
together in the algorithm, the four-population system
\eqref{eq:sec5_four_population_system} corresponds to the practical
implementation of UGKWP \cite{GuoZhuXu_CiCP2026}.
From this viewpoint, the UGKWP flux decomposition becomes transparent. In the
four-population formulation, the interface flux can be written as
\begin{equation}
	\bm{\mathcal F}_{ij}=\bm{\mathcal F}_{ij}^{p}+\bm{\mathcal F}_{ij}^{wr}+\bm F_{ij}^{c},
	\label{eq:sec5_flux_threeparts}
\end{equation}
where \(\bm{\mathcal F}_{ij}^{p}\) is the contribution from simulation particles
including \(f_{pu}\) and the sampled \(f_{wf}\) part, \(\bm{\mathcal F}_{ij}^{wr}\) is
the deterministic contribution from the pre-collision wave population, and
\(\bm{\mathcal F}_{ij}^{c}\) is the contribution from the collided population. The first
part is tallied stochastically, whereas the latter two are evaluated
deterministically from the wave or macroscopic representation.

Operationally, the UGKWP update may therefore be viewed as a two-level
algorithm built on the four-population system. The particle populations are
transported and tallied by particle tracking, the wave and collided
contributions are evaluated deterministically, the total conservative variables
are updated by the finite-volume balance, and the wave--particle decomposition
is reconstructed for the next step through the survival factor \(\beta\).
Accordingly, UGKP corresponds to the exact two-population decomposition
\(f=f_u+f_c\), while UGKWP first introduces particle and wave phases and then
resolves the collision behavior within each phase. This is why UGKWP admits a
more efficient computational format than UGKP while remaining tied to a
well-defined kinetic population structure.

\subsection{Section remarks}
\label{subsec:sec5_transition}
The developments reviewed in this section show that wave--particle formulations
have become a major branch of multiscale kinetic computation. Historically, this
line can be traced from first-collision source methods and
collided--uncollided decompositions in transport theory, through
time-dependent collision-based hybrids, to the UGKP and UGKWP methods developed
under the UGKS framework. In the latter case, collision history is not used
merely as a numerical correction, fidelity-switching device, or external
hybridization criterion. It is incorporated into a conservative finite-volume
evolution in which particle transport, collisional relaxation, and hydrodynamic
limiting behavior are coordinated over the numerical time step.

This progression clarifies an important methodological distinction. In early
collision-based hybrids, the decomposition mainly separates components that can
be advanced with different numerical resolutions or temporal strategies. In
UGKWP, by contrast, the decomposition becomes an adaptive representation of the
kinetic solution itself within each control volume and time step. The same
method can therefore behave as a particle solver in highly rarefied regimes, as
a gas-kinetic hydrodynamic solver in continuum regimes, and as a wave--particle
method in transitional regimes, without relying on static domain decomposition
or buffer-zone coupling.

The kinetic representation discussed in Sec.~\ref{subsec:sec5_kinetic_representation}
further explains why this behavior is more than an algorithmic convenience. The
uncollided and collided populations underlying UGKP are derived from the two terms
of the BGK integral solution, and the more detailed particle--wave population
structure of UGKWP is obtained by repartitioning these populations over one
numerical time step. Thus the practical wave--particle algorithm is tied to a
derived kinetic population structure, rather than to a heuristic split imposed
from outside the kinetic evolution.

The simplified and related branches discussed above show that this
representation principle is not confined to the original UGKWP construction.
Wave--particle ideas have been adapted to simplified CFD--particle couplings,
DUGKS-based formulations, internal-mode nonequilibrium, and transport systems
beyond monatomic gas dynamics. These variants differ in how closely they inherit
the full UGKS-based wave construction, but they share the same central insight.
The numerical representation should follow the local transport and relaxation
behavior of the kinetic state, rather than remain fixed in advance as purely
deterministic, purely stochastic, or purely macroscopic.

These developments also raise a broader question that applies to all methods
reviewed so far. When a numerical scheme is used across kinetic, transitional,
and continuum regimes, stability and efficiency under coarse discretization are
not sufficient. The scheme should also recover the correct macroscopic limit
when the kinetic scale is unresolved and, more generally, preserve the relevant
asymptotic structure of the underlying kinetic model at the numerical scale
being used. This question motivates the transition from method construction to
multiscale assessment, which is the subject of the next section.

\section{Asymptotic-preserving (AP) and unified-preserving (UP) properties}
\label{sec:ap_up}
The previous sections reviewed how multiscale kinetic methods are constructed. The present section turns to a different question, namely how such methods should be assessed when the kinetic scale is not resolved by the numerical mesh and time step. The issue is not only stability under stiff collisions, but also whether the computed solution represents the correct macroscopic physics and the correct asymptotic structure of the kinetic model.

AP and UP are therefore properties of multiscale fidelity, not solver families. 
The AP property concerns the leading limiting model \cite{Hu_Jin_Li_AP_2017,Jin2022AN}.  
It requires that, as the small parameter tends to zero while the numerical mesh and time step are kept fixed, a kinetic
discretization degenerates into a consistent discretization of the limiting
macroscopic equations. The UP property concerns the asymptotic hierarchy seen
by the fully discrete scheme \cite{GuoLiXu2023UP}. It requires the discrete
evolution to preserve the relevant Chapman--Enskog coefficients, transport
coefficients, and transport--collision balance at finite numerical resolution.

This viewpoint provides the organizing principle of the present section. We
start from the asymptotic setting of stiff kinetic equations, use AP as the
minimal requirement for recovering the leading macroscopic limit, and then
explain why a stronger UP criterion is needed to assess finite-resolution
multiscale fidelity.

\subsection{Asymptotic regime and numerical resolution}
\label{subsec:sec6_why_ap}
Consider the scaled collisional kinetic equation Eq.~\eqref{eq:scaled_boltzmann_intro}. When \(\varepsilon\) is small, the collision process acts on a scale much shorter than the macroscopic evolution. Away from initial layers, boundary layers, and strongly nonequilibrium regions, the distribution is expected to remain close to a local equilibrium \(g_M(\bm W)\). Formally, this approach to equilibrium may be described by the Chapman--Enskog expansion
\begin{equation}
	f=f^{(0)}+\varepsilon f^{(1)}+\varepsilon^2 f^{(2)}+\cdots .
	\label{eq:sec6_CE_expansion}
\end{equation}
At the leading order, the distribution function is constrained to the equilibrium, i.e., $f^{(0)}=g_M(\bm W)$, and the corresponding moment system gives the Euler limit. The first-order correction $f^{(1)}$ produces viscous stress and heat flux and therefore gives the Navier--Stokes behavior when the kinetic model has the correct transport coefficients. Higher-order terms contain further nonequilibrium corrections. Thus the asymptotic structure of a kinetic equation is a hierarchy, not only a single limiting equation \cite{DimarcoPareschi2014,Jin2022AN,Hu_Jin_Li_AP_2017}.

This hierarchy becomes a numerical issue when the mesh size and time step are no longer tied to the mean free path and collision time. A classical kinetic solver that requires
\begin{equation}
	\Delta x=O(\varepsilon),\qquad \Delta t=O(\varepsilon)
	\label{eq:sec6_kinetic_resolution}
\end{equation}
in order to obtain continuum-flow behavior remains accurate on resolved kinetic scales, but becomes inefficient in the continuum regime. A multiscale kinetic method should instead remain stable and consistent on hydrodynamic meshes, where \(\Delta x\) and \(\Delta t\) may be much larger than the kinetic scales while still resolving the macroscopic gradients.

It is useful to distinguish three objects. The first is the continuous kinetic model and its asymptotic hierarchy. The second is the numerical scheme, including transport reconstruction, collision treatment, velocity discretization, and time integration. The third is the numerical scale \(h=(\Delta x,\Delta t)\). AP and UP assess how these three objects interact. AP examines the leading limiting model of the discrete scheme as \(\varepsilon\to0\) with \(h\) fixed. UP examines the asymptotic hierarchy of the discrete scheme at finite \(h\), especially when \(h\) is larger than the kinetic scale.

This distinction also prevents a common confusion. Kinetic methods may have very different algorithmic forms, but AP and UP are not labels for these forms. They are criteria applied to the resulting discrete evolution. The same method family may be AP under one scaling and not under another; likewise, two AP schemes may preserve different amounts of finite-\(\varepsilon\) asymptotic information.

\subsection{Leading-limit consistency and the AP criterion}
\label{subsec:sec6_ap}
The asymptotic-preserving concept was introduced to formalize the minimal asymptotic requirement for numerical methods for stiff kinetic equations. When the small
parameter tends to zero, the kinetic scheme should become a consistent and
stable scheme for the limiting macroscopic equations without requiring the
numerical mesh or time step to resolve the kinetic scale. Let \(P_h^\varepsilon\) denote a kinetic discretization with numerical scale \(h=(\Delta x,\Delta t)\). The AP requirement may be written schematically as
\begin{equation}
	P_h^\varepsilon \longrightarrow P_h^0,
	\qquad \varepsilon\to0,
	\label{eq:sec6_ap_limit}
\end{equation}
with \(h\) fixed, where \(P_h^0\) is a consistent discretization of the limiting macroscopic model. Thus the asymptotic limit and the numerical discretization commute in the leading-order sense: applying the scheme first and then taking the limit should lead to a valid discretization of the
same macroscopic problem obtained by taking the asymptotic limit at the
continuous level. This idea has guided the development of IMEX, micro--macro, penalization, exponential, and related AP schemes for kinetic and transport equations \cite{Jin_1999_AP,Klar_1998,Bennoune_Lemou_Mieussens_2008,Filbet_Jin_2010,DimarcoPareschi2014,Hu_Jin_Li_AP_2017,Jin2022AN}.

A simple relaxation example illustrates the mechanism. For a scaled BGK model, a first-order IMEX discretization may be written as
\begin{equation}
	\frac{f^{n+1}-f^n}{\Delta t}
	+\bm{\xi}\cdot\nabla f^n
	=
	\frac{1}{\varepsilon}\bigl(g(\bm W^{n+1})-f^{n+1}\bigr),
	\label{eq:sec6_ap_bgk_discrete}
\end{equation}
where the transport term is explicit and the stiff relaxation term is implicit. Equivalently,
\begin{equation}
	f^{n+1}
	=
	\frac{\varepsilon}{\varepsilon+\Delta t}
	\left(f^n-\Delta t\,\bm{\xi}\cdot\nabla f^n\right)
	+
	\frac{\Delta t}{\varepsilon+\Delta t}g(\bm W^{n+1}).
	\label{eq:sec6_ap_bgk_solver}
\end{equation}
For fixed \(\Delta t\), the limit \(\varepsilon\to0\) enforces \(f^{n+1}\to g(\bm W^{n+1})\). Taking moments of Eq.~\eqref{eq:sec6_ap_bgk_discrete} removes the relaxation term. If the previous state is already on or close to the equilibrium manifold, the moment flux reduces to the equilibrium flux, and the limiting scheme becomes a discretization of the Euler system under the corresponding fluid scaling. More refined AP constructions are designed so that this limiting process is stable and consistent without kinetic-scale restrictions on \(\Delta x\) and \(\Delta t\) \cite{Jin_1999_AP,Bennoune_Lemou_Mieussens_2008,Filbet_Jin_2010,Jin2022AN}.

The strength of AP is twofold. First, AP addresses stiffness by allowing the
method to remain stable when collisions become frequent, without imposing
$\Delta t=O(\varepsilon)$ or $\Delta x=O(\varepsilon)$ solely for stability
or consistency. Second, AP addresses limiting consistency by ensuring that the
discrete kinetic solution approaches a valid discrete macroscopic solution
rather than an unphysical approximation contaminated by kinetic-scale numerical
artifacts. In this sense, AP is a foundational requirement for all-regime
kinetic computation. A method that fails to be AP cannot be expected to cross
reliably from kinetic to macroscopic regimes on meshes that do not resolve the
kinetic scale.

An AP statement is always tied to a specified scaling and limiting equation. In diffusive transport, the limiting model is typically diffusion. In fluid-dynamic, acoustic, or incompressible scalings, the limiting model may be Euler, incompressible Euler, or Navier--Stokes-type equations, depending on the model and on the asymptotic order retained. Consequently, AP should not be interpreted as a universal statement independent of the asymptotic regime. It states that the discrete kinetic method captures the prescribed leading macroscopic limit under unresolved kinetic scales \cite{Hu_Jin_Li_AP_2017,Jin2022AN}.

This leading-limit nature also marks the
boundary of the AP criterion. In gas-dynamic applications, recovery of the
prescribed macroscopic limit does not by itself guarantee that the
finite-\(\varepsilon\) hydrodynamic corrections are captured with the correct
discrete transport--collision balance.

\subsection{Finite-resolution gap beyond AP}
\label{subsec:sec6_finite_resolution_gap}

The AP property is necessary, but it is not a complete description of multiscale fidelity. Its leading-order nature means that different schemes may receive the same AP classification while producing different results in the near-continuum regime. This is the regime in which \(\varepsilon\) is small but finite, the solution is close to local equilibrium, and Navier--Stokes-level corrections may still determine the accuracy of stress, heat flux, slip, temperature jump, and other finite-rarefaction effects.

The reason is that AP mainly examines the limit $\varepsilon\to 0$ with fixed $h$.
Once the leading limiting equation is recovered, the AP test does not by itself determine whether the first Chapman--Enskog correction is correct at the numerical scale. A scheme may recover the Euler limit and still introduce an incorrect numerical viscosity, an incorrect heat flux, or excessive dissipation when \(\varepsilon\) is small but not zero.

This finite-resolution gap is closely related to the coupling between transport and collision in the discrete evolution. If the interfacial distribution is reconstructed from a purely collisionless transport process, and the relaxation is handled only as a cell-centered source, the leading equilibrium limit may still be correct. However, the first nonequilibrium correction depends on how particles travel, collide, and relax over one numerical space--time step. A discretization that separates these effects too strongly may lose the Navier--Stokes-level asymptotic structure even when it remains AP at leading order.

The issue is especially important for unified gas-kinetic methods. In UGKS and DUGKS, the interface state is not obtained from a collisionless reconstruction followed by an independent relaxation update. Instead, transport and collision are coupled in the local evolution used to build the numerical flux. Their continuum-regime behavior is therefore a property of the fully discrete space--time construction, not only of an implicit collision discretization. The same observation is relevant to wave--particle methods, synthetic methods, and particle AP methods whenever their coarse-step behavior depends on how transport, interaction, and macroscopic information are combined over one step.

The comparison of Chen and Xu~\cite{ref:ChenXu-2015} provides a concrete example of this finite-resolution gap. For the BGK equation, an IMEX AP scheme may drive the cell-centered distribution close to the Chapman--Enskog form, but its interface flux is still reconstructed mainly from collisionless transport. As a result, the flux moments can contain excessive numerical dissipation in the Navier--Stokes regime. In UGKS, the flux is instead obtained from the local integral solution, so that transport and relaxation are coupled during the interfacial evolution. In the continuum limit, the initially reconstructed non-equilibrium part is suppressed by a factor of order \(\tau/\Delta t\), and the leading flux recovers the Chapman--Enskog Navier--Stokes correction. This explains why a scheme may be Euler-AP but still inaccurate for continuum Navier--Stokes flows.

Thus AP identifies whether a scheme reaches the correct leading macroscopic model. It does not fully identify which hydrodynamic model the scheme effectively approximates at finite \(h\) and finite small \(\varepsilon\). The UP property was introduced to address this second question by applying asymptotic analysis to the discrete or modified equation itself \cite{GuoLiXu2023UP}.

\subsection{Discrete-asymptotic fidelity and the UP criterion}
\label{subsec:sec6_up}

We now formulate the UP property as a discrete-asymptotic criterion. Let \(f_h\) denote the solution associated with a fully discrete scheme, or equivalently with its modified equation. Suppose that \(f_h\) admits a Chapman--Enskog-type expansion,
\begin{equation}
	f_h=f_h^{(0)}+\varepsilon f_h^{(1)}
	+\varepsilon^2 f_h^{(2)}+\cdots .
	\label{eq:sec6_discrete_CE}
\end{equation}
The coefficients \(f_h^{(k)}\) are then compared with the corresponding coefficients \(f^{(k)}\) of the continuous expansion in Eq.~\eqref{eq:sec6_CE_expansion}. The UP order specifies how many levels of this coefficient hierarchy are preserved by the discrete evolution \cite{GuoLiXu2023UP}.

This comparison is most transparent at the level of the modified equation. Schematically, a fully discrete kinetic scheme may be represented by
\begin{equation}
	\partial_t f_h+{\bxi}\cdot\nabla f_h
	=
	\frac{1}{\varepsilon}Q_h(f_h)
	+R_T(h;f_h)+\frac{1}{\varepsilon}R_Q(h;f_h),
	\label{eq:sec6_modified_equation}
\end{equation}
where \(h\) denotes the numerical resolution, \(Q_h\) is the discrete collision operator, and \(R_T\) and \(R_Q\) denote the transport and collision truncation effects, respectively. These residual terms are not merely local consistency errors. They enter the Chapman--Enskog analysis of the discrete equation, and therefore may modify the balance equations for \(f_h^{(k)}\), even when the scheme is consistent with the original kinetic equation for fixed \(\varepsilon\). Consequently, the resulting Chapman--Enskog coefficients can be altered and affect the effective Euler, Navier--Stokes, or high-order hydrodynamic behavior of the scheme.

\begin{figure}
	\centering
	\includegraphics[width=0.45\textwidth]{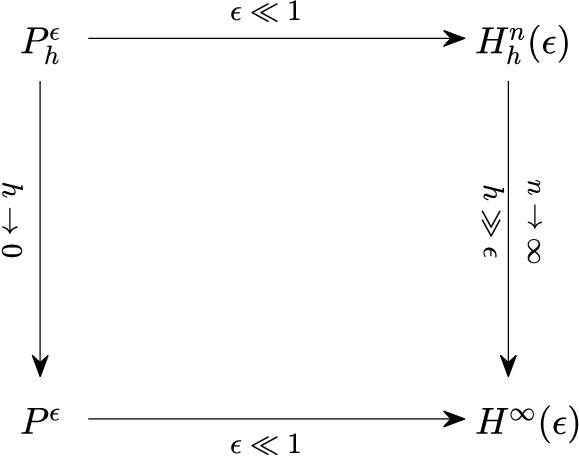}
	\caption{Schematic diagram of a kinetic scheme with UP order \(n\). Here \(P^\varepsilon\) denotes the kinetic equation, \(P_h^\varepsilon\) is a consistent kinetic scheme with numerical resolution \(h\), \(H^\infty(\varepsilon)\) is the Chapman--Enskog hierarchy of \(P^\varepsilon\), and \(H_h^n(\varepsilon)\) is the hierarchy retained through order $n$ for \(P_h^\varepsilon\). For an \(n\)th-order UP scheme, \(f_h^{(k)}=f^{(k)}\) for \(0\le k\le n\).  Reproduced from Fig.~1 of Ref.~\cite{GuoLiXu2023UP}.
		Copyright (2023) by the American Physical Society.}
	\label{fig:upFig}
\end{figure}

The idea can be expressed in terms of two coefficient hierarchies. The continuous kinetic model generates
\begin{equation}
	H^{\infty}(\varepsilon)
	=
	\{f^{(0)},f^{(1)},f^{(2)},\ldots\},
	\label{eq:sec6_Hinf}
\end{equation}
whereas the numerical method generates
\begin{equation}
	H_h^{\infty}(\varepsilon)
	=
	\{f_h^{(0)},f_h^{(1)},f_h^{(2)},\ldots\}.
	\label{eq:sec6_Hh}
\end{equation}
In this setting, a scheme is called \(n\)th-order UP if, under the prescribed relation among \(\varepsilon\), \(\Delta x\), and \(\Delta t\),  the balance equations generated from \eqref{eq:sec6_modified_equation} agree with those of the original kinetic model up to the \(n\)th level of the Chapman--Enskog hierarchy. Equivalently,
\[
f_h^{(k)}=f^{(k)}, \qquad 0\le k\le n,
\]
in the sense of the corresponding asymptotic balance equations, while the first discrepancy appears at the next level. Thus first-order UP preserves the leading equilibrium-level behavior, whereas higher-order UP requires preservation of the nonequilibrium corrections that determine dissipative and higher-order hydrodynamic effects. In particular, Navier--Stokes-level fidelity requires preservation of the first nonequilibrium correction, not merely the equilibrium limit.

It is useful to introduce the ratios between the numerical scales and the kinetic time and length scales,
\begin{equation}
	\delta_t=\frac{\Delta t}{\varepsilon},
	\qquad
	\delta_x=\frac{\Delta x}{\varepsilon}.
	\label{eq:sec6_delta_resolution}
\end{equation}
These quantities measure whether the time step and mesh size resolve the collision time and mean free path in the scaled variables. The UP property is therefore not determined by the formal order of the discretization alone. It also depends on the asymptotic relation among \(\varepsilon\), \(\Delta t\), and \(\Delta x\), as well as  on the discrete coupling between transport and collision. This relation is illustrated in Fig.~\ref{fig:upFig}.

\begin{figure}
	\centering
	\includegraphics[width=0.5\textwidth]{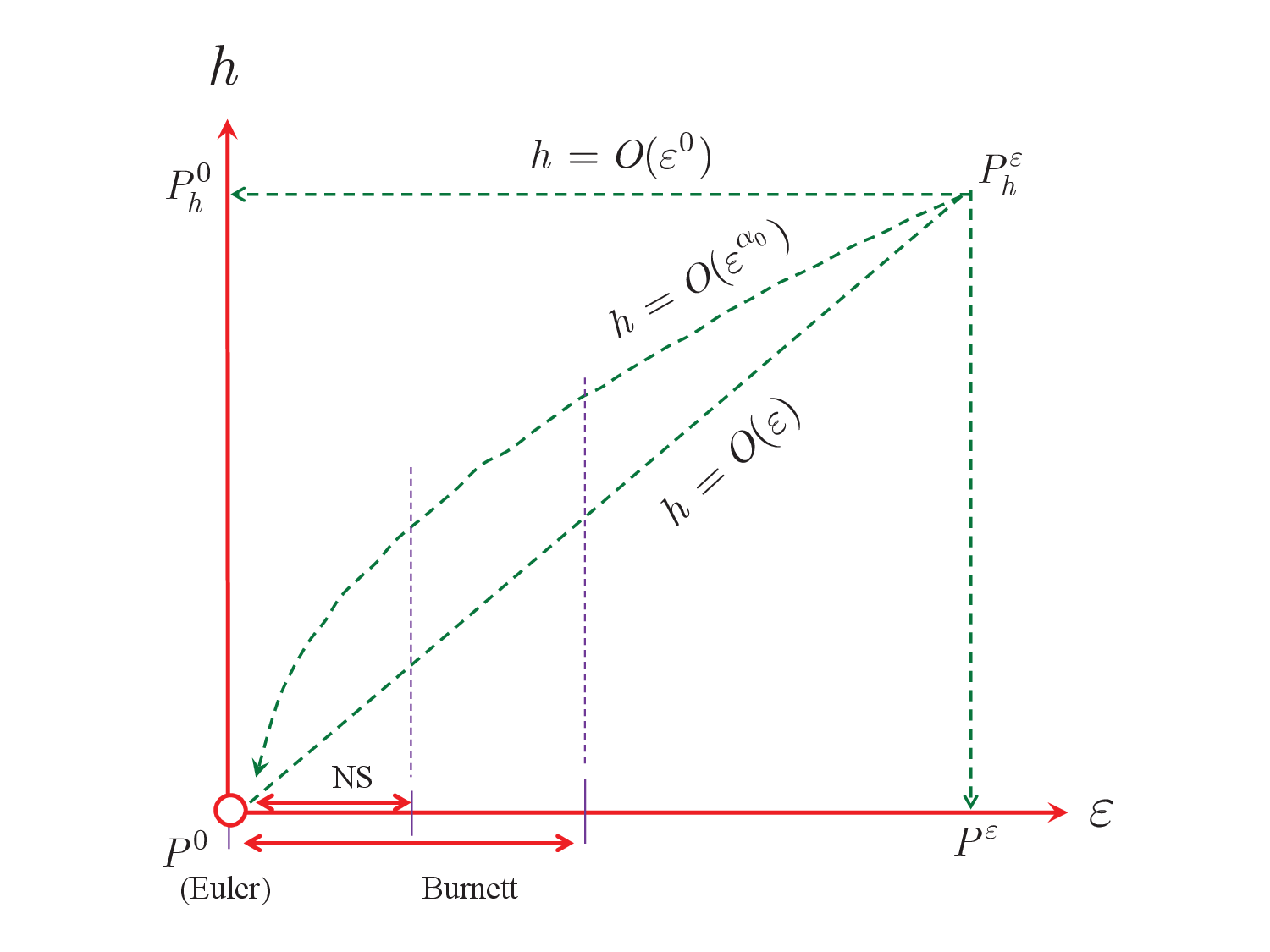}
	\caption{Schematic diagram of admissible asymptotic paths toward hydrodynamic regimes. The region below \(h=O(\varepsilon)\) corresponds to kinetic-scale resolution. The line \(h=O(\varepsilon^{\alpha_0})\) represents the upper admissible numerical scale for a prescribed UP order, with \(\alpha_0\) understood as the infimum of the admissible exponents. The region between \(h=O(\varepsilon)\) and \(h=O(\varepsilon^{\alpha_0})\) represents the parameter range in which the scheme preserves the prescribed asymptotic hierarchy without fully resolving the kinetic scale. Reproduced from Fig.~2 of Ref.~\cite{GuoLiXu2023UP}.
		Copyright (2023) by the American Physical Society.}
	\label{fig:upPath}
\end{figure}

The UP property also gives a scale-dependent description of admissible paths toward hydrodynamic regimes. As shown in Fig.~\ref{fig:upPath}, the region below \(h=O(\varepsilon)\) corresponds to kinetic-scale resolution. A UP scheme may preserve a prescribed asymptotic hierarchy on a coarser numerical scale. If the upper admissible scale is denoted by \(h=O(\varepsilon^{\alpha_0})\), then \(\alpha_0\) characterizes the largest numerical scale allowed by the corresponding UP requirement. Since \(0<\varepsilon<1\), a smaller \(\alpha_0\) permits a larger admissible \(h\). Therefore, a scheme with a smaller \(\alpha_0\) can reach the same hydrodynamic asymptotic regime with larger mesh sizes or time steps.

The formulation of Guo \emph{et al.}~\cite{GuoLiXu2023UP} is based on Chapman--Enskog analysis and is therefore naturally suited to Euler, Navier--Stokes, and higher-order hydrodynamic limits. In this form, UP can be understood as the preservation of the asymptotic coefficients relevant to the target regime. This coefficient-preservation viewpoint is particularly useful for direct-modeling schemes, where the numerical flux itself represents a finite-space--time kinetic evolution. Other representations of the continuous and numerical distributions may also be used for the same purpose. For example, Hermite polynomial expansions provide an alternative basis for assessing the agreement between the original kinetic equation and the modified equation of a numerical scheme through comparison of their expansion
coefficients.

\subsection{Implications for multiscale method assessment}
\label{subsec:sec6_ap_up_methods}
The AP and UP viewpoints lead to a layered assessment of the methods reviewed in Secs.~\ref{sec:deterministic_methods}--\ref{sec:wave_particle_collision}. The first layer is leading-limit consistency. IMEX, penalized, exponential, micro--macro, particle AP, and synthetic schemes can all be AP when designed under the appropriate scaling. This property is essential because it guarantees that unresolved kinetic computations reduce to a meaningful macroscopic discretization \cite{Bennoune_Lemou_Mieussens_2008,Filbet_Jin_2010,DimarcoPareschi2014,Jin2022AN}.

The second layer is finite-resolution hydrodynamic fidelity. Here one must examine the discrete flux, the time integration, the collision treatment, and the coupling between them. A scheme whose AP property is obtained mainly through implicit relaxation may still require additional analysis to determine whether the first-order nonequilibrium correction is preserved. This distinction was demonstrated numerically by Chen and Xu~\cite{ref:ChenXu-2015}. In the transitional lid-driven cavity test at \(\mathrm{Kn}=0.1\), the IMEX AP scheme and UGKS give nearly indistinguishable velocity profiles. In the continuum case at \(\mathrm{Re}=1000\), where the mesh size is much larger than the local mean free path, UGKS recovers the classical Navier--Stokes benchmark profiles much more accurately, whereas the IMEX AP scheme remains excessively dissipative. 

This example clarifies why UGKS- and DUGKS-type methods are naturally aligned with the UP question. Their interfacial distributions are built from local kinetic evolutions that couple transport and relaxation. The numerical flux therefore contains a discrete space--time model of the transport--collision balance, rather than a collisionless flux corrected only by an implicit cell-centered relaxation step \cite{XuHuang2010UGKS,GuoXuWang2013DUGKS,GuoWangXu2015DUGKS,ref:ChenXu-2015}. The UP analysis of Guo \emph{et al.}~\cite{GuoLiXu2023UP} makes this distinction quantitative. For DUGKS, the modified equation shows that, when
\begin{equation}
	\Delta t=O(\varepsilon^\alpha),\qquad
	\Delta x=O(\varepsilon^\beta),\qquad
	\frac{1}{2}<\alpha,\beta<1,
	\label{eq:sec6_dugks_up_scaling}
\end{equation}
the discrete Chapman--Enskog balance equations agree with those of the BGK model through the Navier--Stokes order. DUGKS is therefore second-order UP under this scaling. If the interface distribution is reconstructed instead from a collisionless transport equation, the UP order degenerates under the same numerical resolution, and the resulting scheme cannot recover the Navier--Stokes solution without imposing a stronger time-scale restriction. This gives a precise theoretical counterpart to the earlier numerical observation that AP schemes with different flux constructions may behave very differently in continuum and near-continuum regimes.

The same UP-based perspective can be applied beyond deterministic UGKS and DUGKS. For particle AP methods, one should examine whether stochastic transport, relaxation sampling, and variance-reduction corrections preserve not only the limiting diffusion or fluid model, but also the finite-\(\varepsilon\) transport coefficients. For GSIS-type methods, one should examine how the synthetic macroscopic equations modify the kinetic iteration and whether the coupled solver preserves the desired NSF asymptotics on coarse meshes. For UGKWP and related wave--particle methods, the relevant question is how the sampled particle part, the analytic wave part, and the collision-history repartition together reproduce the asymptotic hierarchy. 

This perspective also affects benchmark design. Testing only the collisionless limit and the leading continuum limit is insufficient. More discriminating tests should include near-continuum configurations where Navier--Stokes-level physics is important but \(\Delta x\) and \(\Delta t\) remain much larger than the mean free path and collision time. The lid-driven cavity comparison of Chen and Xu~\cite{ref:ChenXu-2015} is a useful example because the shear-driven Navier--Stokes structure exposes excessive numerical dissipation that is invisible in a purely Euler-limit AP test. The Taylor-vortex tests in the UP study provide a complementary and more controlled validation. With \(\Delta x\sim\varepsilon^{0.501}\), DUGKS captures the analytic incompressible Navier--Stokes decay for several small \(\varepsilon\), even though the mesh size is much larger than the kinetic length scale. When a coarser \(\Delta x\sim\varepsilon^{0.4}\) is used, visible deviations appear, consistent with the predicted UP scaling. The same study also shows that a collisionless-reconstruction scheme is too dissipative under the DUGKS resolution, while a second-order IMEX--RK scheme can lose its nominal time-asymptotic property once practical spatial reconstruction and its numerical dissipation are included \cite{GuoLiXu2023UP}. These tests show that AP/UP assessment should examine the coupled time--space discretization, not only the time discretization or the limiting equation.

\subsection{Section remarks}
The discussion in this section shows that AP and UP provide complementary criteria for assessing multiscale kinetic schemes. AP establishes the minimal
requirement that a kinetic discretization recover the prescribed leading macroscopic limit when the kinetic scale is unresolved. UP goes further by examining whether the fully discrete method preserves the asymptotic hierarchy and the associated transport--collision balance at finite numerical resolution. This broader viewpoint clarifies the different behavior of AP schemes in continuum and near-continuum regimes, but it also shows that UP analysis is still far from complete.

Several open issues remain. First, UP theory has so far been developed explicitly only for a limited set of representative schemes. Extending it to UGKS, UGKWP, DUGKWP, GSIS-type solvers, AP Monte Carlo methods, and more complex collision models would clarify its generality. Second, adaptive velocity discretization, implicit acceleration, synthetic coupling, and wave--particle representation all change the discrete kinetic hierarchy and therefore need their own UP-level analysis. Third, the present UP formulation is mainly Chapman--Enskog based. For strongly nonlinear nonequilibrium regimes, multiple relaxation pathways, plasma transport, phonon transport, and turbulence-related kinetic models, the relevant coefficient hierarchy may not be a classical Chapman--Enskog sequence.

AP and UP therefore provide complementary levels of assessment. AP determines whether the correct leading macroscopic model is recovered when the kinetic scale is unresolved. UP determines whether the discrete method preserves the deeper asymptotic structure needed for accurate continuum and near-continuum computation. Together they connect numerical design with the physical scale at which transport and collision are observed. This viewpoint prepares the transition to Sec.~\ref{sec:ugkf}, where the observation scale is built directly into a kinetic framework through collision-history populations.

\section{Unified gas-kinetic framework (UGKF)}
\label{sec:ugkf}
Secs.~\ref{sec:deterministic_methods}--\ref{sec:ap_up} examined multiscale kinetic computation from the viewpoints of method construction, coupling strategy, wave--particle representation, and asymptotic assessment. The unified gas-kinetic framework (UGKF) takes a further step by moving the discussion from scheme-level design to scale-dependent physical description \cite{GuoZhuXu_AA2026}. Its central idea is to introduce the observation scale into the kinetic formulation itself, so that molecular transport, collision, and accumulated collision history are described relative to a finite observation window.

In UGKF, the relevant description of a gas is scale dependent \cite{GuoZhuXu_AA2026}. At kinetic scales, individual free transport and collision events remain explicit; at hydrodynamic scales, the cumulative effect of many collisions produces continuum behavior; between these two limits, molecular populations with different collision histories coexist over the observation window. UGKF makes this population structure explicit and uses it to connect kinetic, transitional, and continuum descriptions within one formulation.

This viewpoint also clarifies the scientific position of UGKF within multiscale kinetic theory. UGKS, UGKP, and UGKWP demonstrate how free transport and collision can be coupled in a numerical evolution over a finite space--time scale. AP and UP properties provide criteria for assessing whether such discrete evolutions recover the correct macroscopic limits and preserve the relevant asymptotic structures. UGKF addresses the same multiscale issue at the level of physical formulation, where gas dynamics is described relative to
an observation scale that is large enough for collision histories to accumulate but not necessarily large enough for a purely hydrodynamic closure. In this sense, UGKF provides a framework-level interpretation of unified gas-kinetic methodology and connects numerical multiscale construction with scale-dependent physical description.

With this motivation, the section first formulates the shift from multiscale algorithms to a scale-dependent kinetic framework. It then reviews the UGKF construction from the Boltzmann equation, including the collision-history population system and its reformulated representation. The final part discusses how this observation-scale viewpoint may contribute to the broader
kinetic-to-continuum problem.

\subsection{From methods to framework}
\label{subsec:ugkf_from_methods}

The motivation for UGKF can be stated in simple physical terms. The Boltzmann equation describes a gas through a single distribution function at the kinetic level. However, when the system is observed over a finite time interval, not all molecules have the same collision history. Some molecules experience no collision during the observation window, some remain free only up to an intermediate time and then collide, and others have already entered a strongly collided population. A single distribution function contains the full kinetic information, but it does not explicitly separate these collision-history classes.

UGKF makes this hidden structure explicit. Let \(h_{\rm obs}\) denote an observation time scale. Over the interval \((0,h_{\rm obs}]\), molecules are classified according to their collision histories. The resulting description is neither a pure kinetic limit theory nor a conventional hydrodynamic closure. It is a scale-dependent gas-dynamic formulation in which kinetic, transitional, and hydrodynamic descriptions appear as different observation-scale regimes \cite{GuoZhuXu_AA2026}.

This is the key conceptual shift. In a conventional numerical-method viewpoint, one starts from a prescribed kinetic equation and determines how it should be discretized. In UGKF, one first identifies which molecular populations are distinguishable at a given observation scale, and then formulates the corresponding scale-dependent kinetic description. At kinetic scales, individual free transport and collision events remain explicit. At hydrodynamic scales, the cumulative effect of many collisions produces continuum behavior. Between these two limits, free, transitional, and collided populations coexist and must be represented together.

This viewpoint is closely related to the kinetic representation of UGKWP reviewed in Sec.~\ref{subsec:sec5_kinetic_representation}. In UGKWP, particle and wave populations are separated in order to construct an efficient multiscale algorithm. In UGKF, the same collision-history idea is elevated from an algorithmic representation to a scale-dependent kinetic formulation. This connection explains why UGKF can be viewed as a theoretical extension of the direct-modeling and wave--particle ideas developed earlier \cite{GuoZhuXu_AA2026,GuoZhuXu_CiCP2026}.

The significance of this shift is both numerical and theoretical. Numerically, it explains why a unified gas-kinetic method can remain meaningful across regimes without resolving the kinetic scale everywhere, since the numerical update is constructed on the same scale over which free transport and collisions are observed. Theoretically, it connects multiscale computation with the broader problem of relating molecular motion, kinetic evolution, and macroscopic continuum laws. UGKF should therefore be understood not as another solver family, but as a scale-dependent kinetic description in which the observation scale becomes part of the formulation.

\subsection{Observation-scale collision-history decomposition}
\label{subsec:ugkf_from_boltzmann}
We now show how the observation-scale collision-history decomposition can be
derived from the gain--loss form of the Boltzmann equation
\eqref{eq:boltzmann_intro}. Writing the collision operator as
\begin{equation}
	Q_B(f,f)=Q^+(f,f)-\nu f ,
	\label{eq:sec7_gainloss_revised}
\end{equation}
where \(Q^+\) is the gain term and \(\nu\) is the collision frequency, the
Boltzmann equation admits a characteristic representation along the backward
trajectory ending at \((\x,\bxi,t)\). This yields the formal mild form
\begin{equation}
	f(\x,\bxi,t)
	=
	e^{-\bar{\nu}(t,0;\x,\bxi)} f_0(\x-\bxi t,\bxi)
	+
	\int_0^t
	e^{-\bar{\nu}(t,s;\x,\bxi)}
	Q^+(\x-\bxi(t-s),\bxi,s)\,ds ,
	\label{eq:ugkf_boltzmann_mild}
\end{equation}
where
\begin{equation}
	\bar{\nu}(t,s;\x,\bxi)
	=
	\int_s^t
	\nu(\x-\bxi(t-t'),\bxi,t')\,dt' .
	\label{eq:ugkf_accumulated_frequency}
\end{equation}
Here \(\bar{\nu}(t,s;\x,\bxi)\) is the accumulated collision frequency along
the characteristic over the time interval \([s,t]\). Equivalently, it measures
the expected number of collisions accumulated along that trajectory over this
time interval. Therefore \(e^{-\bar{\nu}(t,s;\x,\bxi)}\) is the survival
probability that no collision occurs between \(s\) and \(t\). The representation
\eqref{eq:ugkf_boltzmann_mild} is formal because both \(Q^+\) and \(\nu\) are
determined by the distribution itself. Nevertheless, it separates the solution
into two physically distinct contributions. The first term represents molecules
that have not collided before time \(t\), while the integral term represents
the population generated by collisions before time \(t\).

The specific contribution of UGKF is to refine this structure after an
observation horizon \(h_{\rm obs}\) is introduced. For \(0\le t\le h_{\rm obs}\),
the uncollided population at time \(t\) can be further separated into molecules that
remain collisionless over the whole observation interval and molecules that
are still free at time \(t\) but will collide before \(h_{\rm obs}\). 
For a molecule located at \(\x\) at time \(t\), the accumulated collision frequency over the whole observation window is
\begin{equation}
	\bar{\nu}_h(\x_0,\bxi) =\bar{\nu}(h_{\rm obs},0;\x_0+\bxi h_{\rm obs},\bxi)
\end{equation}
where $\x_0=\x-\bxi t$ is the starting point of the molecule. The corresponding survival probability over the whole observation window is
\begin{equation}
	\beta_h(\x_0,\bxi)
	=
	\exp[-\bar{\nu}_h(\x_0,\bxi)] .
	\label{eq:ugkf_beta_h}
\end{equation}
By contrast, \(e^{-\bar{\nu}(t,0;\x,\bxi)}\) is the survival probability only
up to the current time \(t\). The difference between these two survival factors
therefore identifies molecules that have not collided by time \(t\), but are
not collisionless over the whole observation window.
This leads to the observation-scale decomposition
\begin{align}
	f_F(\x,\bxi,t;h_{\rm obs})
	&=
	\beta_h(\x_0,\bxi) f_0(\x_0,\bxi),
	\label{eq:ugkf_fF}
	\\
	f_T(\x,\bxi,t;h_{\rm obs})
	&=
	\left[
	e^{-\bar{\nu}(t,0;\x,\bxi)}
	-
	e^{-\bar{\nu}_h(\x_0,\bxi)}
	\right] f_0(\x_0,\bxi),
	\label{eq:ugkf_fT}
	\\
	f_C(\x,\bxi,t;h_{\rm obs})
	&=
	\int_0^t
	e^{-\bar{\nu}(t,s;\x,\bxi)}
	Q^+(\x-\bxi(t-s),\bxi,s)\,ds .
	\label{eq:ugkf_fC}
\end{align}
Here \(f_F\) is the free-transport population that remains collisionless
throughout the observation interval, \(f_T\) is the transitional population
that is still uncollided at time \(t\) but will collide before
\(h_{\rm obs}\), and \(f_C\) is the collision-generated population accumulated
before time \(t\). It is noted that $f_C$ has no explicit dependence on \(h_{\rm obs}\), but it is included in the
same observation-scale decomposition for \(0\le t\le h_{\rm obs}\).

For \(0\le t\le h_{\rm obs}\), this decomposition satisfies the exact identity
\begin{equation}
	f(\x,\bxi,t)=f_F(\x,\bxi,t;h_{\rm obs})
	+f_T(\x,\bxi,t;h_{\rm obs})
	+f_C(\x,\bxi,t;h_{\rm obs}).
	\label{eq:ugkf_three_population_split}
\end{equation}
Indeed, \(f_F+f_T\) recovers the uncollided term in
Eq.~\eqref{eq:ugkf_boltzmann_mild}, while \(f_C\) is the collision-generated
integral term. Thus Eq.~\eqref{eq:ugkf_three_population_split} is not an
additional closure approximation. It is an exact repartition of the formal
Boltzmann mild solution over the prescribed observation window.

The conservative variables associated with the three populations are defined by
\begin{equation}
	\W_k=\int \bm{\psi} f_k\,d\bxi,\qquad
	k=F,T,C,
	\label{eq:ugkf_population_moments}
\end{equation}
so that
\begin{equation}
	\W=\W_F+\W_T+\W_C .
	\label{eq:ugkf_moment_split}
\end{equation}
This repartition provides the population variables used in the UGKF system. The
next step is to write the corresponding evolution equations for
\(f_F\), \(f_T\), and \(f_C\).

\subsection{UGKF system}
\label{subsec:ugkf_system}
\begin{figure}
	\centering
	\includegraphics[width=0.5\textwidth]{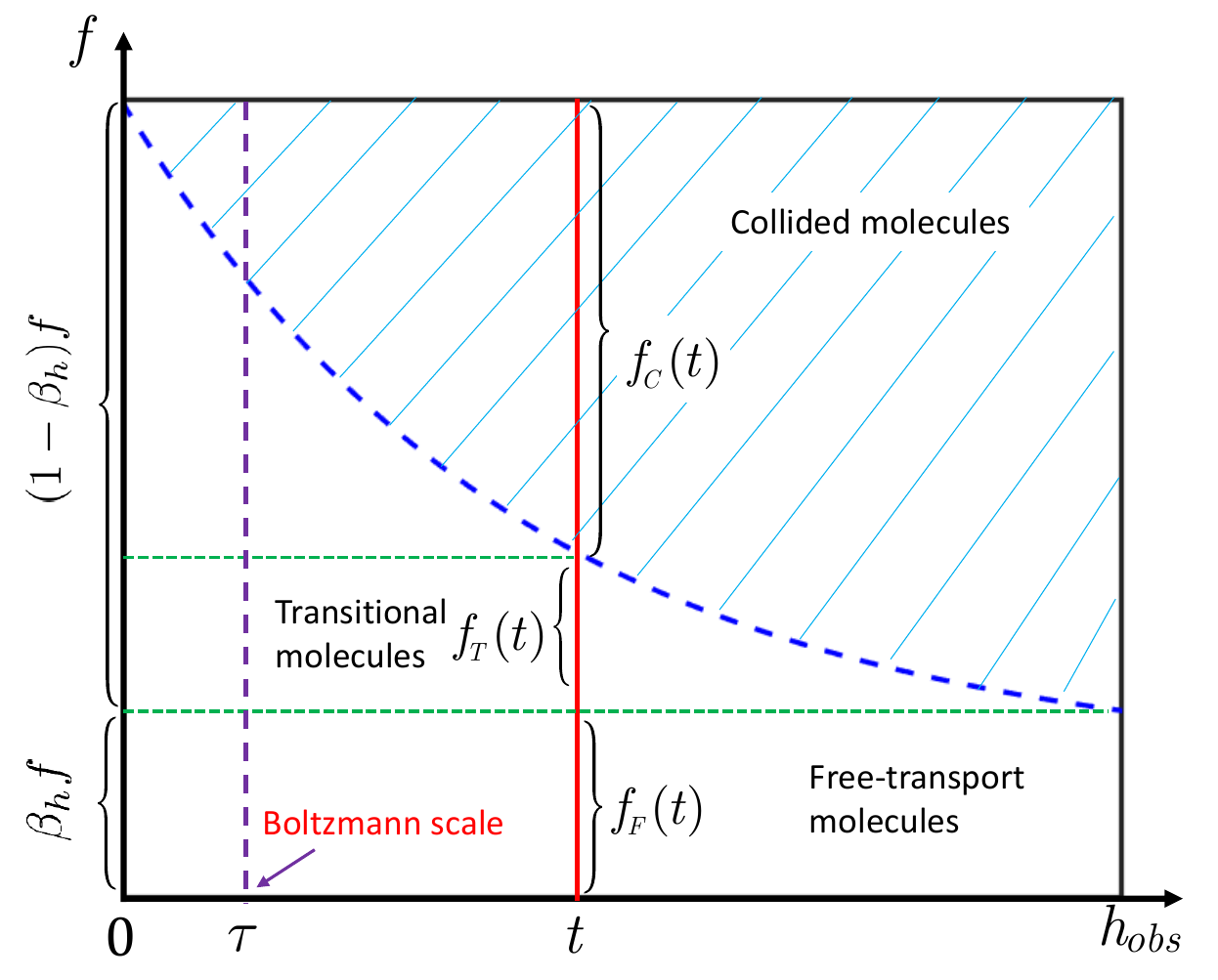}    
	\caption{Classification of gas molecules based on their collisional history over the observation time scale, and the time evolution of the distribution functions of the three populations. The blue dashed curve separates the molecules that have collided by time $t$ from those that remain uncollided. The hatched and unhatched regions represent these two populations, respectively, and $t\in [0, h_{\mbox{obs}}]$ is the elapsed time. Adapted from Fig.~1(b) of Guo et al.~\cite{GuoZhuXu_AA2026}, licensed under		\href{https://creativecommons.org/licenses/by/4.0/}{CC BY 4.0}.	Copyright (2026) The Author(s).}
	\label{fig:UGKF} 
\end{figure}

The observation-scale decomposition introduced above leads to a population
system by differentiating the three components along characteristics. With the
initial repartitioning determined by \(\beta_h\), the three populations satisfy
\begin{subequations}
	\label{eq:sec7_ugkf_system}
	\begin{align}
		&\partial_t f_F+\bm{\xi}\cdot\nabla f_F=0,
		\qquad
		f_F(\bm{x},\bm{\xi},0;h_{\rm obs})
		=\beta_h(\bm{x},\bm{\xi})f_0(\bm{x},\bm{\xi}),
		\label{eq:sec7_ugkf_fF}\\
		&\partial_t f_T+\bm{\xi}\cdot\nabla f_T
		=-\nu(f_T+f_F),
		\qquad
		f_T(\bm{x},\bm{\xi},0;h_{\rm obs})
		=\bigl[1-\beta_h(\bm{x},\bm{\xi})\bigr]f_0(\bm{x},\bm{\xi}),
		\label{eq:sec7_ugkf_fT}\\
		&\partial_t f_C+\bm{\xi}\cdot\nabla f_C
		=-\nu f_C+Q^+,
		\qquad
		f_C(\bm{x},\bm{\xi},0;h_{\rm obs})=0.
		\label{eq:sec7_ugkf_fC}
	\end{align}
\end{subequations}
It is clear that the sum of the three equations recovers the original Boltzmann equation
\cite{GuoZhuXu_AA2026}. Thus the UGKF population system is not a separate kinetic model imposed from
outside, but an observation-scale repartition of the Boltzmann dynamics. 
The system \eqref{eq:sec7_ugkf_system} clearly describes the transport
processes of the three types of molecules, which are sketched in Fig. \ref{fig:UGKF}.

The role of the observation scale can be measured by the collision number accumulated over the observation window,
\begin{equation}
	\delta_h=h_{\rm obs}\nu_m
	\sim \frac{h_{\rm obs}}{\tau_m},
	\label{eq:sec7_delta}
\end{equation}
where \(\nu_m\) and \(\tau_m\) denote representative collision frequency and collision time along the characteristic. The limiting regimes are then identified directly from the population structure. When \(\delta_h\ll 1\), most molecules remain collisionless
over the observation interval, and the free-transport population dominates ($f\approx f_F$, $f_T\to 0$, $f_C\to 0$);
when \(\delta_h=O(1)\), free, transitional, and collided populations may all
contribute; when \(\delta_h\gg 1\), the full-window survival probability is small. The collided population dominates after the initial layer, when the elapsed collision hazard is also large ($f\approx f_C$, $f_F\to0$, $f_T\to0$). Note that a large observation window alone does not imply $f_T\to0$ near $t=0$. In this sense, the same scale-dependent population system describes free-molecular, transitional, and
hydrodynamic regimes through different population balances, without switching
among separate governing equations.

The emergence of continuum behavior can also be seen at the population level. When the collision frequency is frozen locally and the collision-gain state $f^+$ is smooth and slowly varying along a characteristic, the collided population may be expanded as
\begin{equation}
	f_C
	\approx
	\bigl(1-e^{-\nu_m t}\bigr)f^+
	-
	\left[
	\frac{1}{\nu_m}\bigl(1-e^{-\nu_m t}\bigr)
	-te^{-\nu_m t}
	\right]
	D_t f^+,
	\qquad
	f^+=\frac{Q^+}{\nu_m},
	\label{eq:sec7_fC_asympt}
\end{equation}
where \(D_t=\partial_t+\bm{\xi}\cdot\nabla\). For \(\nu_m t\gg1\) after the initial layer, this gives
\begin{equation}
	f_C\approx f^+-\frac{1}{\nu_m}D_t f^+ .
	\label{eq:sec7_fc_continuum}
\end{equation}
For a BGK relaxation model, $f^+=g$, and Eq.~\eqref{eq:sec7_fc_continuum} has the form of the first Chapman--Enskog
correction. For the full Boltzmann operator, $f^+=Q^+/\nu$ generally differs from $g$ at first nonequilibrium order. Its contribution must therefore be retained when determining the Boltzmann transport coefficients; replacing
	$f^+$ by $g$ is an additional model approximation. The population expansion
	identifies the collisional origin of the continuum contribution but is not,
	by itself, a derivation of the full Boltzmann Navier--Stokes closure \cite{GuoZhuXu_AA2026}.

This observation motivates a reformulated UGKF in which the collided
population is represented by its conservative moments rather than by a full
kinetic distribution. This gives a reformulated UGKF of the form
\begin{subequations}
	\label{eq:sec7_reform_ugkf}
	\begin{align}
		\partial_t f_F+\bm{\xi}\cdot\nabla f_F
		&=0, \label{eq:rfF}\\
		\partial_t f_T+\bm{\xi}\cdot\nabla f_T
		&=-\nu(f_T+f_F),\label{eq:rfT}\\
		\partial_t\bm{W}_C+\nabla\cdot\bm{\mathcal F}_C
		&=\left\langle \bm{\psi}\,\nu(f_F+f_T)\right\rangle \label{eq:rfC}.
	\end{align}
\end{subequations}
Here \(\bm{\mathcal{F}}_C=\langle \bm{\xi}\bm{\psi} f_C\rangle\)
is the flux of the collided population. The source terms on the right-hand side
represent the transfer of conservative quantities from the free-transport and
transitional populations into the collided population. Thus \(f_F\) and \(f_T\)
retain the kinetic information associated with collisionless and pre-collision
transport, while \(\bm{W}_C\) carries the conservative content of the collided
population. 

\begin{figure}
	\centering
	\includegraphics[width=0.45\textwidth]{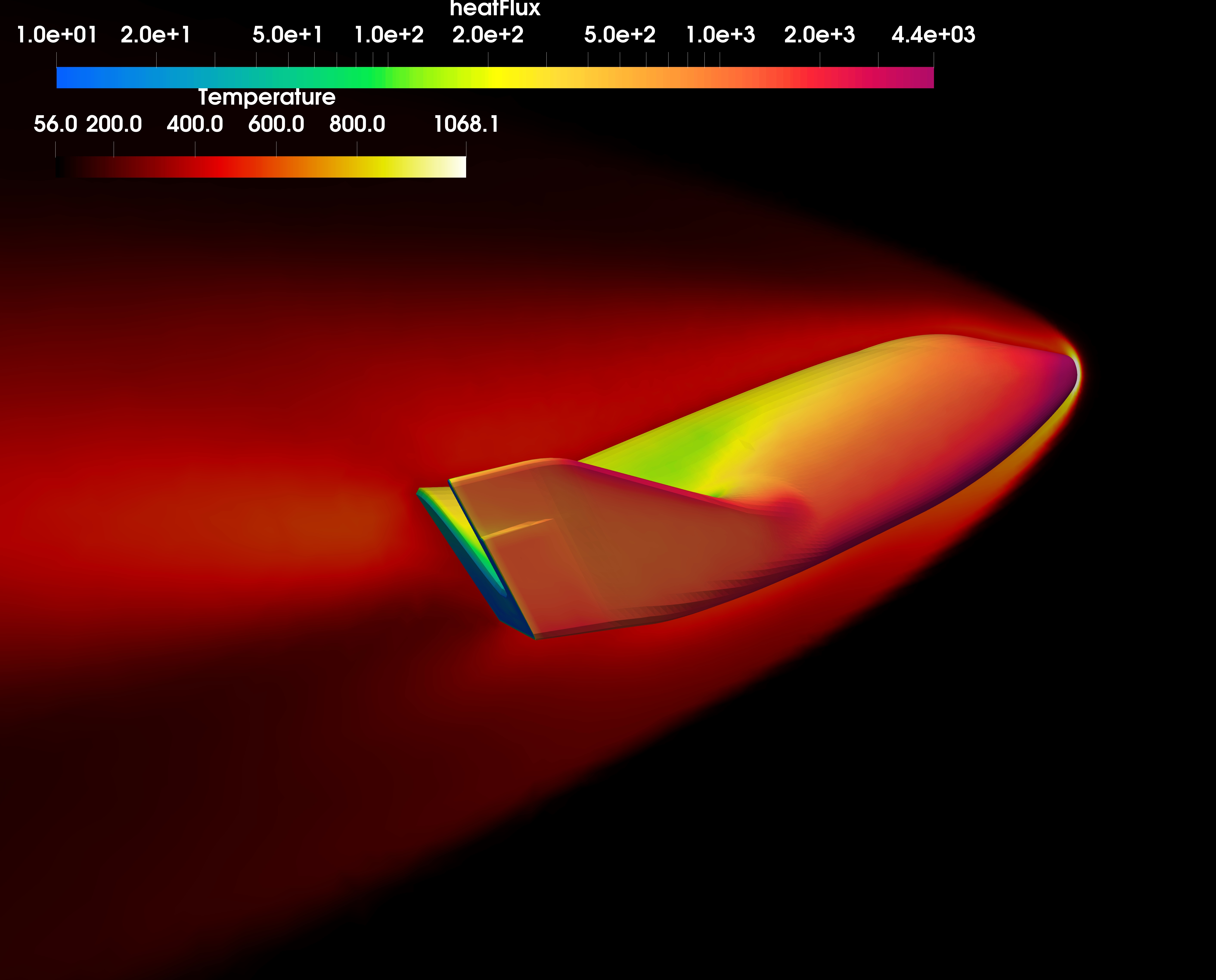}
	\caption{Heat flux and temperature fields of Argon gas around the X38-like vehicle with $\mbox{Kn}=2.75\times 10^{-3}$ and $\mbox{Ma} = 8.0$ at an angle of attack of $20^{\circ}$. Reproduced from Fig.~6(a) of Guo et al.~\cite{GuoZhuXu_AA2026}, licensed under \href{https://creativecommons.org/licenses/by/4.0/}{CC BY 4.0}. Copyright (2026) The Author(s).}
	\label{fig:X38-1} 
\end{figure}

The closure of \(\bm{\mathcal{F}}_C\) is determined by the collisional character
of this population. Since \(f_C\) represents molecules generated by collisions,
its flux can be approximated by the hydrodynamic form suggested by the
asymptotic expression of the collided population in
Eq.~\eqref{eq:sec7_fc_continuum}. It should be noted that this flux closure is an additional approximation, distinct from the exact population decomposition, and its accuracy in transitional flows requires assessment. In this way, the reformulated UGKF states
explicitly which part of the dynamics must remain kinetic and which part may be
advanced through conservative moment equations \cite{GuoZhuXu_AA2026}. 

\begin{figure}
	\centering
	\subfloat[]{\includegraphics[width=0.32\linewidth]{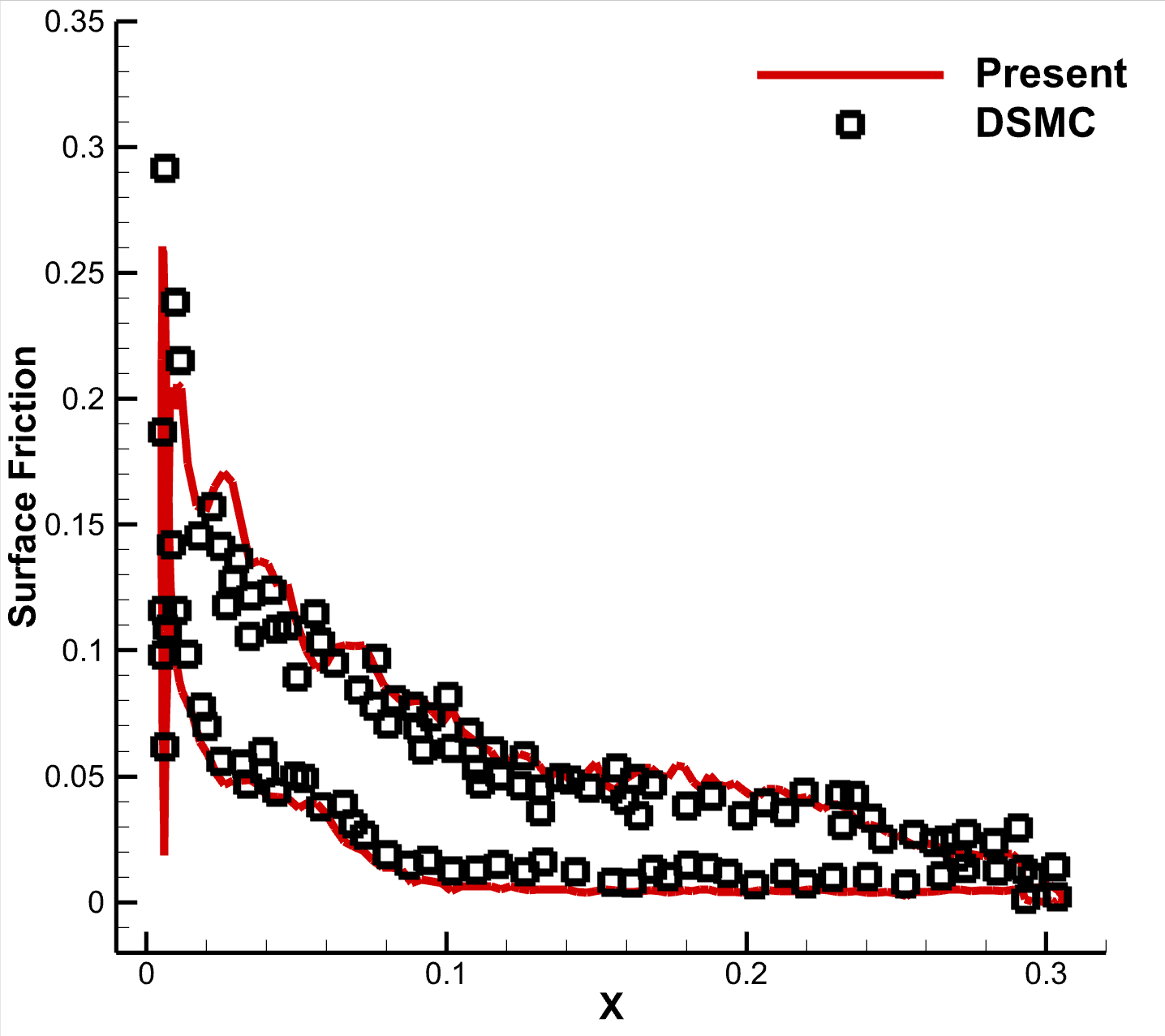}}
	\hfill
	\subfloat[]{\includegraphics[width=0.32\linewidth]{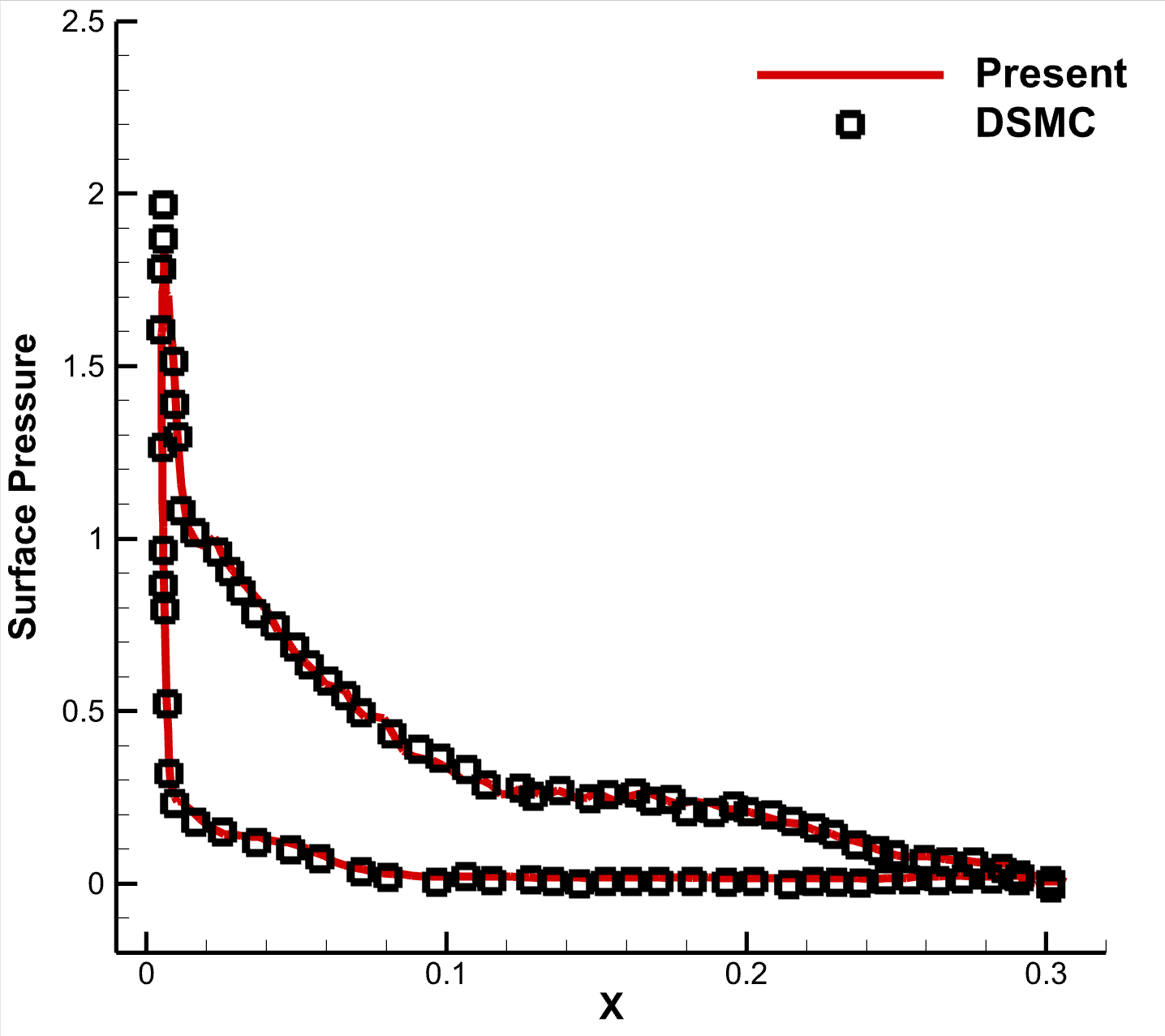}}
	\hfill
	\subfloat[]{\includegraphics[width=0.32\linewidth]{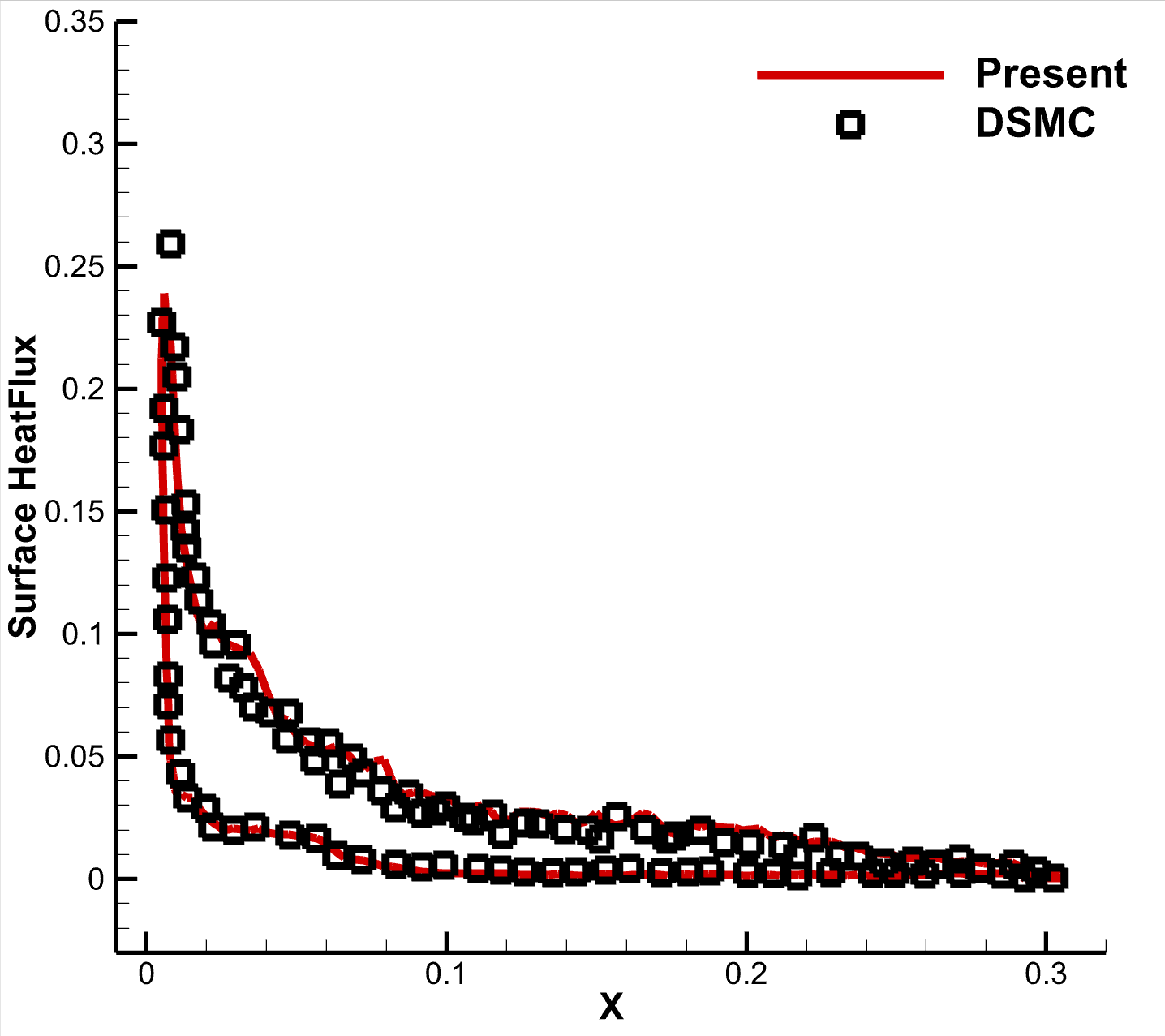}}
	\caption{Aerothermodynamic surface quantities of the flow around the X38-like vehicle with $\mbox{Kn}=2.75\times 10^{-3}$. (a) skin friction, (b) surface pressure, and (c) surface heat flux. DSMC data are from \cite{ref:X38_DSMC}.
	Reproduced from Fig.~7 of Guo et al.~\cite{GuoZhuXu_AA2026}, licensed under \href{https://creativecommons.org/licenses/by/4.0/}{CC BY 4.0}. Copyright (2026) The Author(s).}
	\label{Sfig:X38-2} 
\end{figure}

This reformulation makes explicit the multiscale structure of UGKF. Kinetic
resolution is retained for populations whose collision histories remain
distinguishable over the observation window, whereas the strongly collided
population is described through moment variables and continuum-type closure.
This is the main difference between UGKF and a conventional kinetic--fluid
switching strategy. The separation is not prescribed by an external domain
partition, but follows from collision histories accumulated over the
observation scale. The observation-scale population structure can thus
be used to design kinetic–moment algorithms. Two such methods were developed from  
the reformulated UGKF system \eqref{eq:sec7_reform_ugkf} in Ref.~\cite{GuoZhuXu_AA2026}. The first is a deterministic method, in which Eqs.~\eqref{eq:rfF} and \eqref{eq:rfT} are solved by a discrete velocity method, while Eq.~\eqref{eq:rfC} is solved by the gas-kinetic scheme \cite{ref:GKS}. The second is a hybrid wave--particle method designed for high-speed compressible flows, in which the uncollided and transitional populations are represented by stochastic particles, whereas the collided population is solved by the gas-kinetic scheme. 
As an application, the flow around an X38-like vehicle was simulated. Figure \ref{fig:X38-1} presents the distributions of surface heat flux and temperature field, and Fig.~\ref{Sfig:X38-2} compares skin friction, surface pressure, and heat flux with the DSMC data of Ref. \cite{ref:X38_DSMC}. 

\subsection{Section remarks}
\label{subsec:ugkf_remarks_open}

The significance of UGKF lies in its formulation of multiscale gas dynamics
through observation-scale population structure. The collision-history
decomposition, which appears algorithmically in UGKS, UGKP, and UGKWP as a
representation and efficiency device, is elevated in UGKF to a scale-dependent
kinetic description. In this sense, UGKF should not be viewed as another
solver family or as a replacement for existing multiscale numerical methods.
Rather, it provides a framework-level interpretation of why free transport,
collision, relaxation, and continuum response can be organized coherently over
a finite observation scale.

This viewpoint also gives a useful theoretical perspective on the
kinetic-to-continuum connection. Classical kinetic theory usually relates
molecular dynamics and continuum mechanics through limiting procedures, such
as hydrodynamic limits of the Boltzmann equation. UGKF does not constitute a
rigorous solution of this problem, nor should it be presented as such. Its
contribution is more specific in that it inserts an explicit observation scale between
kinetic and hydrodynamic descriptions and organizes molecular populations
according to their collision histories over that scale
\cite{GuoZhuXu_AA2026}. This provides a constructive bridge between molecular
transport and continuum behavior. The dominant population changes continuously
with the observation scale, rather than being selected by an abrupt switch
between separate governing equations. In this limited sense, UGKF offers a
scale-dependent perspective on the type of kinetic-to-continuum connection associated with
Hilbert's sixth problem \cite{Hilbert1901,Gorban2018}.

Several issues remain open. These include the practical selection of the
observation scale, the discrete consistency of source transfer among different
populations, the robust closure of the collided moment system in intermediate
regimes, and the extension of the framework to polyatomic gases, mixtures,
reactive flows, plasmas, and other nonequilibrium transport systems. These
questions indicate that UGKF is still developing as a theoretical and
computational framework, but they do not weaken its central message that an
observation-scale population structure provides a systematic way to organize
multiscale kinetic descriptions.

\section{Extensions to other transport systems}
\label{sec:extensions_other_transport}
The preceding sections have focused mainly on rarefied-gas dynamics, where the Boltzmann equation and its model equations provide the central kinetic description. The underlying multiscale difficulty, however, is not tied only to molecular gases. Many nonequilibrium transport systems can be described in terms of carrier distributions whose free propagation, interaction, relaxation, and macroscopic response coexist across different physical scales. The carriers may be photons, neutrons, charged particles, phonons, electrons, droplets, or grains, and their interaction mechanisms and limiting macroscopic equations may differ substantially from those of molecular gases. Nevertheless, they raise closely related numerical and modeling questions concerning how to couple transport with interaction, how to recover the correct limiting behavior in unresolved regimes, and how to retain finite-scale nonequilibrium effects.

This section reviews representative extensions to radiative transfer, neutron
transport, plasma transport, phonon and electron--phonon transport, and
gas--particle or granular flows. The aim is not to impose a single
gas-kinetic formulation on all these systems, but to identify which structural
principles can be transferred and how they must be adapted to the physics of
each transport problem.
\subsection{Common structural features}
\label{subsec:sec8_common_structure}
The systems considered in this section do not share one universal governing
equation. Photons, neutrons, phonons, charged particles, dispersed particles,
and molecules have different phase variables, interaction laws, equilibrium or
relaxation states, conserved quantities, source terms, and asymptotic limits.
The diffusion limit of radiative or neutron transport is not the
Navier--Stokes limit of molecular gases. Plasma transport involves
electromagnetic fields and multiple characteristic scales. Phonon transport
leads to Fourier or generalized heat-conduction behavior rather than fluid
momentum equations. Gas--particle systems may involve a dispersed phase whose
local collisional regime differs from that of the carrier gas. The portability
of multiscale kinetic methodology must therefore be understood as the transfer
of structural principles, not as the literal reuse of one model or algorithm.

Despite these differences, a common multiscale transport structure can often
be identified. A carrier distribution is transported in physical space and
possibly in an additional phase space, while interaction terms drive it toward
a lower-dimensional macroscopic behavior. A generic form may be written as
\begin{equation}
	\partial_t f
	+\bm{c}(\bm{z})\cdot\nabla f
	+\nabla_{\bm{z}}\cdot\bigl(\bm{A}f\bigr)
	=
	\frac{1}{\varepsilon}{\mathcal{C}}(f)
	+S(f).
	\label{eq:sec8_generic_transport}
\end{equation}
Here \(f=f(\bm{x},\bm{z},t)\) is a carrier distribution, \(\bm{z}\) denotes
velocity, direction, frequency, internal state, species label, or other phase
variables, \(\bm{c}(\bm{z})\) is the transport velocity, and \(\bm{A}\)
represents phase-space drift such as electromagnetic acceleration. The
operator \(\mathcal{C}\) denotes the stiff interaction mechanism, including
collision, scattering, relaxation, absorption--emission balance, or
inter-carrier exchange. The term \(S\) denotes nonstiff sources,
external forcing, or coupling effects. The parameter \(\varepsilon\) measures
the ratio between the interaction scale and the macroscopic observation scale.

The corresponding macroscopic variables are obtained from moments of \(f\).
Extending the bracket notation to the relevant microscopic variables gives
\begin{equation}
	\bm{U}(\bm{x},t)
	=
	\left\langle \bm{\phi}(\bm{z})f\right\rangle,
	\qquad
	\left\langle \cdot\right\rangle
	=
	\int(\cdot)\,d\bm{z},
	\label{eq:sec8_generic_moments}
\end{equation}
where \(\bm{\phi}\) denotes the relevant moment functions. Depending on the
system, \(\bm{U}\) may represent radiation energy, neutron scalar flux, fluid
mass and momentum, charge density, phonon energy, electron and phonon
temperatures, or dispersed-phase mass and momentum. If phase-space boundary
terms vanish, taking moments of Eq.~\eqref{eq:sec8_generic_transport} gives
\begin{equation}
	\partial_t\bm{U}
	+\nabla\cdot
	\left\langle \bm{c}\bm{\phi}f\right\rangle
	=
	\frac{1}{\varepsilon}
	\left\langle \bm{\phi}\mathcal{C}(f)\right\rangle
	+
	\left\langle \bm{\phi}S(f)\right\rangle
	+
	\left\langle
	\bm{A}\cdot\nabla_{\bm{z}}\bm{\phi}\,f
	\right\rangle .
	\label{eq:sec8_generic_moment_balance}
\end{equation}
The limiting macroscopic equation depends on the scaling, the null space or
balance manifold of \(\mathcal{C}\), and the moments retained in
\(\bm{U}\). In a strongly interacting regime one often has
\begin{equation}
	\mathcal{C}(f^{\rm eq})=0,
	\qquad
	f\approx f^{\rm eq}(\bm{U}),
	\label{eq:sec8_generic_equilibrium}
\end{equation}
or a more general local balance state when the interaction does not conserve
all moments.

This structure explains why similar numerical ideas reappear in different
transport fields. When interactions are weak, ballistic or trajectory-resolving
descriptions are natural. When interactions are frequent, continuum,
diffusion, fluid, or heat-conduction descriptions become dominant. Between
these limits, the main computational difficulty is to couple transport and
interaction on the relevant numerical or physical observation scale. AP and UP
properties assess whether the discrete system recovers the correct limiting
behavior without resolving the small interaction scale. UGKS- and DUGKS-type
fluxes couple transport and interaction over a finite time step. Synthetic
iterative schemes use macroscopic equations to accelerate slow hydrodynamic or
diffusive information transfer. Wave--particle methods represent ballistic and
strongly interacting components differently within one computation. These
principles must be adapted to the physics of each carrier system, but their
common role is to organize the transition between ballistic, intermediate, and
continuum descriptions.
\subsection{Radiative transport}
\label{subsec:sec8_radiation}
Radiative transfer is one of the clearest examples in which multiscale kinetic
ideas can be transferred beyond molecular gases. For a gray radiative transfer equation with isotropic scattering, a representative model is
\begin{equation}
	\frac{1}{c}\partial_t I+\bm{\Omega}\cdot\nabla_x I
	=
	\sigma_s\left(\frac{1}{4\pi}\int_{\mathbb S^2}I\,d\bm{\Omega}-I\right)
	-\sigma_a I+G,
	\label{eq:sec8_rte}
\end{equation}
where \(I(\bm{x},\bm{\Omega},t)\) is the radiation intensity, \(c\) is the
speed of light, \(\bm{\Omega}\) is the propagation direction, \(\sigma_s\) and
\(\sigma_a\) are the scattering and absorption coefficients, and \(G\) denotes
a source or emission term. The relevant small scale is the photon mean
interaction length, determined by scattering and absorption. In optically thin
regions, photons travel over long distances before interaction, and
trajectory-based or angularly resolved descriptions are natural. In optically
thick regions, repeated scattering produces a diffusion-type macroscopic
limit, and direct particle tracking or standard discrete-ordinates transport
becomes inefficient unless the interaction scale is resolved.

This thin-to-thick transition makes radiative transfer a natural setting for
AP and unified gas-kinetic type methods. The numerical difficulty is not merely
to solve Eq.~\eqref{eq:sec8_rte} accurately on an optically resolved mesh, but
to recover the correct diffusion behavior when the cell size and time step are
much larger than the photon mean interaction length and time. Thus the same
basic design principle reappears. Transport and interaction must be coupled
over the numerical evolution scale, rather than treated as two unrelated
processes whose accuracy depends on resolving each scattering event.

A deterministic realization of this idea was first developed by Mieussens
\cite{Mieussens2013UGKSDiffusion}, who applied UGKS to a linear kinetic model
of radiative transfer and proved its AP property in the diffusion limit. This
work is important because it showed that the UGKS flux construction is not
limited to the hyperbolic fluid limits of rarefied gas dynamics. The same
finite-volume interface evolution can also recover a parabolic diffusion limit
when the collision or scattering process dominates. Sun \emph{et al.}
\cite{SunJiangXu2015GrayRTE} extended this strategy to gray radiative transfer
equations coupled with the material thermal energy equation. In that setting,
the interface radiation intensity is constructed from a local integral solution,
while the radiation energy and material temperature are updated through coupled
macroscopic equations. The resulting scheme captures both photon free transport
in optically thin regimes and nonlinear equilibrium diffusion in optically
thick regimes.

The UGKS approach was further extended to frequency-dependent radiative
transfer \cite{SunJiangXuLi2015FrequencyRTE}. This case is more demanding than
gray radiation because the opacity depends on frequency. A single spatial cell
may be optically thick for low-frequency photons and optically thin for
high-frequency photons. The multigroup UGKS treats the frequency-dependent
mean free path as part of the local transport modeling and thereby provides a
continuous transition from ballistic photon motion to diffusive radiation
propagation across both physical and frequency spaces.

DUGKS provides a simplified but closely related route for radiative transfer.
Luo \emph{et al.}~\cite{LuoWangZhangYiTan2018DUGKSRTE} developed a DUGKS for
radiative heat transfer in participating media. By integrating the radiative
transfer equation along characteristics and coupling transport with the
radiative source term in the interface flux, the method gives accurate
solutions from optically thin to optically thick regimes on relatively coarse
meshes. Song \emph{et al.}~\cite{SongZhangZhouGuo2020AnisotropicRTE} extended
this line to anisotropic scattering media by representing the scattering phase
function through Legendre polynomial expansions. More recently, Song
\emph{et al.}~\cite{SongSunGuoGao2025NonlinearGrayRTE} proposed an
AP-DUGKS for nonlinear gray radiative transfer equations, in which the
microscopic radiative transfer equation and the macroscopic material energy
balance equation are discretized consistently. These developments show that
the UGKS/DUGKS idea can be adapted to the main physical complications of
radiative transfer, including nonlinear radiation--material coupling,
frequency-dependent opacity, and anisotropic scattering.

The UGKP method for photon transport developed by Li \emph{et al.}
\cite{Li_UGKP_Photon_2018} provides a stochastic realization of the same
multiscale principle. In optically thin regimes, photon transport is represented by particles. In
optically thick regimes, the frequent-interaction limit is recovered without
requiring the time step to be smaller than the photon mean interaction time.
The thermal radiative transfer extension of Shi \emph{et al.}\
\cite{ref:UGKP_TRT2020} further couples particle transport with finite-volume
macroscopic equations for radiation and material energy, thereby treating
nonlinear radiation--material energy exchange within an AP framework.

The wave--particle extension gives a more efficient representation of the same
multiscale mechanism. In the UGKWP method for photon transport
\cite{ref:UGKWP_Photon2020}, only the collisionless or weakly interacting part
of the radiation field is sampled by particles, while the strongly scattered
part is retained through a wave or analytic representation. As a result, the
method behaves like a Monte Carlo transport method in optically thin regions
and reduces to a nearly particle-free diffusion solver in optically thick
regions. This is the radiative analogue of the wave--particle balance discussed
for gas dynamics, although the continuum limit is diffusive rather than
Navier--Stokes.

Recent developments have broadened this branch in two directions. One concerns
physical generality. Frequency-dependent UGKWP formulations extend the method
from gray radiation to multigroup or frequency-resolved transport
\cite{Yang_UGKWP_FrequencyRTE_2025}. The other concerns computational
efficiency in strongly interacting regimes. Implicit UGKWP and implicit UGKP
methods remove severe time-step restrictions in steady or highly scattering
radiation transport problems
\cite{Liu_ImplicitUGKWP_Radiation_2023,Hu_ImplicitUGKP_Gray_2024}. These
developments show that the wave--particle and implicit multiscale ideas are
not tied to molecular relaxation models, but can be adapted to scattering,
absorption, emission, and radiation--material coupling. 

Beyond the UGKS family, radiative transfer is also closely connected with synthetic acceleration.
In optically thick media, macroscopic radiation energy or diffusion equations
carry the slow large-scale information that would otherwise converge slowly
through repeated transport sweeps. This is the same structural reason why
GSIS-type strategies accelerate near-continuum gas-kinetic computations.
Although the physical closure is different, the numerical role of the
macroscopic synthetic equation is analogous: it propagates the dominant
diffusive information efficiently while the kinetic or angular transport solver
supplies higher-order nonequilibrium corrections
\cite{ZengZhangLiSuWu2026GSISRev}.

\subsection{Neutron transport}
\label{subsec:sec8_neutron}
Neutron transport is closely related to radiative transfer at the level of
streaming--interaction dynamics, but it has its own physical and computational
features. A neutron distribution depends on position, direction, energy group,
and time, while the interaction operator may include scattering, absorption,
fission, and external sources. In highly scattering media, the transport
equation approaches a diffusion-type macroscopic limit; in weakly interacting
or highly heterogeneous regions, an angularly resolved transport description
remains necessary. The multiscale difficulty is therefore to preserve the
transport solution where angular and energy dependence are important, while
recovering the correct diffusion behavior on meshes much larger than the
neutron mean free path.

This class of problems has long been connected with multiscale numerical ideas.
First-collision source methods and diffusion synthetic acceleration were
developed originally in neutron and radiative transport to treat localized
sources, optically thick regimes, and slow convergence of source iteration.
From the viewpoint of the present review, these classical techniques already
contain two ingredients that later reappear in unified kinetic methods, namely
collision-history organization and macroscopic acceleration of strongly
interacting transport.

More recently, UGKS- and DUGKS-type ideas have been introduced into neutron
transport. Zhou \emph{et al.}~\cite{Zhou_Guo_DUGKS_Neutron_2020} developed a
DUGKS for steady multidimensional and multigroup neutron transport. In this
method, the interface flux is constructed by coupling streaming and collision
over the numerical evolution scale, so that the scheme can recover the
diffusion limit on coarse meshes while remaining applicable in transport
regimes. This is the neutron-transport analogue of the coupled
transport--interaction flux used in gas-kinetic schemes, although the limiting
macroscopic equation is diffusion rather than Euler or Navier--Stokes.

Efficiency in optically thick multigroup problems requires additional
acceleration. Zhou \emph{et al.}~\cite{ZhouGuo2023AccelSDUGKS} extended the
steady DUGKS by introducing a macroscopic coarse-mesh acceleration strategy
combined with preconditioned Krylov iteration. This development is important
because it connects unified kinetic flux construction with the long tradition
of diffusion-based acceleration in neutron transport. More recently, Zhong
\emph{et al.}~\cite{ZhongGuoZhou2026PRCDUGKS} proposed a predictor--corrector
DUGKS for Boltzmann transport models with non-conservative collision operators,
with applications to multigroup neutron transport. In parallel,
three-dimensional multigroup UGKS formulations have been developed for
large-scale neutron calculations \cite{Shuang_UGKS_3DNeutron_2019}.

These developments show that multiscale kinetic methodology is not restricted
to relaxation models for molecular gases. In neutron transport, the role of
``collision'' is played by scattering, absorption, fission, and source
coupling, and the relevant continuum behavior is diffusive. 

Nevertheless, the same structural requirement persists. Streaming and interaction must be
coordinated at the numerical scale, and macroscopic diffusion information
should be used efficiently in strongly scattering regimes. The neutron
extensions therefore provide a useful example of how unified kinetic ideas can
be incorporated into established transport discretizations while improving
asymptotic behavior and coarse-mesh performance.
\subsection{Plasma transport}
\label{subsec:sec8_plasma}
Plasma and charged-particle transport provide a more demanding extension of
multiscale kinetic methodology. In neutral-gas dynamics, the main transition is
often organized by the ratio between the molecular mean free path and the
macroscopic length scale. In plasma problems, collisionality is only one part
of the scale structure. The Debye length, plasma frequency, cyclotron period,
Larmor radius, electromagnetic wave speed, charge separation scale, and
inter-species relaxation time may all influence the effective description.
Accordingly, the macroscopic limits are also more diverse. Depending on the
regime and scaling, the kinetic model may approach neutral-fluid, multi-fluid,
two-fluid, drift-kinetic, gyrokinetic, or magnetohydrodynamic descriptions.

A representative kinetic form for charged particles may be written as
\begin{equation}
	\partial_t f_s
	+\bm{\xi}\cdot\nabla f_s
	+\frac{q_s}{m_s}
	\left(
	\bm{E}+\bm{\xi}\times\bm{B}
	\right)\cdot\nabla_{\bm{\xi}} f_s
	=
	\mathcal{C}_s(f)
	+S_s ,
	\label{eq:sec8_plasma_kinetic}
\end{equation}
where \(f_s\) is the distribution of species \(s\), \(q_s\) and \(m_s\) are
its charge and mass, \(\bm{E}\) and \(\bm{B}\) are electromagnetic fields,
\(\mathcal{C}_s\) denotes collision or relaxation effects, and
\(S_s\) denotes sources, ionization, recombination, or other coupling
terms. The fields may be prescribed, determined by Poisson's equation in
electrostatic models, or coupled to Maxwell's equations in fully
electromagnetic regimes. Compared with the generic structure in
Eq.~\eqref{eq:sec8_generic_transport}, the phase-space drift induced by
electromagnetic forcing is no longer a secondary correction. It is often a
dominant part of the multiscale dynamics.

DUGKS-type methods have been extended to electrostatic plasma transport. Liu
\emph{et al.}~\cite{LiuQuanChenZhouCao2020PlasmaDUGKS} developed a
finite-volume DUGKS for electrostatic plasma and compared it with
particle-in-cell methods. In this formulation, the finite-volume kinetic evolution
couples particle transport, relaxation, and electric-field acceleration, while
the electrostatic field is obtained from the charge distribution. Related
BGK--Vlasov--Poisson formulations show that the transport--interaction coupling
idea remains useful in regimes where kinetic effects, field dynamics, and
continuum behavior coexist.

Wave--particle methodology provides another route for plasma and multispecies
charged-particle systems. Liu and Xu
\cite{Liu_Xu_UGKWP_MixturePlasma_2021} extended UGKWP to multispecies gas
mixtures and plasma transport. A key point of this development is the
asymptotic-complexity-diminishing property, through which the number of active
kinetic particles is reduced automatically as the local cell Knudsen number
decreases, while the continuum component carries an increasing fraction of the
solution. More recently, Pu and Xu~\cite{Pu_Xu_UGKWP_PIP_2025} developed a UGKWP method for partially
ionized plasma, connecting particle transport in kinetic regimes with
two-fluid and magnetohydrodynamic behavior in near-continuum regimes. These
works illustrate how a wave--particle representation can organize multispecies
transport, interspecies relaxation, and electromagnetic coupling across
different plasma regimes.

Synthetic acceleration is also beginning to show its portability in plasma
applications. Wen \emph{et al.}~\cite{Wen_Zhang_Wu_GSIS_PlasmaEdge_2026}
developed a GSIS-type method for neutral particle flows in the plasma edge.
There, a neutral kinetic equation is coupled with macroscopic synthetic
equations so that slow hydrodynamic information can be propagated efficiently,
while the kinetic solver supplies higher-order nonequilibrium corrections.
The method preserves asymptotic behavior on meshes larger than the neutral
mean free path and accelerates convergence when the kinetic
neutral subsystem is embedded in a larger multiphysics plasma-edge
environment.

These developments show both the promise and the additional difficulty of
extending multiscale kinetic ideas to plasma transport. The transferable
principle is the same as in neutral gases: transport, interaction, relaxation,
and macroscopic response should be organized on the numerical or physical
observation scale. The implementation, however, must account for electromagnetic
phase-space dynamics, multiple species, charge neutrality or charge separation,
and several competing small scales. Plasma transport therefore provides a
particularly stringent test of whether unified kinetic methodology can be
extended from collisional carrier transport to fully coupled multiphysics
systems.
\subsection{Phonon and electron--phonon transport}
\label{subsec:sec8_phonon}
Phonon transport is a natural extension of multiscale kinetic methodology to
nanoscale heat transfer. When the characteristic length of a device or
microstructure becomes comparable with phonon mean free paths, Fourier's law
is no longer sufficient to describe heat conduction. Ballistic transport,
quasi-ballistic transport, boundary scattering, dispersion, polarization, and
multiple relaxation processes may all influence the effective thermal response.
A kinetic description based on the phonon Boltzmann transport equation is then
needed, but direct deterministic or Monte Carlo solutions face the same
multiscale difficulty as gas dynamics and radiative transfer: ballistic
resolution is required in weakly scattering regimes, whereas diffusion or
generalized heat-conduction behavior should be recovered efficiently in
strongly scattering regimes.

A representative phonon kinetic equation can be written schematically as
\begin{equation}
	\partial_t e
	+\bm{v}_{g}(\bm{k},p)\cdot\nabla e
	=
	\mathcal{C}_{\rm ph}(e),
	\label{eq:sec8_phonon_bte}
\end{equation}
where \(e=e(\bm{x},\bm{k},p,t)\) denotes a phonon energy distribution or
deviation distribution, \(\bm{k}\) is the wave vector, \(p\) denotes the
polarization branch, \(\bm{v}_{g}\) is the group velocity, and
\(\mathcal{C}_{\rm ph}\) represents phonon scattering and relaxation. In
relaxation or Callaway-type models, the interaction operator may contain
several relaxation channels, for example normal and resistive processes. The
macroscopic heat flux is obtained from moments of the phonon distribution, and
the strongly scattering limit leads to Fourier heat conduction or to a
generalized hydrodynamic heat-transport model, depending on the scaling and
collision model.

The deterministic branch of multiscale phonon computation developed naturally
through UGKS- and DUGKS-type ideas. Guo and Xu
\cite{Guo_Xu_DUGKS_Phonon_2016} developed a DUGKS for multiscale heat transfer
based on the phonon Boltzmann transport equation. In this method, phonon
advection and scattering are coupled in the finite-volume interface evolution,
so that ballistic and diffusive heat transport can be captured within one
scheme. Luo and Yi \cite{Luo_Yi_DUGKS_Phonon_2017} incorporated phonon
dispersion and polarization, and Zhang and Guo
\cite{Zhang_Guo_DUGKS_Phonon_2019} extended the DUGKS framework to multiscale
heat transfer with arbitrary temperature differences. These developments show
that the coupled transport--scattering flux idea is not restricted to
molecular velocity space, but can also be adapted to frequency- and
branch-dependent heat carriers.

Particle and wave--particle formulations provide a complementary route for
phonon transport. Liu \emph{et al.}~\cite{LiuYangZhangJiXu2025UGKWPPhonon}
developed a UGKWP formulation for gray phonon transport, in which ballistic or
nonequilibrium phonons are represented by particles, while the near-equilibrium
and strongly scattered contribution is carried by a wave or macroscopic
component. The same wave--particle idea has been further extended to
multiscale phonon transport with dispersion and polarization effects through
frequency-space grouping and adaptive particle sampling
\cite{LiuYangZhangJiXu2025UGKWPIUGKPPhononDispersion}. For steady multiscale
phonon transport, Liu \emph{et al.}~\cite{Liu_IUGKP_Phonon_2026} developed an
implicit UGKP method. In that formulation, the steady integral solution is used
to organize particle sampling, transport, and relaxation, so that individual
phonon scattering events need not be resolved in the diffusive regime. The
underlying principle is the same as in radiative and gas-dynamic UGKWP methods,
but the continuum limit is thermal diffusion rather than fluid dynamics.

Synthetic acceleration has also been extended to phonon transport. Liu
\emph{et al.}~\cite{Liu_Zhang_Yuan_GSIS_Phonon_2022} proposed a fast-converging
scheme for the phonon Boltzmann equation with dual relaxation times. The
method couples the kinetic phonon equation under Callaway's model with
synthetic macroscopic equations for temperature and heat flux. The macroscopic
system propagates the slow diffusive information efficiently, while the
kinetic solver supplies higher-order nonequilibrium corrections. As in gas
dynamics, the purpose is not to replace the kinetic equation by a heat equation
everywhere, but to use the correct macroscopic structure to accelerate
convergence and preserve the small-Knudsen-number behavior on meshes larger
than the phonon mean free path.

Coupled electron--phonon transport introduces another level of nonequilibrium.
Under ultrafast heating or nanoscale energy deposition, electrons and phonons
may not share a common temperature, and the inter-carrier energy exchange
becomes an essential part of the thermal relaxation process. Zhang
\emph{et al.}~\cite{Zhang_DUGKS_ElectronPhonon_2024} developed a DUGKS for
electron--phonon nonequilibrium heat conduction in which electron and phonon
advection, scattering, and inter-carrier coupling are treated within one
finite-volume kinetic evolution. This extension shows that multiscale kinetic
methods can also address coupled carrier systems in which several relaxation
pathways and nonequilibrium temperatures coexist.

Overall, phonon and electron--phonon transport demonstrate the portability of
multiscale kinetic methodology to thermal systems. The relevant carrier is no
longer a gas molecule, the equilibrium state is thermal rather than
Maxwellian in molecular velocity, and the continuum limit is heat conduction
rather than Euler or Navier--Stokes dynamics. Nevertheless, the same numerical
principles remain central. Transport and scattering should be coupled over the
numerical evolution scale, macroscopic heat-conduction information should be
used in strongly scattering regimes, and the representation should adapt
between ballistic, quasi-ballistic, and diffusive transport.
\subsection{Gas--particle and granular flows}
\label{subsec:sec8_multiphase}
Gas--particle and granular flows provide a different type of extension from
radiative, neutron, plasma, and phonon transport. In these systems, the carrier
gas and the dispersed phase may belong to different flow regimes. The gas
phase is often well described by a continuum Navier--Stokes or gas-kinetic
hydrodynamic solver, while the dispersed phase may range from nearly
collisionless particle motion to collision-dominated collective behavior. The
relevant multiscale parameter is therefore not only the rarefaction of the gas,
but also the particle-phase Knudsen number, Stokes number, volume fraction,
and inter-particle collision time.

A schematic kinetic equation for the dispersed phase may be written as
\begin{equation}
	\partial_t f_p
	+\bm{v}\cdot\nabla f_p
	+\nabla_{\bm{v}}\cdot
	\left(\bm{a}_p f_p\right)
	=
	\mathcal{C}_p(f_p)+S_{gp},
	\label{eq:sec8_particle_phase_kinetic}
\end{equation}
where \(f_p(\bm{x},\bm{v},t)\) is the particle distribution function,
\(\bm{v}\) is the particle velocity, \(\bm{a}_p\) denotes acceleration due to
drag, gravity, or other forces, \(\mathcal{C}_p\) represents inter-particle
collisions or granular relaxation, and \(S_{gp}\) represents
gas--particle exchange and other coupling terms. In dilute regimes, particle
trajectory crossing and free transport are important, and an
Eulerian--Lagrangian description is natural. In dense or collision-dominated
regimes, the dispersed phase can approach a hydrodynamic or two-fluid
description. Between these limits, neither a purely trajectory-based method nor
a purely continuum particle-phase model is uniformly efficient.

This makes gas--particle flow a natural setting for wave--particle
representations. In the coupled GKS--UGKWP framework, the gas phase is evolved
by a gas-kinetic Navier--Stokes solver, while the solid-particle phase is
represented by UGKWP \cite{ref:UGKWP_Multiphase2022_CiCP,Yang_Shyy_Xu_GKS_UGKWP_POF2022}. The wave--particle
decomposition is applied to the dispersed phase. Collisionless or weakly
collisional particle motion is represented by particles, whereas the
collision-dominated part is described through a wave or continuum contribution.
As a result, the same formulation can connect the Eulerian--Lagrangian
description in dilute regimes with the Eulerian--Eulerian two-fluid
description in dense regimes.

This framework has been extended to more complex gas--solid configurations,
including three-dimensional gas--particle fluidized beds and polydisperse
gas--solid particle flows \cite{Yang_Shyy_Xu_JFM2024}. These applications are
important because they show that the wave--particle idea is not limited to
molecular or photon transport. It can also serve as an adaptive representation
for a dispersed phase whose local behavior changes between trajectory
transport, inter-particle collision, and collective hydrodynamic response.
A related branch is the UGKP method for dilute granular flow developed by Wang
and Yan \cite{Wang_Yan_UGKP_Granular_2020}, where the particle method is
constructed to describe granular transport while retaining the multiscale
spirit of unified gas-kinetic particle evolution.

The gas--particle extension broadens the meaning of multiscale unification.
The transition is not simply from a rarefied gas to a continuum gas, nor from
ballistic transport to diffusion. Instead, the method must unify different
computational descriptions of the dispersed phase. Depending on local particle
collisionality and coupling strength, the same framework should behave as an
Eulerian--Lagrangian particle method, an Eulerian--Eulerian two-fluid method,
or an intermediate wave--particle method. This illustrates a more general
lesson for nonequilibrium transport, namely that multiscale kinetic methodology is not
defined by one particular carrier or one particular macroscopic limit, but by
the scale-adaptive organization of transport, interaction, and continuum
response.
\subsection{Section remarks}
\label{subsec:sec8_portability}
The extensions reviewed above show that multiscale kinetic methodology has
become a broader approach to nonequilibrium transport beyond rarefied-gas
dynamics. Its portability is not based on a universal collision operator or a universal
equilibrium distribution, but on a common structural requirement. A transported carrier distribution must be evolved together with
interaction, relaxation, scattering, field coupling, or interphase exchange,
and the numerical representation must remain valid when the interaction scale
is much smaller than the macroscopic observation scale.

Several transferable principles can be identified. First, transport and
interaction should be coupled over the numerical evolution scale whenever the
cell size and time step are not tied to the microscopic interaction length and
time. This is the role played by UGKS- and DUGKS-type fluxes in gas dynamics,
and by their analogues in radiative, neutron, phonon, plasma, and
electron--phonon transport. Second, AP and UP viewpoints provide criteria for
testing whether a scheme recovers the correct limiting behavior and preserves
the relevant asymptotic structure under coarse kinetic resolution. Third,
synthetic and moment-coupled strategies use lower-dimensional macroscopic
equations to propagate slow continuum or diffusive information efficiently.
Fourth, wave--particle representations provide an adaptive way to separate
ballistic or weakly interacting components from strongly interacting
components within one computation. Finally, the observation-scale viewpoint
suggested by UGKF offers a physical interpretation of why different
representations should be used for different collision or interaction histories.

At the same time, these extensions also reveal clear limitations. The meaning
of equilibrium, conservation, dissipation, and continuum closure is
problem-dependent. In radiative and neutron transport, the continuum limit is
usually diffusive. In phonon transport, it is thermal and may involve multiple
relaxation channels. In plasma transport, electromagnetic phase-space dynamics,
charge neutrality, multiple species, and several characteristic scales must be
handled simultaneously. In gas--particle systems, the multiscale transition may
belong primarily to the dispersed phase rather than to the carrier gas. Thus a
method that is successful for molecular gases cannot be transferred by simply
replacing symbols in the governing equation.

The interaction operators also differ substantially. Some conserve mass,
momentum, and energy; some conserve only selected moments; others contain
absorption, emission, fission, ionization, recombination, or inter-carrier
energy exchange. As a result, the macroscopic variables
\(\bm{U}=\langle\bm{\phi}f\rangle\), the balance relation
\(\langle\bm{\phi}\mathcal{C}(f)\rangle\), and the closure of the flux
\(\langle\bm{c}\bm{\phi}f\rangle\) must be redesigned for each physical system.
Boundary conditions, material interfaces, source terms, positivity, entropy or
energy consistency, and discrete conservation are likewise system-specific.
These issues become especially delicate in strongly heterogeneous media and in
problems with multiple relaxation pathways.

Another limitation concerns assessment. For gas dynamics, AP and UP analysis
can be connected to Euler, Navier--Stokes, and Chapman--Enskog structures. In
other transport systems, the corresponding asymptotic hierarchy may be
diffusive, thermal, electromagnetic, or multiphase, and may not have a direct
analogue of the gas-dynamic Chapman--Enskog expansion. Meaningful multiscale
assessment therefore requires problem-dependent limiting models and benchmark
tests. A scheme should be examined not only in the two extreme limits, but also
in intermediate regimes where ballistic and continuum effects coexist and where
the numerical scale is larger than the microscopic interaction scale.

The main lesson is therefore balanced. Multiscale kinetic ideas are highly
portable at the level of structure, but not at the level of a single formula.
They provide a framework for organizing transport, interaction, macroscopic
response, and scale-adaptive representation across different carrier systems.
Their successful use depends on respecting the physical content of each
transport problem, including its conserved quantities, relaxation mechanisms,
limiting equations, and observable scales. This perspective closes the
discussion of carrier-based nonequilibrium transport systems.
\section{Kinetic approaches to turbulence}
\label{sec:kinetic_turbulence}
Secs.~\ref{sec:deterministic_methods}--\ref{sec:extensions_other_transport} have reviewed multiscale kinetic methods for
nonequilibrium transport systems in which carrier distributions evolve through
transport, interaction, relaxation, and macroscopic response. Turbulence
presents a related but distinct multiscale problem. Its unresolved dynamics are
not associated with a separate physical carrier population, but with
hydrodynamic fluctuations generated by the nonlinear Navier--Stokes equations.
Moreover, turbulent flows generally lack a clean separation between the mean or
resolved motion and the fluctuating eddies. This feature is one of the central
reasons why local closures formulated directly at the Navier--Stokes moment
level may become insufficient for complex turbulent flows.

A kinetic viewpoint offers a different way to organize this difficulty. Instead of closing only a few low-order moments through an immediate
constitutive relation, a distribution function for turbulent or fluid-element
velocities may be introduced, so that the Reynolds stresses, turbulent kinetic
energy, turbulent transport fluxes, and higher-order correlations are
interpreted as its moments. This provides a natural framework for representing finite-scale transport, relaxation,
nonequilibrium response, nonlocality, and history dependence. However, the
transfer of kinetic ideas to turbulence requires a reinterpretation of the
kinetic variables. The relevant ``particles'' are fluid elements, eddy-like
parcels, or subgrid turbulent structures rather than gas molecules.

This section reviews how kinetic ideas have been used to develop alternative
perspectives on turbulence modeling and computation.
The aim is not to replace established direct numerical simulation (DNS), large eddy simulation (LES), Reynolds-averaged Navier-Stokes (RANS),
or Reynolds-stress-transport methodologies, but to clarify the potential of
kinetic variables, moment closures, relaxation models, nonequilibrium
corrections, and wave--particle representations for describing turbulent
fluctuations and their effects on mean or resolved flows.

\subsection{Multiscale turbulence and kinetic viewpoints}
\label{subsec:turbulence_kinetic_viewpoints}
Turbulence is intrinsically multiscale. The range of dynamically active scales
increases rapidly with the Reynolds number, so DNS becomes prohibitively expensive when all relevant scales must be resolved. LES
and RANS reduce this cost by filtering or averaging the flow field, but they
introduce unclosed subgrid stresses, Reynolds stresses, turbulent
kinetic-energy fluxes, and higher-order correlations. In this respect,
turbulence shares the central difficulty discussed throughout this review, namely that a
coarse observation scale must represent the cumulative effect of unresolved
transport, interaction, and relaxation. The distinction is that the unresolved
motions are hydrodynamic fluctuations and eddy-like structures, rather than
physical carriers.

The appeal of a kinetic viewpoint is that turbulence closure is no longer
restricted to a local relation between the Reynolds stress and the mean strain
rate. Instead, a distribution function carrying information beyond the lowest
moments is employed, so that linear eddy viscosity appears only as a
leading-order approximation, while nonlinear stress response, finite relaxation,
and higher-order turbulent transport can be organized systematically. The
classical analogy between molecular momentum transport and turbulent mixing
already points in this direction. Extended Boltzmann and BGK-type formulations
make this analogy more explicit by treating turbulent relaxation at the kinetic
level and then projecting the resulting model onto macroscopic moments
\cite{ChenHD_2003Science,chen2004expanded}.

\begin{figure}[t]
	\centering
	\includegraphics[width=0.95\textwidth]{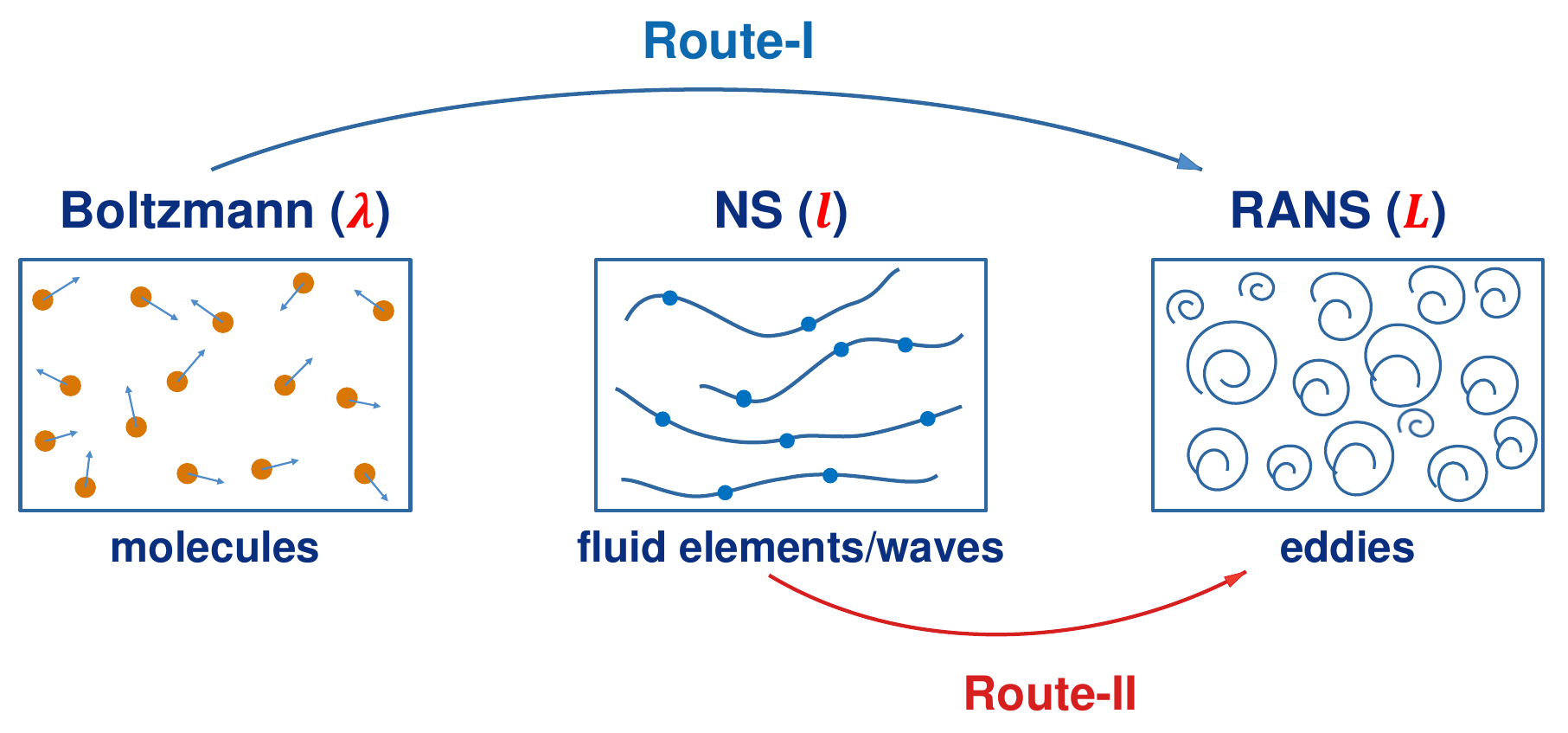}
	\caption{Two routes for kinetic modeling of turbulence. Route-I constructs turbulent kinetic models from molecular kinetic theory or its analogy. Route-II starts from the Navier--Stokes dynamics of fluid elements. $\lambda$, $l$, and $L$ are typical lengths of the kinetic, hydrodynamic, and ensemble scales, respectively.}
	\label{fig:turbulence_routes}
\end{figure}

Two conceptual routes can be distinguished for kinetic modeling of turbulence, as illustrated in Fig. \ref{fig:turbulence_routes}. The first starts from molecular kinetic
theory or from an analogy with molecular gases. Turbulent effects are then
encoded through modified equilibria, turbulent relaxation times, filtered
kinetic equations, lattice-Boltzmann closures, or Fokker--Planck-type models.
This route is valuable because it connects turbulence closures with the structure of
kinetic theory \cite{ChenHD_2003Science,chen2004expanded,Girimaji2007PRL,MalaspinasSagaut2012JFM,luan2025fokkerplanck}.
Its limitation is also clear because molecular thermal fluctuations and turbulent
hydrodynamic fluctuations have different physical origins \cite{chen2023average}, so a molecular
Boltzmann equation is not automatically a first-principles theory of turbulent
fluctuations.

The second route starts from the dynamics of fluid elements governed by the
Navier--Stokes equations. In this route, the distribution function describes
fluid-element velocities, eddy-like parcel velocities, or subgrid turbulent
structures rather than molecular velocities. Lundgren's one-point distribution
model and Klimontovich-type kinetic formulations of averaged turbulence dynamics
belong to this category \cite{Lundgren1969POF,chen2023average,xin2026kinetic}. In these models,
turbulent kinetic energy plays a role analogous to temperature, but it is not
identified with molecular thermal energy. Averaged turbulent flow is instead
viewed as a finite-relaxation or finite-Knudsen-number system, for which
nonequilibrium corrections beyond the linear eddy-viscosity approximation can
be organized through kinetic moments and asymptotic analysis.

Wave--particle turbulent simulation (WPTS) can be viewed as a direct modeling following this second viewpoint. Its particles represent unresolved fluid
elements or subgrid turbulent structures, while the wave component carries the
resolved flow field. The wave--particle decomposition adapts to local grid
resolution and turbulence intensity, so that well-resolved or laminar regions
are advanced mainly by the continuum wave component, whereas under-resolved
turbulent regions use stochastic particles to transport subgrid
nonequilibrium information \cite{YangXu2025WPTSGeneral,yang2025wpts,yang2026wpts}. In this sense,
WPTS connects the direct-modeling philosophy of UGKS and UGKWP with turbulence
simulation, while giving the particle representation a hydrodynamic rather than
molecular interpretation.

\subsection{Kinetic analogies and closure-generating models}
\label{subsec:turbulence_closure_models}
One line of kinetic turbulence modeling starts from molecular kinetic theory,
or from an analogy with molecular kinetic theory, and uses the resulting
structure to generate turbulence closures. A representative model has the
relaxation form
\begin{equation}
	\partial_t f+\bm{\xi}\cdot\nabla f
	=
	-\frac{1}{\tau_t}\left(f-f^{\rm eq}\right),
	\label{eq:turbulence_bgk_analogy}
\end{equation}
where \(f=f(\bm{x},\bm{\xi},t)\) is no longer interpreted literally as a
molecular distribution in the usual thermodynamic sense. Instead, its local
spread in velocity space is associated with turbulent fluctuations, and
\(\tau_t\) is a turbulent relaxation time. This idea underlies the extended
Boltzmann kinetic equation of Chen \emph{et al.}\
\cite{ChenHD_2003Science}, where complex
turbulent physics is modeled at the kinetic level and then projected to
macroscopic variables. In the near-equilibrium limit, the kinetic model
recovers eddy-viscosity behavior. Away from this limit, the finite relaxation
time provides a way to encode nonequilibrium corrections that are difficult to
represent by a purely local constitutive law.

The expanded analogy of Chen \emph{et al.} \cite{chen2004expanded} made this connection more explicit. The Reynolds stress, $\sigma_{ij}=-\left\langle C_i C_j f\right\rangle$, is estimated from the Chapman--Enskog expansion of a BGK-type turbulent kinetic model. At first order, 
\begin{equation}
	\sigma^{(1)}_{ij}=2\nu_{t}S_{ij},
	\qquad
	\nu_{t}=\frac{2}{3}K\tau_{t},
	\label{eq:turbulence_first_order_stress}
\end{equation}
where \(\bm{C}=\bm{\xi}-\bm{U}\) is now the fluid-element velocity fluctuation relative to the mean flow $\bm{U}$, \(K\) is the
turbulent kinetic energy, and  $S_{ij}=\left(\partial_i U_j+\partial_j U_i\right)/2$ is the mean strain-rate tensor. Thus the linear eddy-viscosity model appears
as the first nonequilibrium correction of the kinetic description. At second
order, the same expansion yields nonlinear and non-Newtonian contributions, for example
\begin{align}
	\sigma^{(2)}_{ij}
	=&
	-2\nu_tD_t\!\left(\tau_tS_{ij}\right)
	-\frac{6\nu_t^2}{K}
	\left(
	S_{ik}S_{kj}
	-\frac{1}{3}\delta_{ij}S_{kl}S_{kl}
	\right)
	\nonumber\\
	&+
	\frac{3\nu_t^2}{K}
	\left(
	S_{ik}\Omega_{kj}+S_{jk}\Omega_{ki}
	\right),
	\label{eq:turbulence_second_order_stress}
\end{align}
where $D_t=\partial_t+\bm{U}\cdot\nabla$, and $\Omega_{ij}=\left(\partial_i U_j-\partial_j U_i\right)/2$.
These terms contain a memory effect through \(D_t(\tau_t S_{ij})\)
and nonlinear strain--rotation interactions. Their tensorial structure is
closely related to that of established nonlinear eddy-viscosity models. This
is an important message of the kinetic analogy: nonlinear turbulence closures
may be interpreted as higher-order nonequilibrium corrections in a kinetic
expansion, rather than only as empirical tensorial additions.

Filtered kinetic approaches form a related but distinct branch. Girimaji
\cite{Girimaji2007PRL} derived a Boltzmann-type equation for filtered
turbulence and emphasized that direct filtering of the molecular distribution
places the unresolved turbulent effect implicitly inside the averaged collision
operator. This makes the kinetic equation difficult to reconcile with the
Reynolds-stress structure of the filtered Navier--Stokes equations. By
transforming the velocity variable relative to the unresolved fluctuation,
Girimaji obtained a filtered kinetic equation in which the Reynolds or subgrid
stress appears explicitly through additional drift terms. This formulation
clarifies how turbulence effects enter kinetic and macroscopic descriptions,
although the Reynolds or subgrid stress itself still requires closure.

In the lattice-Boltzmann and filtered-BGK formulation, Malaspinas and Sagaut
\cite{MalaspinasSagaut2012JFM} developed a consistent subgrid-scale modeling
framework based on Hermite expansion of the filtered distribution. Their
analysis identifies a hierarchy of subgrid terms and clarifies the common
practice of modifying the relaxation time by an eddy viscosity. This provides
a systematic kinetic embedding of LES closures, while also showing that a
single scalar relaxation-time correction is generally insufficient for more
general compressible nonequilibrium effects.

Fokker--Planck formulations provide another closure-generating route. Luan
\emph{et al.}~\cite{luan2025fokkerplanck} proposed a turbulent Fokker--Planck
model by analogy with Brownian motion. In schematic form,
\begin{equation}
	\partial_t f
	+\bm{\xi}\cdot\nabla f
	+\bm{a}\cdot\nabla_{\bm{\xi}}f
	=
	\frac{1}{\tau_t}
	\nabla_{\bm{\xi}}\cdot
	\left(\bm{C}f\right)
	+
	\frac{2K}{3\tau_t}
	\nabla_{\bm{\xi}}\cdot(\nabla_{\bm{\xi}}f) .
	\label{eq:turbulence_fokker_planck}
\end{equation}
The first moment recovers the RANS momentum equation, while higher
moments yield Reynolds-stress transport and constitutive relations. Through
Chapman--Enskog analysis, the model recovers linear eddy viscosity at first
order and produces quadratic nonlinear stress expressions at second order.
This again shows how kinetic formulations can generate conventional and
nonlinear turbulence closures from a unified moment structure.

Overall, these closure-generating models demonstrate the usefulness of
kinetic theory as a structured modeling framework. They clarify how
eddy-viscosity, nonlinear stress response, memory effects, and subgrid
corrections can arise from kinetic moments and relaxation. Their limitation is
also clear. When the starting point is the molecular Boltzmann equation or a
molecular analogy, one must ensure that molecular thermal fluctuations are not
confused with turbulent hydrodynamic fluctuations. This concern motivates a
second route, in which the kinetic description is built directly from the
dynamics of fluid elements.

\subsection{Kinetic description of turbulent fluid elements}
\label{subsec:turbulence_fluid_elements}

A more intrinsic kinetic route is to describe the statistics of fluid elements
rather than molecules. This idea has a long history. Lundgren
\cite{Lundgren1969POF} derived and modeled equations for turbulent distribution
functions, and closed the one-point equation by introducing a relaxation model
for the pressure-fluctuation term. The resulting model resembles a BGK
equation for turbulent velocity fluctuations and was applied to idealized
nonhomogeneous turbulent flows. Although the model required additional
information such as the dissipation rate, it already contained the key idea
that turbulent transport can be described through a velocity distribution of
fluid elements.

A recent more systematic formulation was developed by Chen
\emph{et al.}~\cite{chen2023average}. The starting point is a
Klimontovich-type kinetic representation of incompressible fluid elements,
\begin{equation}
	\partial_t f
	+\bm{\xi}\cdot\nabla f
	+\bm{a}\cdot\nabla_{\bm{\xi}}f
	=0,
	\qquad
	\bm{a}=-\nabla p+\nu_0\nabla^2\bm{u}.
	\label{eq:turbulence_klimontovich}
\end{equation}
Here the fine-grained distribution is monokinetic, $f=\delta(\bm\xi-\bm u(\bm x,t))$, with $\langle f\rangle=1$ and $\langle\bm\xi f\rangle=\bm u$. $p$ is the pressure, $\nu_0$ is the molecular kinematic viscosity, and $\nabla\cdot\bm u=0$. With this monokinetic constraint, the zeroth and first moments of Eq.~\eqref{eq:turbulence_klimontovich} give the incompressible Navier--Stokes equations exactly. Thus the kinetic equation is not imposed as a molecular model from outside; rather, it is a phase-space
representation of the motion of fluid elements governed by the incompressible Navier--Stokes
dynamics, and in this sense it provides a first-principles kinetic representation at the incompressible
continuum level.

The closure problem enters only after this exact fine-grained representation is
ensemble-averaged. Averaging Eq.~\eqref{eq:turbulence_klimontovich} gives a one-point
kinetic equation for the averaged velocity distribution \(F=\overline f\),
\begin{equation}
	\partial_t F
	+\bm{\xi}\cdot\nabla F
	+\overline{\bm{a}}\cdot\nabla_{\bm{\xi}}F
	=\mathcal{C},
	\label{eq:turbulence_averaged_kinetic}
\end{equation}
where \(\overline{\bm{a}}=-\nabla{\bar{p}}+\nu_0\nabla^2\bm{U}\) denotes the mean acceleration induced by the
Navier--Stokes dynamics, and \(\mathcal C\) represents the unclosed effect of correlations between the
fluctuating acceleration and the fluctuating distribution. This collision-like
term conserves mass and momentum,
\begin{equation}
	\langle \mathcal{C}\rangle=0,
	\qquad
	\langle \bxi \mathcal{C}\rangle=0,
	\label{eq:turbulence_collision_conservation}
\end{equation}
but it is not conservative with respect to turbulent kinetic energy. This is
consistent with the fact that turbulent kinetic energy is produced, transported,
and dissipated by the turbulent dynamics.

The mean velocity, turbulent kinetic energy, Reynolds stress, and turbulent
energy flux are then obtained from moments of \(F\),
\begin{equation}
	\bm{U}=\overline{\bm{u}}=\langle \bxi F\rangle, \quad K=\dfrac{1}{2}\langle|\bm{C}|^2 F\rangle,\quad \bm{\sigma}=-\langle \bm{C}\bm{C} F\rangle, \quad \bm{Q}=	\dfrac{1}{2}\langle \bm{C}|\bm{C}|^2 F\rangle.
\end{equation}
The corresponding moment system is the incompressible RANS-type system,
\begin{subequations}
	\label{eq:turbulence_rans_moment_system}
	\begin{align}
		\nabla\cdot\bm{U}
		&=0,
		\\
		\partial_t\bm{U}
		+\bm{U}\cdot\nabla\bm{U}
		&=
		-\nabla\overline{p}
		+\nu_0\nabla^2\bm{U}
		+\nabla\cdot\bm{\sigma},
		\\
		\partial_t K
		+\bm{U}\cdot\nabla K
		&=
		-\nabla\cdot\bm{Q}
		+\bm{\sigma}:\bm{S}
		+\mathcal{D}_K-\epsilon .
	\end{align}
\end{subequations}
Here \(\epsilon\) is the dissipation rate, and \(\mathcal{D}_K\) denotes additional transport effects
such as pressure diffusion and molecular diffusion. The closure problem is now
shifted from an assumed Reynolds-stress law to the modeling of the collision-like
operator \(\mathcal C\) in Eq.~\eqref{eq:turbulence_averaged_kinetic}.

Chen \emph{et al.}~\cite{chen2023average} proposed a BGK-type approximation,
\begin{equation}
	\mathcal{C}_{\rm BGK}(F)
	=
	\frac{F^{\rm eq}-F}{\tau_t},
	\label{eq:turbulence_bgk_fluid_element}
\end{equation}
where \(F^{\rm eq}\) is a Gaussian distribution centered at \(\bm{U}\),
\begin{equation}
	F^{\rm eq}
	=
	\left(\frac{3}{4\pi K_{\rm eq}}\right)^{3/2}
	\exp\left[
	-\frac{3|\bm{\xi}-\bm{U}|^2}{4K_{\rm eq}}
	\right].
	\label{eq:turbulence_gaussian_equilibrium}
\end{equation}
For homogeneous high-Reynolds-number turbulence, the energy
moment of the collision term gives
\begin{equation}
	\frac{K_{\rm eq}-K}{\tau_t}=-\epsilon,
	\qquad
	K_{\rm eq}=K-\tau_t\epsilon .
	\label{eq:turbulence_Keq_relation}
\end{equation}
Thus the turbulent kinetic energy enters as a velocity-space variance, but it
is not identified with molecular thermal temperature. This distinction is a
central advantage of the fluid-element route over a direct averaging of the
ordinary molecular Boltzmann equation.

A central modeling issue is the choice of \(\tau_t\). Chen
\emph{et al.}~\cite{Chen2024OneParameterTau} estimated
\(\tau_t=C_\tau K/\epsilon\) with \(C_\tau=6/7\) for stationary homogeneous
isotropic turbulence from a fluctuation--dissipation argument. Combined with
Eq.~\eqref{eq:turbulence_Keq_relation}, the first-order Chapman--Enskog
analysis gives
\begin{equation}
	\nu_T
	=
	\frac{2}{3}\tau_t K_{\rm eq}
	=
	\frac{2}{3}C_\tau(1-C_\tau)\frac{K^2}{\epsilon}.
	\label{eq:turbulence_tau_choice_viscosity}
\end{equation}
Xin \emph{et al.}~\cite{xin2026kinetic} observed that the same eddy viscosity
is obtained from the alternative root \(C_\tau=1/7\), while the associated
turbulent Prandtl number and higher-order transport coefficients become more
consistent with standard turbulence modeling. They therefore adopted
\begin{equation}
	\tau_t=\frac{1}{7}\frac{K}{\epsilon},
	\label{eq:turbulence_tau_xin_choice}
\end{equation}
which preserves the eddy-viscosity level of the earlier estimate but improves
the turbulent diffusion and higher-order nonequilibrium closures.

Xin \emph{et al.}~\cite{xin2026kinetic} further performed a Chapman--Enskog analysis of the turbulent BGK model. It is shown that
the first-order hydrodynamic limit recovers the linear eddy-viscosity model for
the Reynolds stress and the gradient-diffusion model for the turbulent
kinetic-energy flux. At second order, the model produces nonlinear stress and
turbulent-flux corrections, including terms associated with strain history,
finite relaxation, and higher-order gradients. This supports the interpretation
of averaged turbulent flow as a finite-relaxation, or finite-Knudsen-number,
system in which the linear eddy-viscosity model is only the leading
near-equilibrium approximation.

The same work also extends the model to wall-bounded turbulence. In the
near-wall region, turbulent fluctuations are damped and molecular viscosity
becomes important. This is handled by modifying the relaxation time through a
wall damping function and by adding a source term whose moments reproduce
molecular diffusion of turbulent kinetic energy. In schematic form,
\begin{equation}
	\partial_t F
	+\bm{\xi}\cdot\nabla F
	+\overline{\bm{a}}\cdot\nabla_{\bm{\xi}}F
	=
	\mathcal C_{\rm BGK}(F)+S,
	\label{eq:turbulence_wall_bounded_kinetic}
\end{equation}
with
\begin{equation}
	\langle S\rangle=0,
	\qquad
	\langle \bxi S\rangle=0,
	\qquad
	\frac{1}{2} \langle |\bm{C}|^2 S\rangle
	=
	\nu_0\nabla^2K .
	\label{eq:turbulence_wall_source_moments}
\end{equation}
This construction allows both a low-Reynolds-number kinetic model resolving the
viscous sublayer and a high-Reynolds-number version with wall-function
treatment. The reported Couette-flow tests show good agreement for mean
velocity, skin friction, and Reynolds shear stress, while also indicating that
near-wall anisotropy remains a challenging issue \cite{xin2026kinetic}.

The fluid-element route is important because it assigns a clear hydrodynamic
meaning to the kinetic variables. The distribution function represents
fluid-element velocity statistics, the collision-like operator represents the
averaged effect of turbulent fluctuations, and turbulent kinetic energy is a
moment of the distribution rather than a thermodynamic temperature. The main
open issues are the self-consistent closure of the relaxation time,
dissipation rate, pressure diffusion, wall effects, and higher-order
nonequilibrium corrections for general inhomogeneous turbulent flows.

\subsection{Wave--particle turbulent simulation on unresolved grids}
\label{subsec:turbulence_wpts}
A recent computational development along the kinetic turbulence direction is
the wave--particle turbulent simulation (WPTS), which provides a direct-modeling
realization of the fluid-element kinetic viewpoint on unresolved grids. Its motivation is different
from that of a conventional eddy-viscosity closure. Instead of representing the
effect of unresolved turbulence only through a local modification of the
viscous stress, WPTS introduces a scale-adaptive wave--particle representation
for turbulent transport on coarse grids. The wave component describes the
resolved Eulerian flow field, while stochastic particles represent unresolved
fluid elements or subgrid turbulent structures. These particles are not gas
molecules. They are hydrodynamic carriers of unresolved turbulent motion, and
their velocity variance is associated with subgrid turbulent kinetic energy
\cite{YangXu2025WPTSGeneral,yang2025wpts,yang2026wpts}.

The kinetic model underlying WPTS may be written schematically as
\begin{equation}
	\partial_t f+\bxi\cdot\nabla f
	=
	\frac{g-f}{\tau_n},
	\qquad
	\tau_n=\tau+\tau_t ,
	\label{eq:turbulence_wpts_kinetic}
\end{equation}
where \(g\) is an equilibrium state constructed from the resolved macroscopic variables $\bm{W}=(\rho,\rho\bm{U}, \rho E)$ and turbulent kinetic energy $K$, \(\tau\) is the
molecular relaxation time, and \(\tau_t\) is a modeled turbulent characteristic
time. The equilibrium width includes molecular thermal energy as well as unresolved turbulent kinetic energy, so that both contribute to
the velocity-space spread of the distribution.

With $\tau_n$ frozen locally over the integration interval, the integral solution of Eq.~\eqref{eq:turbulence_wpts_kinetic} gives the
same transport--relaxation structure that underlies UGKS and UGKWP,
\begin{equation}
	f(\bm{x},\bxi,t)
	=
	\frac{1}{\tau_n}
	\int_0^t
	g(\bm{x}',\bxi,t')
	e^{-(t-t')/\tau_n}\,dt'
	+
	e^{-t/\tau_n}f_0(\bm{x}-\bxi t,\bxi),
	\label{eq:turbulence_wpts_integral}
\end{equation}
where \(\bm{x}'=\bm{x}+\bxi(t'-t)\). The first term represents the cumulative
relaxation contribution and is evolved through the wave component. The second
term represents finite-distance transport of the initial distribution and is
represented by particles when this nonequilibrium component remains important.
Thus the decomposition is not a fixed domain partition. It is determined by
the local balance between relaxation and transport over the numerical
evolution scale.

The particle implementation additionally uses a modeled relaxation-type
trajectory \cite{YangXu2025WPTSGeneral,yang2026wpts},
\begin{equation}
	\frac{\mathrm{d}\bm{x}_p}{\mathrm{d}t}=\bm{v}_p,
	\qquad
	\frac{\mathrm{d}\bm{v}_p}{\mathrm{d}t}
	=
	\frac{\bm{U}-\bm{v}_p}{\tau_n}+\bm{a},
	\label{eq:turb_wpts_particle}
\end{equation}
where $\x_p$ and $\v_p$ are the particle position and velocity, respectively, and \(\bm{a}\) denotes a modeled
acceleration, often dominated by the pressure-gradient contribution in current
implementations. Particles transport unresolved turbulent information over a
finite distance and then deposit their conservative quantities back into the
wave component. This gives WPTS a nonlocal transport mechanism that is absent
from a purely local eddy-viscosity model.

A key modeling element is the turbulent characteristic time \(\tau_t\), which
controls the survival probability, transport distance, and sampling fraction of
the particle component. In the round-jet formulation based on Prandtl's
mixing-length hypothesis, the turbulent viscosity and characteristic time are
modeled as
\begin{equation}
	\nu_t=(C_{\rm ml}l)^2|\bm{S}|,
	\qquad
	\tau_t=\frac{\rho\nu_t}{p},
	\label{eq:turbulence_wpts_mixing_tau}
\end{equation}
where \(l\) is a mixing length, \(C_{\rm ml}\) is a model coefficient, and
\(|\bm{S}|\) is the magnitude of the resolved strain rate
\cite{yang2026wpts}. This connection with mixing-length theory is physically
natural, since both descriptions involve fluid elements traveling a finite
distance before exchanging momentum with their surroundings. The difference is
that WPTS realizes this idea through explicit nonequilibrium transport of
stochastic particles, rather than only through a local turbulent viscosity.

The adaptive nature of WPTS is central to its multiscale behavior. In laminar
or sufficiently resolved regions, the particle fraction decreases and the
method approaches the underlying continuum gas-kinetic or Navier--Stokes
solver. In under-resolved turbulent regions, particles are generated to carry
subgrid nonequilibrium transport. The wave and particle components therefore
provide a unified computational description from laminar or well-resolved flow
to coarse-grid turbulent flow. This is consistent with the direct-modeling
viewpoint: the model is constructed on the numerical space--time scale rather
than by assuming that one fixed continuum closure is valid at all resolutions.

The reported jet simulations are encouraging. The WPTS calculation of a
spatially developing round jet at \(Re=5000\) reproduced the main qualitative
flow structures and quantitative turbulence statistics on a grid far coarser
than a DNS grid \cite{yang2025wpts}. The later mixing-length-based extension
tested round jets at \(Re=5000\) and \(Re=20000\), and recovered important
similarity features, including centerline velocity decay and self-similar
profiles in the fully developed turbulent region \cite{yang2026wpts}. These
results suggest that the wave--particle representation may provide a practical
route for coarse-grid turbulence simulation, especially in flows where
finite-scale turbulent transport is important.

A related deterministic development is the double time-relaxation kinetic
model (DtrKM) of Cao \emph{et al.}~\cite{cao2025double} for compressible
turbulence on unresolved grids. In DtrKM, the unresolved turbulent kinetic
energy \(K_{\rm utke}\) is introduced as an additional state variable through a
sample-space variable \(k_u\), and the relaxation process is organized as
\begin{equation}
	\partial_t f+\bxi\cdot\nabla f
	=
	\frac{f^{\rm eq}-f}{\tau+\tau_t}
	+
	\frac{g-f^{\rm eq}}{\tau^*}.
	\label{eq:turbulence_dtrkm}
\end{equation}
Here \(f^{\rm eq}\) is an intermediate turbulent equilibrium that carries
unresolved turbulent information, while \(g\) is the final Maxwellian
equilibrium. Through a first-order Chapman--Enskog projection, the model yields
a six-variable macroscopic system in which the first five equations govern
mass, momentum, and total energy, and the sixth equation governs
\(K_{\rm utke}\). DtrKM is therefore best viewed as a deterministic kinetic
SGS model for compressible LES, whereas WPTS is a stochastic wave--particle
transport model for unresolved turbulent structures.

From the perspective of this review, WPTS indicates a common
computational trend. Kinetic turbulence modeling on unresolved grids need not
be limited to adding eddy viscosity to the Navier--Stokes equations. It can
introduce mesoscopic variables, relaxation pathways, and particle transport
mechanisms that carry unresolved turbulent information at the numerical scale.
Among these developments, WPTS is particularly notable because it directly
combines finite-distance turbulent transport with a wave--particle
representation, providing a promising extension of the UGKWP philosophy from
rarefied-gas dynamics to turbulence simulation.

\subsection{Section remarks}
\label{subsec:turbulence_remarks}
The kinetic approaches reviewed in this section show that turbulence can be
studied from a mesoscopic viewpoint without identifying turbulent fluctuations
with molecular thermal motion. Molecular-kinetic analogies, filtered
Boltzmann equations, lattice-Boltzmann closures, Fokker--Planck models,
fluid-element distributions, and WPTS all attempt to organize unresolved
turbulent transport through distribution functions, moments, relaxation
processes, or particle transport. Their common motivation is the same
multiscale difficulty that appears throughout this review. A coarse observation
scale must represent unresolved motion without resolving all active small
scales.

At the same time, the physical status of these models is not uniform.
Molecular Boltzmann, filtered-BGK, and Fokker--Planck formulations are useful
closure-generating frameworks, especially for deriving eddy viscosity,
nonlinear stress corrections, and SGS terms. They do not, by themselves,
constitute a first-principles kinetic theory of turbulence. Fluid-element formulations have a closer connection to Navier--Stokes
turbulence, because the distribution function describes hydrodynamic fluid
elements rather than molecules. WPTS adds a computational realization of this
viewpoint by using particles to represent finite-distance unresolved turbulent
transport on coarse grids.

Several issues remain open. The relaxation time, turbulent dissipation,
pressure diffusion, near-wall behavior, anisotropic nonequilibrium correction,
and boundary conditions for velocity-space distributions still require robust
modeling. For WPTS, the turbulent characteristic time, particle sampling,
deposition, statistical variance, and grid dependence need further systematic
validation. For compressible, reacting, rotating, or multiphysics turbulence,
additional internal variables, energy pathways, and coupling mechanisms may be
needed.

There is also an open question at the level of multiscale assessment. In
rarefied-gas dynamics, AP and UP properties can be related to kinetic, Euler,
Navier--Stokes, and Chapman--Enskog asymptotics. For turbulence, the relevant
small scales are not molecular mean free paths and collision times, but
unresolved eddy-transport lengths, relaxation times, and coarse-graining
windows. A future analogue of AP or UP for kinetic turbulence models would need
to assess whether RANS, LES, rapid-distortion, wall-asymptotic, and
finite-scale nonequilibrium behaviors are preserved as grid scale and modeled
relaxation time change.

The present state of the field is therefore exploratory but promising. Kinetic
theory should not be presented as a completed replacement for DNS, LES, RANS,
or Reynolds-stress-transport models. Its value lies in providing a richer
representation of turbulent fluctuations, a systematic moment structure for
Reynolds stresses and higher-order correlations, and a finite-relaxation
framework for nonequilibrium and nonlocal turbulent transport. These features
make turbulence a natural future frontier for multiscale kinetic modeling and
computation.
\section{Summary and outlook}
\label{sec:summary_outlook}
This review has surveyed multiscale kinetic methods for nonequilibrium flow and
transport from a unified methodological perspective. Instead of treating
deterministic solvers, stochastic particle methods, hybrid decompositions,
synthetic acceleration, wave--particle schemes, asymptotic properties, and
theoretical framework as separate subjects, the discussion has organized
them into a connected structure. At the method level, kinetic states are
represented and evolved by deterministic velocity-space discretization,
stochastic particles, or mixed wave--particle descriptions. At the strategy
level, multiscale information is organized through micro--macro decomposition,
kinetic--fluid coupling, high-order/low-order coupling, GSIS-type synthetic
acceleration, deterministic--stochastic coupling, reduced-order structures, and
collision-history-based representations. At the property assessment level, AP and UP
criteria assess whether a discrete kinetic evolution remains faithful to the
correct limiting model and to the deeper asymptotic hierarchy when the kinetic
scale is not resolved. At the framework level, UGKF introduces the observation
scale into the physical description itself and organizes gas dynamics according
to collision-history populations.

Several broad conclusions emerge from this perspective. First, multiscale
kinetic computation has evolved far beyond its original role as a specialized
tool for rarefied gas dynamics. It now provides a general methodology for
transport problems in which macroscopic constitutive descriptions become
insufficient, while fully resolved kinetic simulation remains too expensive.
The most successful methods are not those that merely switch between kinetic
and continuum solvers, but those that reorganize the numerical evolution
according to the local competition among free transport, collision,
relaxation, and hydrodynamic response. This principle is particularly clear in the UGKS, DUGKS, UGKP,
and UGKWP families, where transport and interaction are coupled over the
numerical space--time scale rather than treated as independent processes.

Second, the distinction among method, strategy, property, and framework is
essential. Deterministic schemes, particle methods, and wave--particle
algorithms specify how the kinetic state is represented. Micro--macro, hybrid,
HOLO, synthetic, and deterministic--stochastic formulations specify how kinetic
and macroscopic information are coupled. AP and UP are not algorithmic families
by themselves, but assessment criteria for asymptotic and physical fidelity.
UGKF operates at a different level again by formulating gas transport in terms of the observation scale that determines which collision-history
populations are distinguishable. Keeping these levels separate helps avoid conceptual confusion and
clarifies the respective contributions of different developments.

Third, AP remains a foundational requirement, but it is not a complete measure
of multiscale quality. AP identifies whether a kinetic scheme
degenerates to a consistent macroscopic solver when the small kinetic parameter
vanishes at fixed mesh and time step. This is indispensable for all-regime
computation. However, many practical near-continuum simulations involve small
but finite Knudsen numbers, where Navier--Stokes-level transport coefficients,
numerical dissipation, and higher-order asymptotic corrections matter. The UP
viewpoint refines the assessment by determining how much of the asymptotic hierarchy
is preserved by the fully discrete scheme. Extending UP-type analysis to
UGKS-type, UGKWP-type, GSIS-type, adaptive-velocity, stochastic, and more
general kinetic solvers remains an important open direction.

Fourth, the wave--particle and direct-modeling developments indicate that
representation itself has become an active part of multiscale modeling. In
classical deterministic and particle methods, the representation is often fixed
in advance. In UGKWP-type methods, however, the kinetic state is represented by
particles or waves according to the local transport and relaxation behavior
over one time step. This adaptive representation is not simply a computational
device; it expresses the physical fact that free-streaming and strongly
collided populations play different roles at different observation scales. The
later kinetic representation of UGKWP and the UGKF formulation make this point
more explicit by linking numerical wave--particle decomposition to
collision-history-based kinetic descriptions.

At the framework level, UGKF suggests a broader way to view the
kinetic-to-continuum connection. Classical kinetic theory often emphasizes
singular limiting procedures from the Boltzmann equation to Euler or
Navier--Stokes equations. UGKF adds an observation-scale perspective in which
free transport, collision history, transitional behavior, and continuum
response are organized as different population balances of the same transport
process. This viewpoint does not replace classical kinetic theory or rigorous
hydrodynamic-limit analysis. Rather, it provides a constructive scale-dependent
structure for describing intermediate regimes between the kinetic and
hydrodynamic limits.

The extensions reviewed beyond rarefied-gas dynamics reinforce the same
message. Radiative transfer, neutron transport, phonon heat conduction,
electron--phonon nonequilibrium, plasma transport, and gas--particle or
granular flows involve different carriers, phase variables, interaction laws,
equilibria, conserved or relaxed quantities, and limiting equations. Yet they
share a common multiscale structure. Ballistic or trajectory-resolving
transport may coexist with diffusion, heat conduction, fluid behavior, or
collective continuum response. Interaction, scattering, relaxation, or
interphase exchange may be stiff. The appropriate representation depends on
the scale at which the transport process is observed and computed. The
portability of multiscale kinetic methodology therefore lies not in a universal
equation, but in transferable principles such as transport--interaction
coupling, asymptotic consistency, synthetic acceleration, and scale-adaptive
representation.

The kinetic approaches to turbulence discussed separately point to a related
but distinct frontier. Turbulence is not a carrier-transport extension in the
usual sense, because its unresolved dynamics arise from hydrodynamic
fluctuations generated by the Navier--Stokes equations rather than from a
separate microscopic population. Nevertheless, kinetic ideas provide a
promising framework for representing turbulent fluctuations, finite-scale
transport, nonlinear stress response, relaxation, memory effects, and
coarse-grid unresolved structures. Molecular-kinetic analogies,
fluid-element distribution functions, and
wave--particle turbulent simulations all indicate that kinetic modeling may
offer richer finite-scale and nonequilibrium information than closures written
solely at the Navier--Stokes moment level. This direction is still developing,
but it illustrates how multiscale kinetic thinking may influence turbulence
modeling and computation beyond traditional rarefied-flow applications.

Looking forward, progress is likely to proceed along three connected axes. The
first is algorithmic. Future methods must become more accurate, adaptive,
scalable, and robust for realistic multiscale applications involving complex
geometry, high-dimensional phase space, internal nonequilibrium, chemical
reaction, multiphysics coupling, stochastic noise, and strong nonequilibrium
boundaries. The second is analytical. AP, UP, modified-equation analysis,
stochastic error analysis, stability theory, conservation and entropy
properties, and discrete consistency need to be extended to broader classes of
schemes, especially wave--particle methods, synthetic solvers, adaptive
representations, and coupled multiphysics algorithms. The third is conceptual.
The observation-scale viewpoint suggests that multiscale kinetic computation is
not only the efficient solution of a kinetic equation across regimes. It is
also the organization of transport physics according to the scales on which
free motion, interaction, relaxation, fluctuation, and continuum response are
observed and represented.

The lasting significance of the developments reviewed here therefore lies not
in any single algorithm, but in a change of viewpoint. Multiscale kinetic
computation has become a scale-aware methodology for connecting microscopic
transport, mesoscopic kinetic evolution, and macroscopic continuum behavior.
Its future impact will depend on how successfully numerical algorithms,
asymptotic analysis, and physical modeling can be combined to describe
nonequilibrium transport and turbulent multiscale dynamics across the
increasingly complex range of scales encountered in science and engineering.

\section*{Acknowledgements}
This work was supported by the National Natural Science Foundation of China (Grant Nos. 12472290 and 92371197).

\bibliographystyle{elsarticle-num}
\bibliography{msRev_final}

\end{document}